\documentclass{IEEEtran}

\usepackage[english]{babel}
\usepackage[utf8]{inputenc}
\newcommand{\subparagraph}{}
\usepackage{titlesec}
\usepackage[T1]{fontenc}
\usepackage{xcolor}
\usepackage{cite}

\usepackage{amsmath,amssymb,mathrsfs,bbm}
\usepackage{graphicx}
\usepackage[colorinlistoftodos]{todonotes}
\usepackage[colorlinks=true, allcolors=blue]{hyperref}
\usepackage{multirow}
\usepackage[lined,linesnumbered,ruled]{algorithm2e}
\usepackage{booktabs}

\usepackage{psfrag,epsfig,graphics}
\usepackage{amsmath,amsthm,amssymb}
\usepackage{mathbbol}
\usepackage{amssymb}   

\usepackage{multirow,diagbox,makecell}

\usepackage{pifont}%
\newcommand{\cmark}{\ding{51}}%
\newcommand{\xmark}{\ding{55}}%

\usepackage[most]{tcolorbox}

\DeclareSymbolFontAlphabet{\amsmathbb}{AMSb}%

\newcommand{\cp}[1]{\ifmmode {\mathcal{#1}}\else ${\mathcal{#1}}$\fi}
	
\newcommand{\bA}{\boldsymbol{A}}

\newcommand{\bM}{\boldsymbol{M}}

\newcommand{\bS}{\boldsymbol{S}}

\newcommand{\bW}{\boldsymbol{W}}
\newcommand{\bX}{\boldsymbol{X}}
\newcommand{\bY}{\boldsymbol{Y}}

\newcommand{\ba}{\boldsymbol{a}}
\newcommand{\bb}{\boldsymbol{b}}

\newcommand{\bd}{\boldsymbol{d}}
\newcommand{\bm}{\boldsymbol{m}}

\newcommand{\be}{\boldsymbol{e}}

\newcommand{\by}{\boldsymbol{y}}

\newcommand{\bx}{\boldsymbol{x}}

\newcommand{\bpsi}{\boldsymbol{\psi}}

\newcommand{\btheta}{\boldsymbol{\theta}}

\newcommand{\bPsi}{\boldsymbol{\Psi}}

\newcommand{\cb}[1]{\boldsymbol{#1}}

\newcommand{\diag}{\operatorname{diag}}

\newcommand{\Mlib}{\boldsymbol{M}_{\!\textsf{Lib}}}

\usepackage{color}  %

\def\cred{\textcolor{red}}

\def\cblue{\textcolor{blue}}
\def\cmag{\textcolor{magenta}}
\definecolor{darkgreen}{rgb}{0.0, 0.85, 0.0}
\newcommand{\cdgreen}{\textcolor{darkgreen}}
\definecolor{darkgray}{rgb}{0.0, 0.0, 0.0}

\definecolor{red}{rgb}{0.0, 0.0, 0.0}

\renewcommand{\cdgreen}{\textcolor{red}}
\renewcommand{\cmag}{\textcolor{red}}

\newcommand{\enhancedcoloredtbox}[2]{
\begin{tcolorbox}[enhanced,attach boxed title to top center={yshift=-3mm,yshifttext=-1mm},
  colback=blue!5!white,colframe=blue!75!black,colbacktitle=orange!5!orange,
  title=#1,fonttitle=\bfseries,
  boxed title style={size=small,colframe=orange!50!black} ] #2
\end{tcolorbox}}

\usepackage{mdframed}

\title{Spectral Variability in Hyperspectral Data Unmixing: A Comprehensive Review}

\author{Ricardo~Augusto~Borsoi, 
Tales~Imbiriba,
Jos\'e~Carlos~Moreira~Bermudez, 
Cédric~Richard,
Jocelyn~Chanussot,
Lucas~Drumetz,
Jean-Yves~Tourneret,
Alina~Zare,
Christian~Jutten}

\begin{document}
\maketitle

\begin{abstract}
\cblue{The final version of this paper can be found in the IEEE Geoscience and Remote Sensing Magazine.} 
The spectral signatures of the materials contained in hyperspectral images, also called endmembers (EM), can be significantly affected by variations in atmospheric, illumination or environmental conditions typically occurring within an image. Traditional spectral unmixing (SU) algorithms neglect the spectral variability of the endmembers, what propagates significant mismodeling errors throughout the whole unmixing process and compromises the quality of its results.
Therefore, large efforts have been recently dedicated to mitigate the effects of spectral variability in SU. This resulted in the development of algorithms that incorporate different strategies to allow the EMs to vary within a hyperspectral image, using, for instance, sets of spectral signatures known \textit{a priori}, Bayesian, parametric, or local EM models. Each of these approaches has different characteristics and underlying motivations.
This paper presents a comprehensive literature review contextualizing both classic and recent approaches to solve this problem. We give a detailed evaluation of the sources of spectral variability and their effect in image spectra. Furthermore, we propose a new taxonomy that organizes existing \cmag{works} according to a practitioner's point of view, based on the necessary amount of supervision and on the computational cost they require. We also review methods used to construct spectral libraries (which are required by many SU techniques) based on the observed hyperspectral image, as well as algorithms for library augmentation and reduction.
Finally, we conclude the paper with some discussions and an outline of possible future directions for the field.
\end{abstract}

\section{Introduction}

Hyperspectral cameras are able to sample electromagnetic spectra at hundreds of contiguous wavelength intervals. 
The high spectral resolution of hyperspectral images makes them an important tool for the precise identification and discrimination of different materials in a scene. Hyperspectral images contribute significantly to different fields and are now at the core of a vast number of applications such as space exploration~\cite{kouyama2016development}, land-use analysis, mineral detection, environment monitoring, field surveillance~\mbox{\cite{Bioucas-Dias-2013-ID307,Manolakis:2002p5224}}, disease diagnosis and \mbox{image-guided surgery~\cite{lu2014medical}.}

Notwithstanding the advantages brought forth by their high spectral resolution, hyperspectral cameras operate on a delicate trade-off between spatial resolution and signal-to-noise ratio. 
This happens since the light observed at the sensor is decomposed into several spectral bands, which in turn demands the pixel size to be large enough to attain an acceptable signal-to-noise ratio.
When combined with a large target-to-sensor distance, which is common in many applications, this leads to images with low spatial resolution~\cite{shaw2003spectralImagRemote}.
The limited spatial resolution of hyperspectral images means that each image pixel is actually a mixture of $P$ different pure materials, whose spectra are termed \emph{endmembers} (EM), present in the scene~\cite{Keshava:2002p5667}.
This mixing process conceals important information about the pure materials and their distribution in an image.
Spectral unmixing (SU) aims to solve this problem by decomposing a hyperspectral image into the spectral signatures of the endmembers and their fractional \emph{abundance} proportions for each pixel~\cite{bioucas2012unmixingReview}.

The simplest and most widely used model to represent the interaction between light and the EMs in the scene is the Linear Mixing Model (LMM)~\cite{Keshava:2002p5667}, which represents a given %
pixel $\by_n$ indexed by $n$ with $L$ spectral bands as:
\begin{align} \label{eq:LMM}
    &\by_n = \bM_0 \, \ba_n + \be_n, %
    \,\, \text{subject to }\,\cb{1}^\top\ba_n = 1 \text{ and } \ba_n \geq \cb{0} %
\end{align}
where $\bM_0 = [\bm_{0,1},\,\ldots, \,\bm_{0,P}]$ is an $L\times P$ matrix whose columns are the $P$ endmembers, $\ba_n$ is a vector containing the abundances of every endmember in the pixel~$\by_n$ and $\be_n$ is an additive noise vector. 
Traditionally, the LMM assumes that the signatures $\bM_0$ of the pure materials are the same for all pixels $\by_n$, $n=1,\ldots,N$ in the image. 
Although this assumption leads to a well-posed and computationally simpler framework, it limits the applicability of the LMM since it can jeopardize the accuracy of estimated abundances in many circumstances due to the spectral variability of the endmembers.

\begin{figure}
    \centering
    \includegraphics[width=\linewidth]{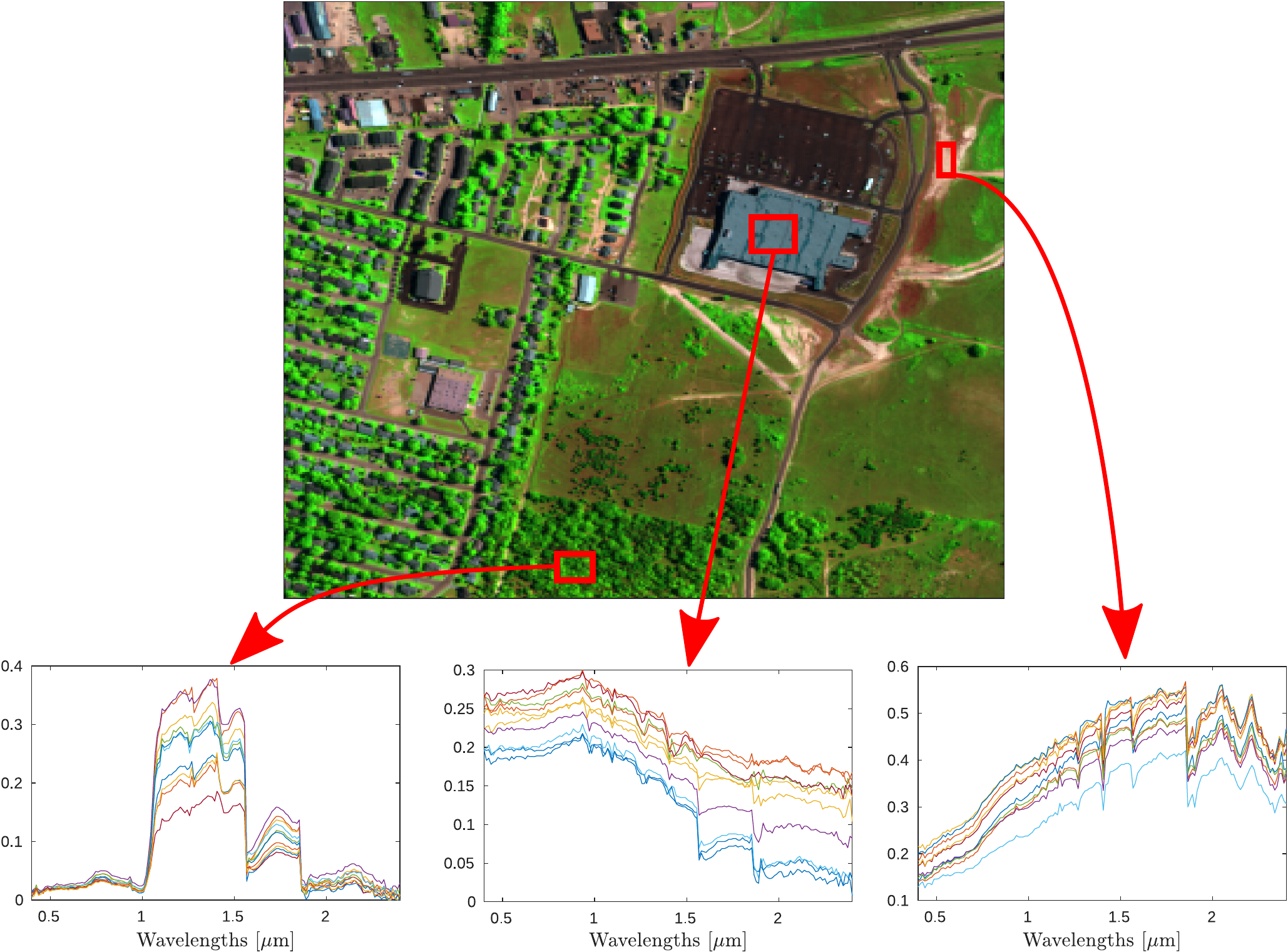}
    \caption{Spectral variability is ubiquitous in hyperspectral images: the pixels in regions composed of a single material (e.g., tree, roof and soil in the image above) can contain very different spectral signatures.}
    \label{fig:illustrative_variability_urban}
\end{figure}
\subsection{\textbf{Spectral variability in SU}}

Spectral variability is an effect commonly observed in many scenes in which the spectral signatures of the pure constituent materials vary across the observed hyperspectral image, as illustrated in Fig.~\ref{fig:illustrative_variability_urban}. It can be caused, for instance, by variable illumination and atmospheric conditions. Variability can also be intrinsic to the very definition of a pure material, such as signatures of a single vegetation species varying significantly due to different growing and environmental conditions~\cite{Zare-2014-ID324-variabilityReview,somers2011variabilityReview}.

In this context, the use of a single matrix $\bM_0$ for all pixels in the LMM~\eqref{eq:LMM} leads to problems such as \emph{proportion indeterminacy}, where errors in the estimation of the endmember spectra at each pixel propagate to the estimated abundances. This results in erroneous abundance estimation and in the selection of too many endmembers to represent the spectrum of each pixel $\by_n$~\cite{Zare-2014-ID324-variabilityReview,garcia2005VMESMA,somers2011variabilityReview}. 
Due to the significant impact of endmember variability on abundance estimation quality, a lot of effort has recently been dedicated to develop algorithms that are able to obtain better abundance estimates in this scenario.

The most general form of the LMM considering spectral variability generalizes~\eqref{eq:LMM} to allow for a different endmember matrix for each pixel, resulting in:
\begin{align} \label{eq:model_variab_general}
    &\by_n = \bM_n \, \ba_n + \be_n, %
    \,\, \text{subject to }\,\cb{1}^\top\ba_n = 1 \text{ and } \ba_n \geq \cb{0}
\end{align}
for $n=1,\dots,N$, where $\bM_n\in\amsmathbb{R}^{L\times P}$ is the $n^{\rm th}$ pixel endmember matrix.

SU considering spectral variability can be generally defined as two complementary problems related, respectively, to the recovery of the abundances and to the recovery of the endmembers. These can be defined as:
\begin{itemize}
    \item[\textsf{P\textsubscript{1}}]:\hskip\fontdimen2\font\relax To mitigate the adverse effects of spectral variability in the abundance estimation; %
    
    \item[\textsf{P\textsubscript{2}}]:\hskip\fontdimen2\font\relax To estimate the spectral signatures of the endmembers present in each pixel of the image.
\end{itemize}
Substantial interest has been recently raised for both of these problems. While all SU methods must deal with \textsf{P\textsubscript{1}} while accounting for spectral variability, not all of them take \textsf{P\textsubscript{2}} into consideration due to the additional difficulty it entails.

\begin{figure*}
    \centering
    \includegraphics[width=1\textwidth]{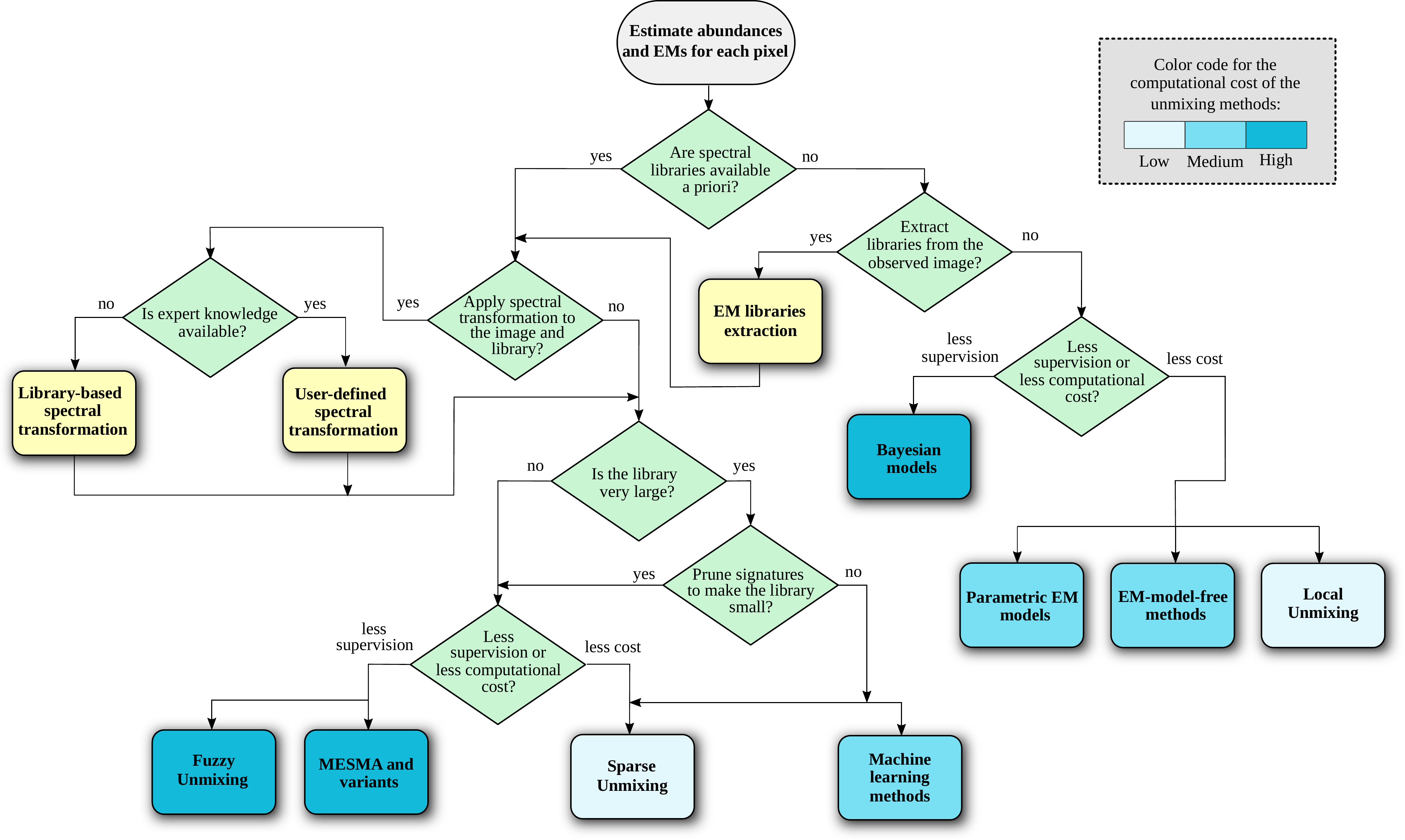}
    \caption{Decision tree for hyperspectral unmixing considering spectral variability. The blue boxes denote families of unmixing algorithms, while the yellow boxes denote additional techniques related to the extraction and processing of spectral libraries.}
    \label{fig:decision_tree}
\end{figure*}

\begin{table*}[th]
\renewcommand{\arraystretch}{2.2}
\centering
\caption{Characteristics of each group of SU techniques and where they are reviewed in the paper}
\begin{tabular}{c|cccccccc}
\hline
& \makecell{MESMA and\\variants} & Fuzzy SU & Sparse SU & Machine Learning & Local SU & \makecell{Parametric\\EM models} & EM-model-free & Bayesian \\\hline

\makecell{Amount of user\\supervision} & $\bullet$ & $\bullet$ & $\bullet\bullet$ & $\bullet\bullet\bullet$ & $\bullet\bullet\bullet$  & $\bullet\bullet\bullet$  & $\bullet\bullet$  & $\bullet$ \\

Computational cost & $\bullet\bullet\bullet$ & $\bullet\bullet\bullet$ & $\bullet$ & $\bullet\bullet$ & $\bullet$  & $\bullet\bullet$ & $\bullet\bullet$ & $\bullet\bullet\bullet$ \\

\makecell{Requires spectral\\libraries?} & \cmark & \cmark & \cmark & \cmark & \xmark & \xmark & \xmark & \xmark \\

\makecell{Estimates pixel-\\\cmag{dependent} endmembers?} & \cmark & \cmark & \cmark & \xmark & \cmark & \cmark & \xmark & \xmark \\\hline

\makecell{Where to find\\in the paper:} & Sec.~\ref{sec:MESMA_and_variants} & Sec.~\ref{sec:MESMA_and_variants} & Sec.~\ref{sec:sparse_SU} & Sec.~\ref{sec:machineLearning_SU} & Sec.~\ref{sec:local_SU} & Sec.~\ref{sec:parametric_mdl_SU} & Sec.~\ref{sec:EM_mdl_free_SU} & Sec.~\ref{sec:Bayesian_SU} \\\hline

\makecell{Illustrative description\\of the key ideas:} & \multicolumn{3}{c}{Fig.~\ref{fig:MESMA_sparseSU_diagram}} & \multicolumn{1}{c}{Fig.~\ref{fig:ML_SU_diagram}} & \multicolumn{1}{c}{Fig.~\ref{fig:local_SU_diagram}} & \multicolumn{2}{c}{Fig.~\ref{fig:param_mdlFree_diagram}} & \multicolumn{1}{c}{Fig.~\ref{fig:bayesian_illustrative_diagram}}\\\hline

\end{tabular}
\label{tab:SU_methods_organization}
\end{table*}

\begin{table*}[th]
\renewcommand{\arraystretch}{2.2}
\setlength{\tabcolsep}{10pt}
\centering
\caption{Characteristics of spectral library extraction and pruning techniques and where they are reviewed in the paper}
\begin{tabular}{c|ccc}
\hline
\makecell{Library extraction\\techniques:} & \makecell{Image-based\\library extraction} & \makecell{Library generation\\from physics models} & \makecell{Spatial interpolation\\of EM signatures} \\\hline

\makecell{\\\\Key idea} 
& \multicolumn{1}{p{4cm}}{Extracts multiple EM signatures from the observed image and cluster them to construct a library} 
& \multicolumn{1}{p{4cm}}{Create synthetic EM signatures using physico-chemical mathematical models describing EM variability} 
& \multicolumn{1}{p{4cm}}{Estimate EM signatures for each pixel by interpolating pure pixels at known spatial locations} \\

Adapted to the HI? & \cmark & \xmark & \cmark \\

\makecell{Amount of user\\supervision} & $\bullet\bullet$ & $\bullet\bullet\bullet$ & $\bullet\bullet$ \\

\makecell{Depends on the existence\\of pure pixels?} & \cmark & \xmark & \cmark \\

\makecell{Where to find\\in the paper:} & Sec.~\ref{sec:constructing_libs_imageBased} & Sec.~\ref{sec:constructing_libs_physicsMdl} & Sec.~\ref{sec:constructing_libs_interp} \\
\hline

\makecell{Library pruning\\techniques:} & Library reduction & Endmember selection & Same-class EM pruning \\\hline

\makecell{\\\\Key idea} 
& \multicolumn{1}{p{4cm}}{Remove redundant signatures from an existing library to reduce the computational complexity of SU} 
& \multicolumn{1}{p{4cm}}{Remove entire EM classes (e.g., water, tree) not present in the observed image from the library}
& \multicolumn{1}{p{4cm}}{Select the signatures from each EM class most closely related to the observed image before SU} \\

Adapted to the HI? & \xmark & \cmark & \cmark \\

\makecell{Amount of user\\supervision} & $\bullet$ & $\bullet\bullet$ & $\bullet\bullet$ \\

\makecell{Improves the computa-\\tional cost of SU?} & \cmark & \cmark & \cmark \\

\makecell{Improves SU quality?} & \xmark & \cmark & \cmark \\

\makecell{Where to find\\in the paper:} & Sec.~\ref{sec:sub_lib_reduction} & Sec.~\ref{sec:lib_pruning_EM_selection} & Sec.~\ref{sec:lib_pruning_same_class} \\\hline
\end{tabular}
\label{tab:spectral_libraries_organization}
\end{table*}

\subsection{\textbf{Contribution, taxonomy and organization}}

Many SU algorithms have been proposed to address problems \textsf{P\textsubscript{1}} and \textsf{P\textsubscript{2}}. Different algorithms follow different methodologies to represent the endmembers in the scene. Existing methods employ Bayesian, parametric, or spatially localized models, as well as libraries containing different instances of material spectra known \textit{a priori}. This multiplicity of models gives rise to solutions presenting different advantages and disadvantages in terms of computational complexity, accuracy and amount of user supervision.

In this paper we categorize the methods according to criteria that are most relevant to the practitioner, such as, e.g., computational complexity, to provide a comprehensive review that complements and updates previous review papers~\cite{somers2011variabilityReview,Zare-2014-ID324-variabilityReview,drumetz2016variabilityReviewRecent,drumetz2020bookchapterVariability}.
Since existing SU methods that address spectral variability have very heterogeneous characteristics, navigating the field can be difficult, especially when taking into account both classical algorithms and recent developments. This difficulty motivated the present review, which presents a novel taxonomy aimed at the practitioner, as well as a comprehensive categorization of existing approaches. The contributions and highlights of the present paper are described in the following.

\subsubsection{\textbf{A new taxonomy, for the practitioner}} 
\begin{figure*}
\begin{mdframed}[backgroundcolor=black!10]
    \centering
    \begin{minipage}{0.3\linewidth}
    \fontsize{9pt}{10.5pt}\selectfont
    \textbf{MESMA} and \textbf{Sparse SU} are the main methods based on spectral libraries. The basic principle behind MESMA is to iteratively search for the combination of EM signatures in the library which, among all possibilities, allows for the closest reconstruction of each observed pixel under the LMM. Sparse SU, on the other hand, performs EM selection and abundance estimation in a single optimization problem using sparsity and structuring constraints and penalties, which allows for faster processing times.
    \end{minipage}%
    \begin{minipage}{0.6999\linewidth}
    \hfill
    \includegraphics[width=0.98\linewidth]{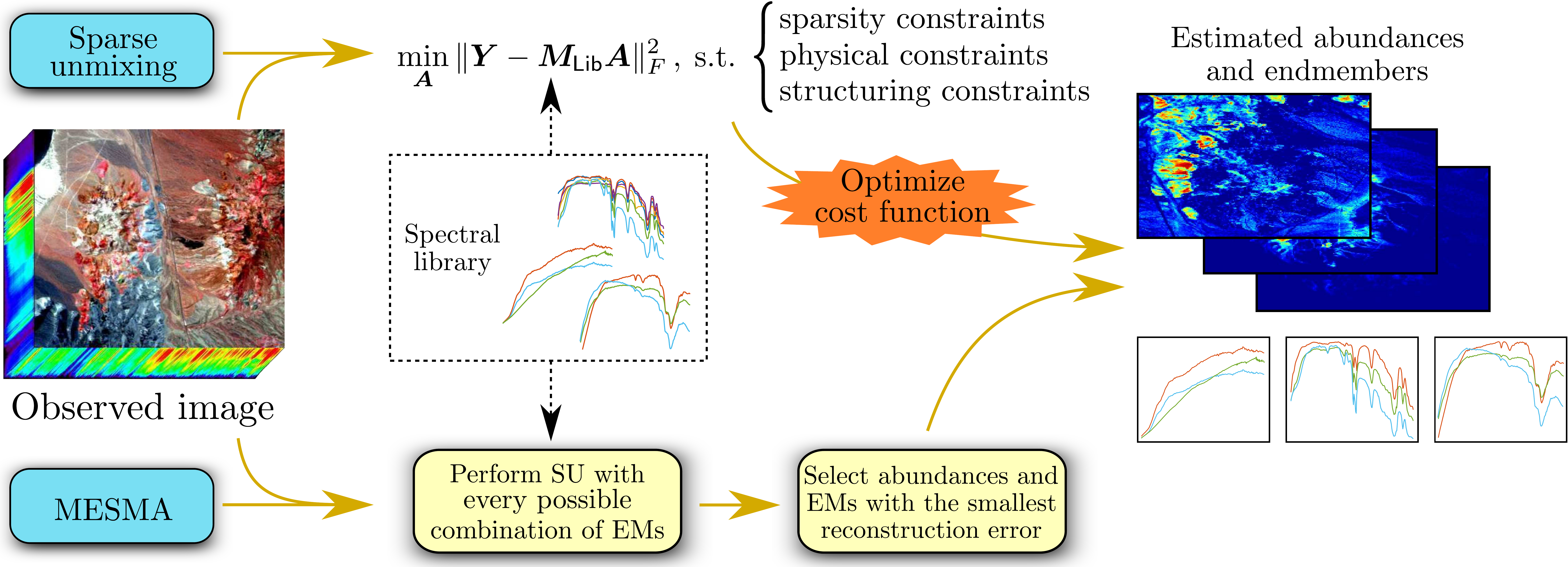}
    \end{minipage}
    \caption{Illustrative description of MESMA, fuzzy and Sparse SU techniques.}
    \label{fig:MESMA_sparseSU_diagram}
\end{mdframed}
\end{figure*}

\begin{figure*}
\begin{mdframed}[backgroundcolor=black!10]
    \centering
    \begin{minipage}{0.3\linewidth}
    \fontsize{9pt}{10.5pt}\selectfont
    The flexibility and representation power of \textbf{Machine Learning} algorithms can be exploited to address spectral variability by formulating SU as a supervised learning problem. One simple approach is to learn a mapping between the mixed pixels and the abundance and EMs based on training data generated synthetically using a spectral library. However, the incorporation of expert knowledge about the SU problem in the design of the machine learning algorithm is important to obtain a better performance and to address spectral variability effectively.
    \end{minipage}%
    \begin{minipage}{0.6999\linewidth}
    \hfill
    \includegraphics[width=0.98\linewidth]{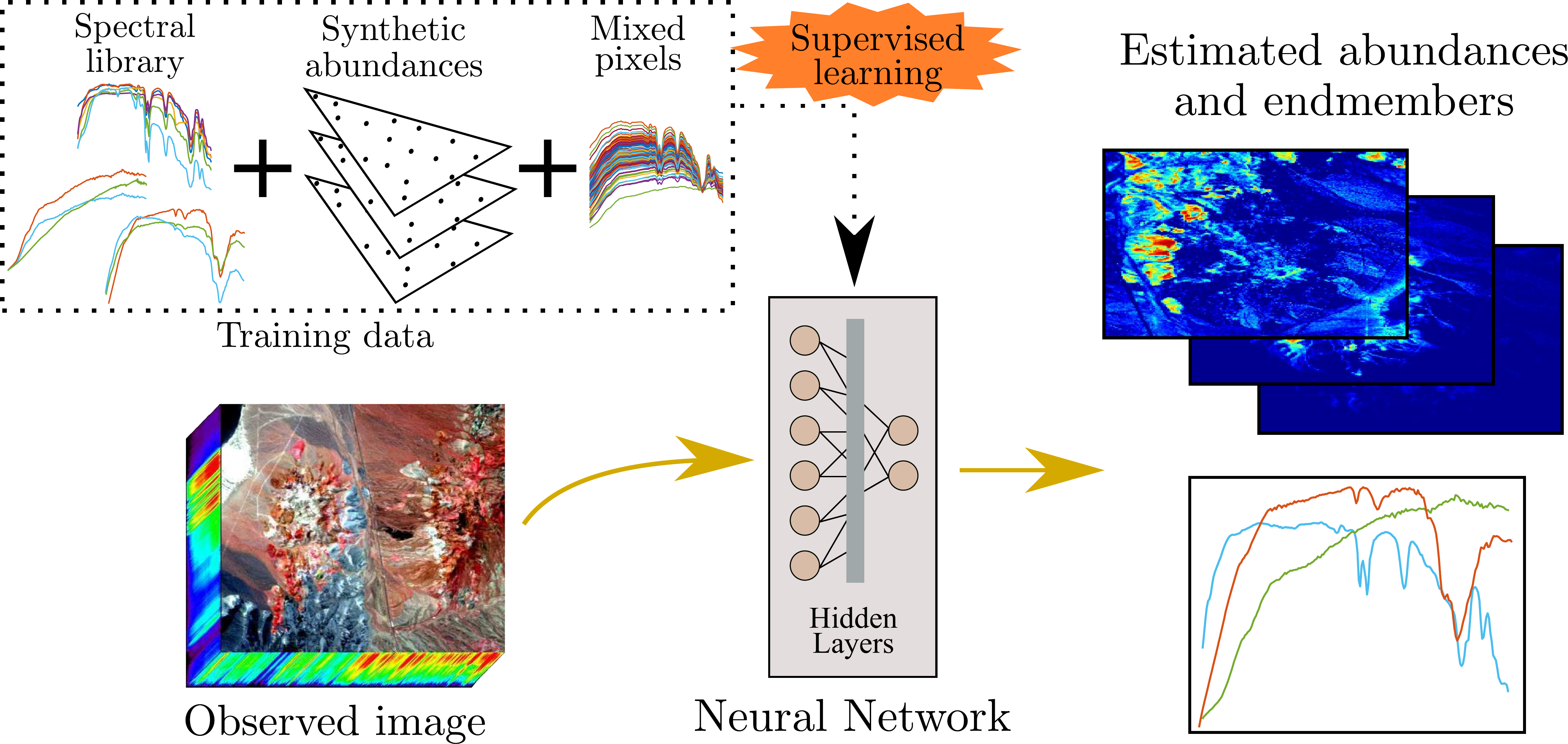}
    \end{minipage}
    \caption{Illustrative description of machine learning-based SU techniques.}
    \label{fig:ML_SU_diagram}
\end{mdframed}
\end{figure*}

\begin{figure*}
\begin{mdframed}[backgroundcolor=black!10]
    \centering
    \begin{minipage}{0.3\linewidth}
    \fontsize{9pt}{10.5pt}\selectfont
    \textbf{Local SU} addresses the variability of endmember signatures across space by performing SU on small, compact spatial regions of the image in which the EMs can be assumed to be approximately constant. The local SU results for each image region are afterwards clustered in order to assemble the global abundance maps and sets of EM spectra. Local SU offers a lot of flexibility in the choice of the segmentation of the image and of the local EM extraction and clustering strategies, which can have a significant impact on the global SU results.
    \end{minipage}%
    \begin{minipage}{0.6999\linewidth}
    \hfill
    \includegraphics[width=0.98\linewidth]{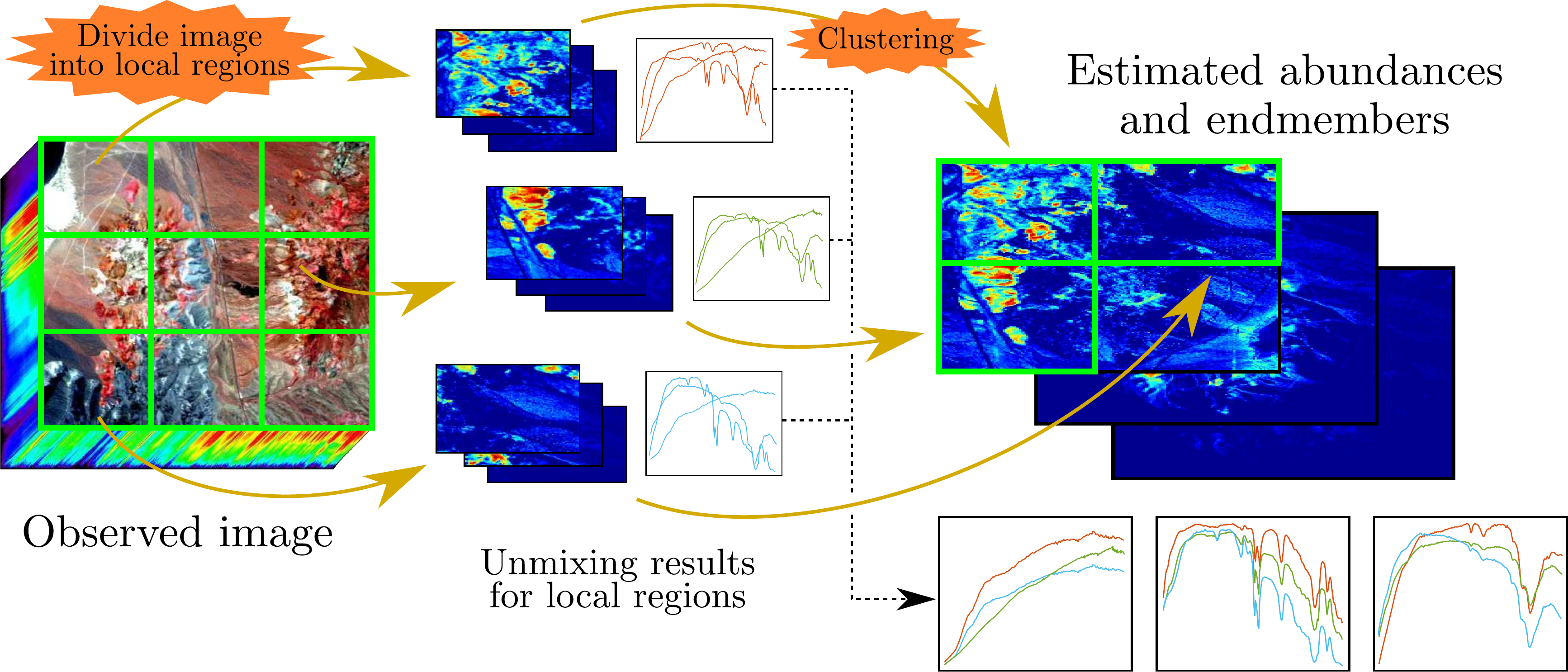}
    \end{minipage}
    \caption{Illustrative description of local SU techniques.}
    \label{fig:local_SU_diagram}
\end{mdframed}
\end{figure*}

\begin{figure*}
\begin{mdframed}[backgroundcolor=black!10]
    \centering
    \begin{minipage}{0.3\linewidth}
    \fontsize{9pt}{10.5pt}\selectfont
    \textbf{Parametric EM models} represent the (variable) signatures of the EMs as a function of a low-dimensional vector of parameters. The abundances and the vector of EM parameters for each pixel are then recovered by solving an optimization problem. \textbf{EM-model-free} methods, on the other hand, generally attempt to mitigate spectral variability indirectly through the design of robust cost functions using, e.g., additive residual terms. The use of regularization terms is important in both cases to incorporate \textit{a priori} knowledge about the problem.
    \end{minipage}%
    \begin{minipage}{0.6999\linewidth}
    \hfill
    \includegraphics[width=0.98\linewidth]{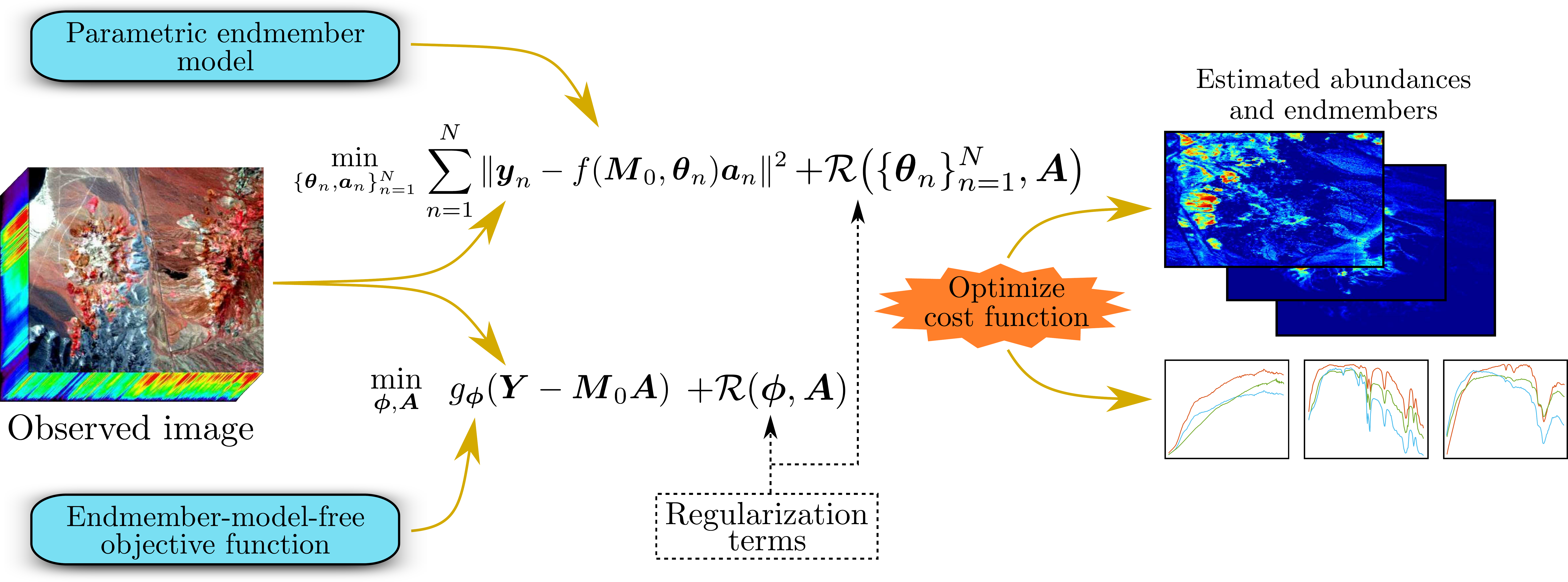}
    \end{minipage}
    \caption{Illustrative description of SU techniques based on parametric EM models and of EM-model-free SU approaches.}
    \label{fig:param_mdlFree_diagram}
\end{mdframed}
\end{figure*}

\begin{figure*}
\begin{mdframed}[backgroundcolor=black!10]
    \centering
    \begin{minipage}{0.3\linewidth}
    \fontsize{9pt}{10.5pt}\selectfont
    \textbf{Bayesian SU} methods represent the EM signatures at each pixel as a realization of a statistical distribution. Statistical distributions are first attributed to the EMs and to the abundances and, possibly, to other variables or to hyperparameters of these distributions. Using the Bayes rule, the SU results are then derived from the posterior distribution in a Bayesian inference problem. The abundances and EM distributions can be computed as, e.g., the mean or as the mode of the posterior distribution.
    \end{minipage}%
    \begin{minipage}{0.6999\linewidth}
    \hfill
    \includegraphics[width=0.98\linewidth]{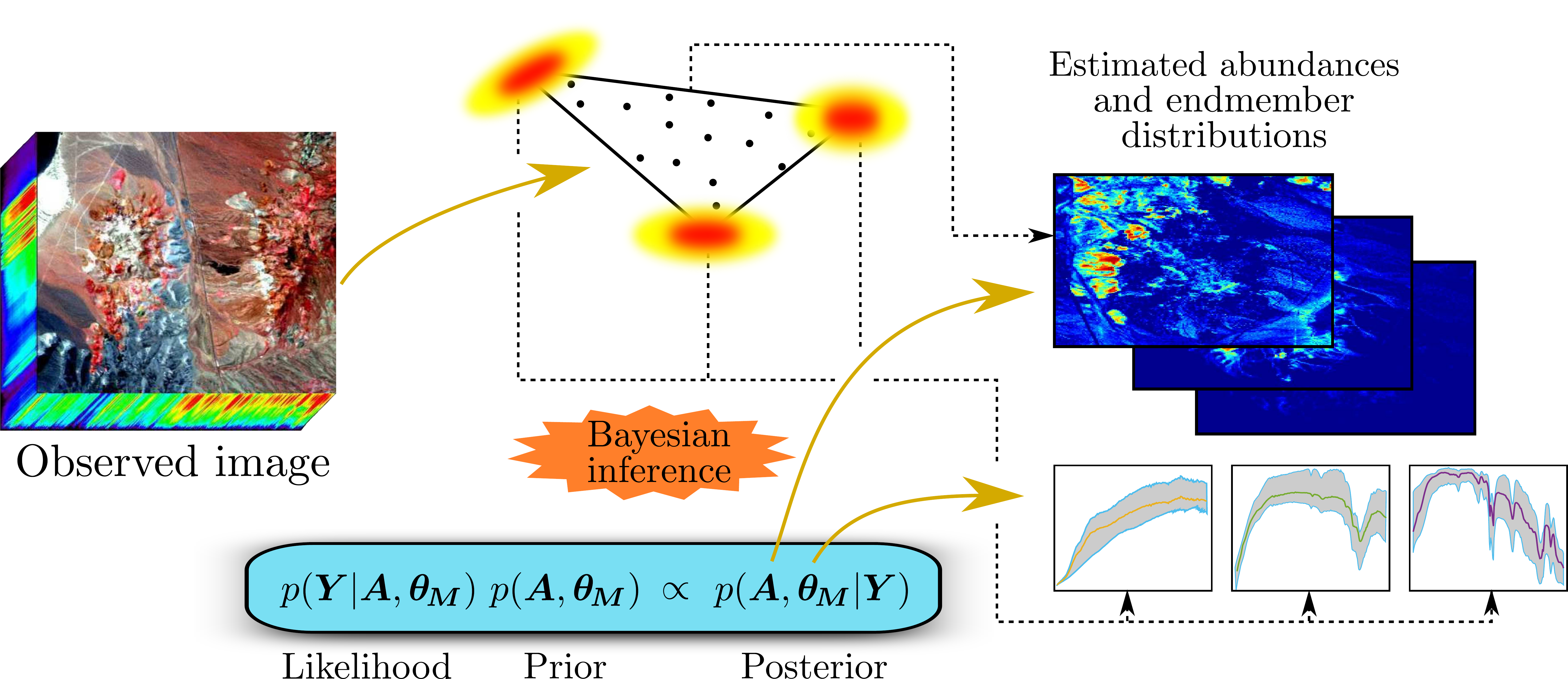}
    \end{minipage}
    \caption{Illustrative description of Bayesian SU techniques.}
    \label{fig:bayesian_illustrative_diagram}
\end{mdframed}
\end{figure*}

We propose a new taxonomy to organize the existing techniques according to a practitioner's point of view, based on the amount of user supervision and on the computational complexity required to solve the SU problem. The resulting taxonomy is summarized in the form of a decision tree shown in Fig.~\ref{fig:decision_tree}, which can be used to guide the choice of a family of SU algorithms. The decision tree also dictates the organization of the rest of the paper. We start from whether a spectral library is known a priori or not, and proceed to different families of SU methods based on the trade-offs they offer regarding the need for user supervision and computational cost. Table~\ref{tab:SU_methods_organization} summarizes the main characteristics of each group of techniques, and points to illustrations with high-level descriptions of the key ideas on which they are based (Figs.~\ref{fig:MESMA_sparseSU_diagram}--\ref{fig:bayesian_illustrative_diagram}).

\subsubsection{\textbf{Comprehensive overview and recent highlights}} 
We provide a comprehensive review of the methods developed to solve the SU problem with EM variability. We encompass and contextualize both the classic strategies that have been reviewed before as well as numerous recent developments in the field. Thus, both classic and recent algorithms are categorized according to the proposed taxonomy, which helps to highlight the recent advances in each area.

\subsubsection{\textbf{Spectral libraries, \textit{ex situ}}} 

A considerable number of SU methods \cmag{addresses} spectral variability using libraries of spectra that originally had to be acquired \textit{a priori} (e.g., through laboratory of \textit{in situ} measurements), which used to limit the applicability of these methods. An important recent development concerns methods that can either extract spectral libraries directly from the observed images, or generate them using physics-based mathematical models of material spectra. This allows for the widespread applicability of library-based SU techniques in situations where spectral libraries are not available or cannot be obtained. Such methods are reviewed in Section~\ref{sec:constructing_libs}, and an illustrative description of these techniques can be \mbox{seen in Fig.~\ref{fig:diagram_library_extraction}.}

Moreover, library pruning techniques, which were originally devised to reduce the size of libraries so as to improve the computational complexity of SU, have evolved to consider also the quality of the unmixing results. Recent library pruning methods aim at removing, before unmixing, either entire EM classes or individual spectral signatures which are not likely to be present in observed image. This reduces the ill-posedness of the SU problem and can improve abundance estimation. These techniques are discussed in Section~\ref{sec:pruning_libs}. 

Table~\ref{tab:spectral_libraries_organization} summarizes the key ideas involved in library extraction and pruning methods, as well as their main characteristics.

\subsubsection{\textbf{Experimental aspects and toolbox}}
The practical aspects related to the evaluation of SU methods when spectral variability is considered are also discussed in Section~\ref{sec:experimental}. This includes the generation of realistic synthetic data and a list of existing software resources that are available to the reader. We also present an illustrative simulation in order to demonstrate the application of a few of the SU techniques reviewed in the paper, which were chosen by selecting different paths in the proposed decision tree. This example is made publicly available in the form of a software toolbox at \url{https://github.com/ricardoborsoi/unmixing_spectral_variability} and in~\cite{borsoi2020codesSUvariabilityRevPaper}.

The paper is organized as follows. In Section~\ref{sec:overall_origins} we present a detailed overview of the physical effects that originate spectral variability and their effects on the endmembers and on the hyperspectral image. In Section~\ref{sec:overall_methods_libs}, we review the SU methods accounting for spectral variability that use spectral libraries. In Section~\ref{sec:overall_methods_blind}, we describe the blind methods that do not require EM signatures to be known \textit{a priori}. We then discuss the construction and pruning of spectral libraries in Section~\ref{sec:overall_concerning_libs}. Section~\ref{sec:experimental} discusses the evaluation of SU algorithms when spectral variability is present, lists existing software resources, and presents an end-to-end illustrative example comparing some existing techniques selected following the proposed decision tree. Finally, we conclude the paper in Section~\ref{sec:conclusions} with some discussion and conclusions about the existing methods and future research directions.

\section{Origins of Spectral Variability and their Effects}
\label{sec:overall_origins}

The variability in the spectral signatures occurs mainly due to (a) atmospheric effects, (b) illumination and topographic changes, and (c) intrinsic variation of the spectral signatures of the materials (i.e., due to physico-chemical variations).
Understanding how these conditions affect the spectral signatures of the materials and the unmixing results is important in order to develop informed models and methods to deal with EM variability.
\cred{Such knowledge can be used, for instance, \cdgreen{to generate physics-based or physically-inspired models that include the effects of spectral signature variability. Such models} can then be directly incorporated into the SU process (as discussed in Sec.~\ref{sec:parametric_mdl_SU}), or used to generate synthetic spectral libraries for library-based SU (as discussed in Sec.~\ref{sec:constructing_libs_physicsMdl}).}

In addition to spectral unmixing, spectral variability also affects other hyperspectral imaging tasks, which prompted extensive investigations into its causes and on how it manifests in the material spectra. In this context, a recent review article by James Theiler and his coworkers provides an excellent overview of spectral variability in hyperspectral target detection~\cite{theiler2019spectralVariabilityTargetDet_GRSM}. In particular, the causes and effects of spectral variability in target detection are reviewed, with a focus on the study of environmentally induced variability (caused by, e.g., atmospheric and topographic changes) through an in-depth view of radiative transfer models. A detailed computer simulation is also included to illustrate how the material spectra are affected by changes in the different parameters of the radiative transfer model.

In the following, we review the causes and effects of spectral variability from a spectral unmixing perspective. Although we also introduce the radiative transfer function interpretation of some atmospheric and topographic effects, we focus our exposition on a more generic analysis of the consequences that spectral variability has on the observed pixel spectra and on the results of spectral unmixing as reported by previous experimental works (i.e., with a stronger focus on the results of, e.g., atmospheric compensation methods as opposed to the interpretation of the imaging models themselves). The interested reader can find a more comprehensive and in-depth analysis from a radiative transfer function standpoint in~\cite{theiler2019spectralVariabilityTargetDet_GRSM}.

\subsection{\textbf{Atmospheric effects}}

One of the main sources of spectral variability is  the interference by the atmosphere when measuring ground reflectance. Atmospheric gases (such as O$_3$, O$_2$, CH$_4$, CO$_2$, etc.), aerosols and, most prominently water vapor, absorb significant amounts of radiation, while other molecules and aerosols scatter incoming light~\cite{griffin2003compensationAtmosphericEffects}. These effects have an impact on the radiance measured at the sensor, which can become significantly different than that corresponding to the desired ground reflectance.
Atmospheric absorption from gases is also heavily wavelength dependent, whereas aerosol absorption varies smoothly in spectra.
These effects must be compensated to achieve an accurate characterization of surface reflectance.

Atmospheric compensation models can be roughly divided into statistical (empirical) and physics-based models~\cite{griffin2003compensationAtmosphericEffects}.
Statistical models are based on additional information about the atmospheric influence, usually obtained by means of reference objects or calibration panels in the scene. This information is used to find a relationship (e.g., linear) between the radiances observed at the sensor and at the surface of the scene~\cite{griffin2003compensationAtmosphericEffects}. This results in a gain and an offset factor for each spectral band, which are then uniformly applied to every image pixel to compensate for the atmospheric effects~\cite{griffin2003compensationAtmosphericEffects}.
Sometimes, when a reference object is not present in the scene, naturally occurring objects can be employed as reference spectra, most commonly consisting of smooth bodies of water, which exhibit low reflectance and can be considered as dark objects~\cite{shaw2003spectralImagRemote}.
The downsides of this approach are that the true reflectance of a reference object must be accurately known, and that it does not account for the spatial variability of the distribution of gases and aerosols. This variability can be very significant, and thus can introduce spatially-dependent residual atmospheric effects. A classical example of statistical methods is the empirical line method (ELM)~\cite{shaw2003spectralImagRemote}.

Physics-based models, on the other hand, are robust alternatives to empirical methods which do not assume additional information about the scene to be known. 
These methods are currently mature and widely used, addressing the limitations of empirical methods by employing a rigorous model that explicitly describes the absorption and scattering effects due to atmospheric gases and aerosols~\cite{gao2009atmosphericCorrectionReview}.
Popular examples include the Atmospheric Removal (ATREM) and the Fast Line-of-Sight Atmospheric Analysis of Spectral Hypercubes (FLAASH) algorithms~\cite{griffin2003compensationAtmosphericEffects}.

\begin{figure}
    \centering
    \includegraphics[width=\linewidth]{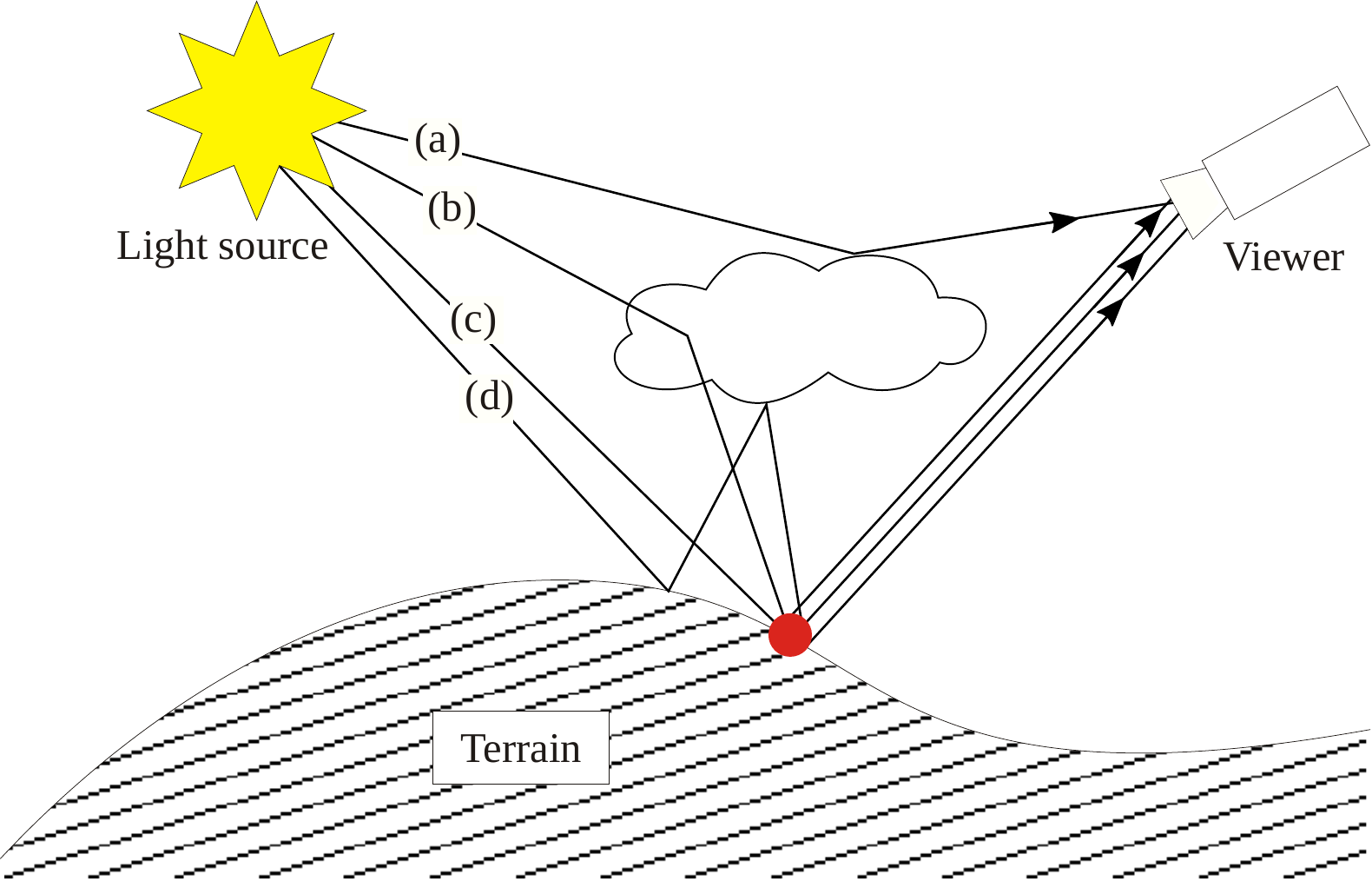}
    \caption{Illustration of the effects of the atmosphere on the acquired hyperspectral image. The sources of radiation are represented by (a) light directly reflected by the atmosphere to the sensor, (b) light scattered by the atmosphere and reflected by the ground, (c) light directly reflected by the ground and (d) light reflected by surrounding regions on the ground and then scattered to the sensor.}
    \label{fig:atmospheric_variab_source_diagr}
\end{figure}

Assuming a ground terrain illuminated by the sun, the light incident on a pixel in the sensor can be roughly characterized by four sources: solar radiation directly reflected off the ground, light directly reflected off the atmosphere into the sensor, light scattered by the atmosphere and reflected off the ground, and light that is reflected off surrounding regions on the ground and then scattered before reaching the sensor (constituting the adjacency effect)~\cite{healey1999hyperAtmosphericConditions,Nascimento-2005-ID325-doesICAplaysArole}.
These effects are illustrated in Fig.~\ref{fig:atmospheric_variab_source_diagr}.
A model for the reflectance at the sensor $y_{\rm sensor}$ is given by~\cite{griffin2003compensationAtmosphericEffects}:
\begin{align} \label{eq:atmospheric_model}
    y_{\rm sensor} = y_{\rm atm} T_{\rm g}
    + \frac{y_{\rm s} T_{\rm g} T_{\downarrow} T_{\uparrow} + (y_{\rm avg}-y_{\rm s}) T_{\rm g}T_{\downarrow}T_{\uparrow}r}
    {1-y_{\rm avg}s} \,,
\end{align}
where $y_{\rm s}$ is the \cmag{reflectance of the surface of interest}, $T_{\rm g}$ is the gaseous transmittance, $y_{\rm atm}$ the reflectance of the atmosphere, ${T}_{\downarrow}$ and ${T}_{\uparrow}$ are the upward and downward scattering transmittances, $r$ is the ratio between diffuse and total transmittance for the ground-to-sensor path, ${s}$ is the spherical albedo of the atmosphere, and $y_{\rm avg}$ is the average surface reflectance in a region around a pixel, which is used to account for scattering (adjacency)~effects~\cite{griffin2003compensationAtmosphericEffects}. %

Physics-based atmospheric correction algorithms then try to obtain the ground reflectance $y_{\rm s}$ from the at-sensor reflectance $y_{\rm sensor}$ by solving~\eqref{eq:atmospheric_model}.
In the overall working of these algorithms the first step for atmospheric compensation consists of retrieving the atmospheric parameters necessary to represent the quantities in~\eqref{eq:atmospheric_model}, mainly consisting of aerosol description (visibility and type of aerosol) and amount of water vapor for each pixel~\cite{matthew2002atmosphericEvaluationFLAASH}.
They are typically based on variations of the so-called three-band ratio technique, which is an important step used to quantify the amount of water vapor for each pixel.
The three-band ratio technique basically compares ratios of radiances measured near the edges of a number of spectral wavelengths which are known to present heavy water-vapor absorption (e.g., at around 0.91~$\mu m$, 0.94~$\mu m$ and 1.14~$\mu m$), using this information to derive the column water vapor information for each pixel~\cite{shaw2003spectralImagRemote,lau2004atmosphericComparissonHymap}.
After the necessary parameters have been estimated,~\eqref{eq:atmospheric_model} can be solved for the ground reflectance and an optional post-processing step can be employed (called spectral polishing) to remove artifacts from the correction process~\cite{matthew2002atmosphericEvaluationFLAASH}.

Physics-based models can represent and account for the interaction between solar radiation and the atmosphere very accurately.
However, for this accuracy to translate into meaningful surface reflectance estimates, these models require precise information about atmospheric properties, which are very difficult to obtain in practice. This is specially true for scattering and absorption by aerosols, which are hard to characterize accurately due to their spatial and temporal variability~\cite{song2001atmospherifEffectsChangeDetection}.
Inaccuracies in the estimation of these parameters (which include the atmospheric visibility, aerosol model type and an atmospheric model) introduce errors in the retrieved surface reflectance spectra that can be significant and \mbox{spectrally non-uniform~\cite{griffin1999sensitivityAtmosphericCompensation}.}

Furthermore, unlike water vapor compensation, which is performed on a pixel-by-pixel basis, most methods assume that individual aerosol and gas concentrations are uniform across the scene (resulting in a single transmittance spectrum being computed for each gas) \cite{griffin1999sensitivityAtmosphericCompensation,matthew2002atmosphericEvaluationFLAASH}.
While this is true for some gases (such as NH$_4$, O$_2$, CH$_4$, CO$_2$, etc.) that are fairly constant in the atmosphere~\cite{lau2004atmosphericComparissonHymap}, it is far from true for aerosols, which may show significant variation in space~\cite{wilson2014spatialVaraibilityAerosol,bhatia2018propagationUncertaintiesAtmosphericUnmixing}.
Aerosol concentration can vary depending on the environment (e.g., in large cities and rural areas), and thus must be informed by the user to the existing algorithms~\cite{lau2004atmosphericComparissonHymap}.
Moreover, standard aerosol types often do not adequately represent the scene being processed, leading to inaccuracies in the retrieved spectra~\cite{bassani2015impactMicrophysicalAerosol}. 
Furthermore, experimental studies have found that aerosol optical thickness has a significant spatial variability within a single scene~\cite{wilson2014spatialVaraibilityAerosol,kaufman2006aerosolVariabilityCloud} and is often correlated with \mbox{cloud concentrations~\cite{kaufman2006aerosolVariabilityCloud}.}

Some works attempted to estimate aerosol optical thickness for smaller patches of the image individually using shadow detection results~\cite{schlapfer2018aerosolEstimationShadow}, which depends on the presence of a large number of shadowed pixels. However, acquiring precise data for an accurate and possibly spatially variable atmospheric correction is generally difficult, which means that the results of common atmospheric compensation methods can be subject to significant errors~\cite{wilson2014spatialVaraibilityAerosol}.
For instance, a number of studies have investigated the residual errors in surface reflectance data after the application of atmospheric compensation methods by comparing the processed results with \textit{in situ} data or using simulations. These studies found that generally there is still an appreciable error in the retrieved reflectances.
As an example, errors in the retrieved reflectance by atmospheric corrections due to the spatial variability of aerosol optical thickness over southern England were found to be of up to 1.7\%, with 5\% errors in the normalized difference vegetation index (NDVI)~\cite{wilson2014spatialVaraibilityAerosol}. This can be significant for practical applications, as it corresponds to errors of up to~30\% in biomass production estimates~\cite{kaufman1993refForBiomassProductinoErrorNDVI,wilson2014spatialVaraibilityAerosol}.
Furthermore, standard methods for column water vapor retrieval loose accuracy when the aerosol optical thickness is high, leading to errors of up to 10\% if aerosol effects are not properly compensated~\cite{larsen2005waterVaporRetrievalHazyAtmosphere}.
Note that experimental measurements in a water quality management application found significant differences between the true and retrieved spectral responses. Errors of up to 15\% in reflectance spectra were found, more prominently concentrated in short ($<$450~nm) and long ($>$750~nm) wavelength intervals~\cite{markelin2016atmosphericEutrophicLake}.
Another study evaluated a number of physics-based atmospheric correction methods in an experiment for a \textit{playa and canola} target and found that although the average relative differences were moderate, ranging between~0.023 and~0.042, larger deviations of up to~0.12 occurred in the near-infrared region~\cite{staenz2002evaluationAtmosphericPlayaCanola}.
A study with simulated data found that incorrectly supplying input parameters to the model used in the FLAASH algorithm can lead to considerable errors in the retrieved reflectance, with an absolute difference of up to~0.11, and a strong sensitivity to moisture/optical depth (visibility) errors~\cite{griffin1999sensitivityAtmosphericCompensation}. Also, very large errors can be introduced by a bad specification of the aerosol model type, with higher errors generally present in short wavelengths where scattering processes are most significant~\cite{griffin1999sensitivityAtmosphericCompensation}.

The influence of uncertainties in column water vapor and aerosol optical depth specification on SU was investigated in~\cite{bhatia2018propagationUncertaintiesAtmosphericUnmixing} (given their influence in the retrieved reflectances). The performance degradation was found to be more severe in abundance than in reflectance estimation, with degradation of up to 30\% in high scattering conditions. The results were more severely affected due to uncertainties in water vapor amount than in aerosol optical thickness, although the latter showed a strong influence on the quality of the reconstructed abundance maps when the endmembers were spectrally similar.

\begin{figure}[ht]
    \centering
    \includegraphics[width=0.9\linewidth]{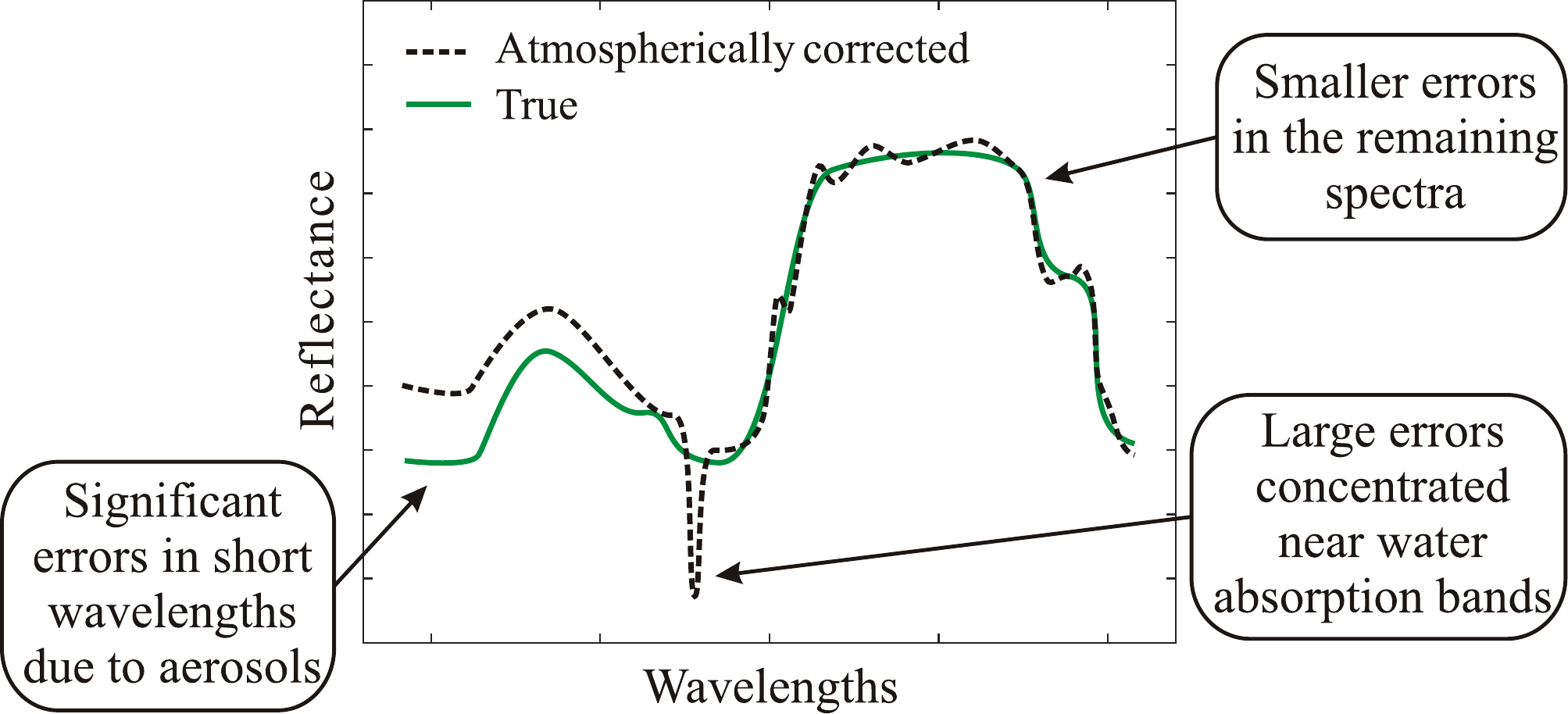}
    \vspace{-0.3cm} %
    \caption{Illustration of variability caused by atmospheric effects.}
    \label{fig:atmospheric_spectra_illustr}
\end{figure}

Finally, it is interesting to highlight that two characteristics were noticed from these studies. First, the errors in the retrieved reflectances are fairly non-uniform in spectral bands, with large spikes often concentrated near bands where there is significant gas/water absorption~\cite{griffin1999sensitivityAtmosphericCompensation,bhatia2018propagationUncertaintiesAtmosphericUnmixing}. Second, errors due to bad aerosol specification are quite significant in short wavelengths (450~nm-750~nm), where they are concentrated~\cite{griffin1999sensitivityAtmosphericCompensation,markelin2016atmosphericEutrophicLake}.
All these effects are illustrated in Fig.~\ref{fig:atmospheric_spectra_illustr}.

\subsection{\textbf{Illumination and topographic effects}}

Varying illumination conditions are one of the main sources of spectral variability in spectral mixture analysis~\cite{murphy2012evaluatingClassificationIlluminationChange}. Illumination changes are mainly due to two effects: varying terrain topography, which affects the angles of the incident radiation, and occlusion of the light source by other objects (leading to shaded areas).

A number of work handled the presence of heavily shaded areas by considering the presence of an additional endmember representing shadow~\cite{plaza2004comparisonEMextraction,Keshava:2000p6547,mcgwire2000hyperspectralSparseVegetationCover,boardman1993automatedUnmixingConvexGeom,roberts1998originalMESMA,dennison2003libraryPruningMESMAcost,roper2013shadowCorrectionTechs}.
Although this approach is very simple, its effectiveness is certainly limited since a single spectral signature can be insufficient to adequately represent all pixels affected by shadow~\cite{zhang2013detectionLIDARdataFusionPhysical}.
For instance, there might be many shadow endmembers since shadows in different regions of the image are influenced by both the material that is being shaded and by the absorption properties of the material that is blocking the light, what might lead to significantly different spectral signatures~\cite{fitzgerald2005multipleShadowFractions}.
Furthermore, besides presenting a lower reflectance amplitude, the shadow EM is also usually significantly affected by nonlinear atmospheric scattering and multipath effects, since these areas are illuminated by a large proportion of diffuse irradiation scattered by the atmosphere (i.e., skylight) and by other nearby objects.
This implies that the shadow endmember is sensitive to the state of the atmosphere and can vary significantly in space depending on the amount of scattered light being reflected from the sky at each position~\mbox{\cite{choi2005investigationDarkEMfeatureSpace,lynch2015shadowsSpatialVariability}}.

When illumination predominantly comes from scattered radiation, the spectrum not only presents a lower amplitude but is also skewed to short (e.g., blue) wavelengths~\cite{adler2002deShadowingAlgorithm,schlapfer2013correctionShadowing}. This means that the signal amplitudes in the shorter (blue) wavelengths are considerably larger than in the rest of \mbox{the spectra~\cite{schlapfer2013correctionShadowing}.}

Furthermore, since the shadow spectral signature is a function of diffuse illumination, it depends on the neighboring image area (where the skylight is scattered)~\cite{schlapfer2013correctionShadowing} and on the cloud cover.
Moreover, variations of ground reflectance may not be easily discernible from atmospheric effects since both effects are observed jointly and are not easily separable~\cite{schlapfer2013correctionShadowing}.
These facts introduce a strong dependence of the shadow signature to the spatial position, and go against the common notion that shadow endmembers can be adequately represented by scaled versions of true endmembers~\cite{shaw2003spectralImagRemote} (that is only true for small illumination variations).

This makes the detection, correction or quantification of shadow a challenging task, since physical-based inversion of these atmospheric effects turns out to be a hard problem. However, this task is still necessary since linear SU with a single dark endmember usually does not successfully quantifies the presence of shadow in the scene~\cite{schlapfer2013correctionShadowing}.

Although the presence of shadows is common in hyperspectral images, a more prominent source of variability comes from the varying topography of the scene, which introduces complex fluctuations of the relative angles between the incoming light source and the sensor for each pixel of the scene.
Topographic variations have been shown to significantly affect spectral reflectance values of soil and green vegetation~\cite{richter2009comparisonTopographicCorrection} as well as rocks in lithologic mapping~\cite{feng2003topographicNormalization}, expanding endmember clusters and causing overlap between classes, hindering the endmember identification and unmixing processes.

\begin{figure}
    \centering
    \includegraphics[width=0.6\linewidth]{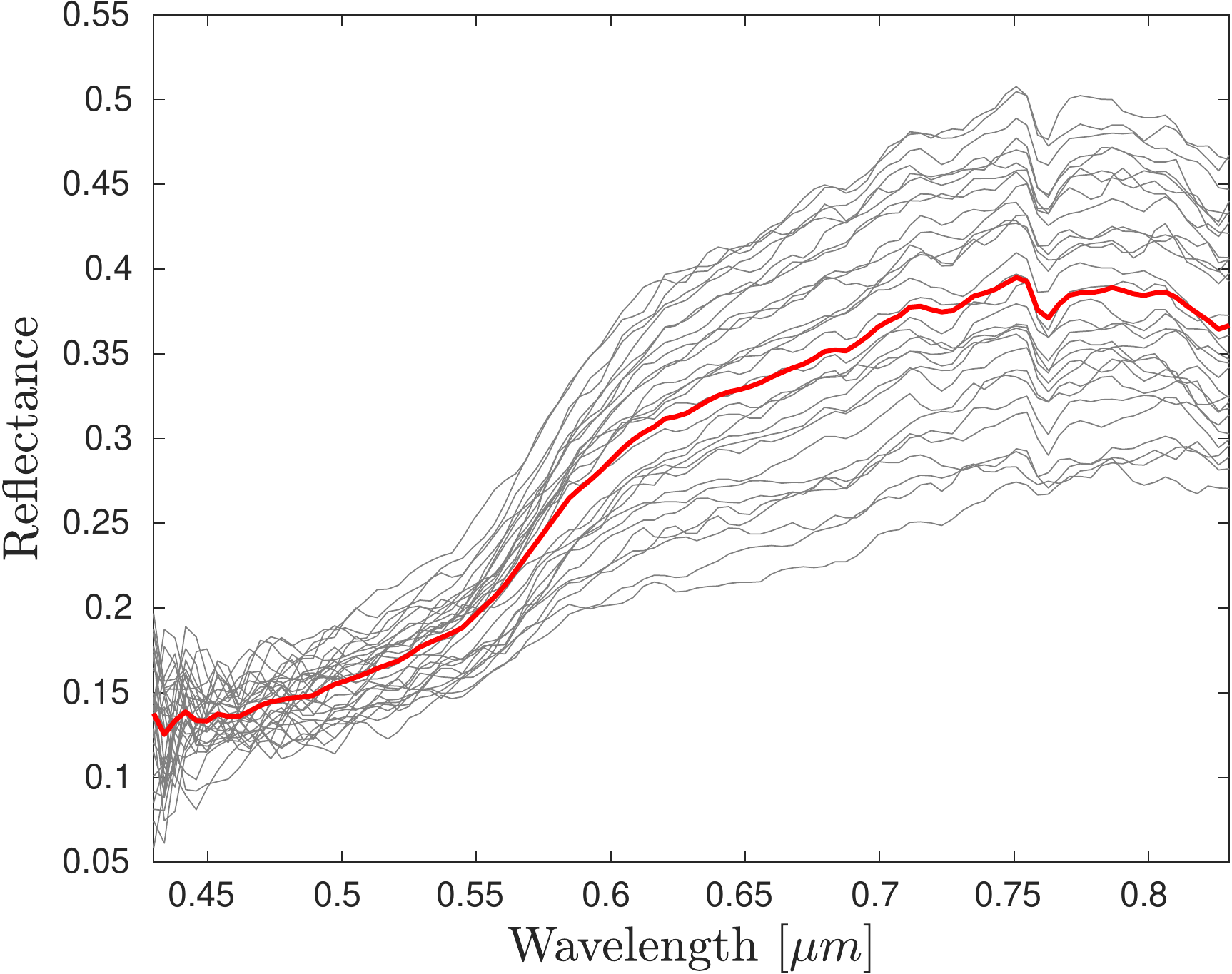}
    \vspace{-0.25cm}
    \caption{Examples of 30 pixel instances classified as red roof in the Pavia image (in gray), which are primarily affected by illumination, and their spectral average (in red). The average Pearson correlation coefficient between each signature and the scaled version of the mean spectra that is closest to it is about~$0.993$, indicating a good agreement between illumination-based spectral variability and the constant scaling model.}
    \label{fig:ref_roofs_paviah_samples}
\end{figure}

Considering that only the amplitude of the incident radiation changes along the scene, the reflectance spectra of the observed pixels in the LMM becomes scaled by a constant positive factor.
This model agrees with the observation that most of the variability in a hyperspectral image can be represented by a constant scaling of reference endmembers~\cite{shaw2003spectralImagRemote}.
As a simple empirical verification, we plot a random subset of 30 pixels of red roofs from the Pavia image, which are pure pixels mostly affected by illumination effects. The results, which are depicted in Fig.~\ref{fig:ref_roofs_paviah_samples}, indicate that these pixels differ mostly by a scaling factor.

\begin{figure}
    \centering
    \includegraphics[width=\linewidth]{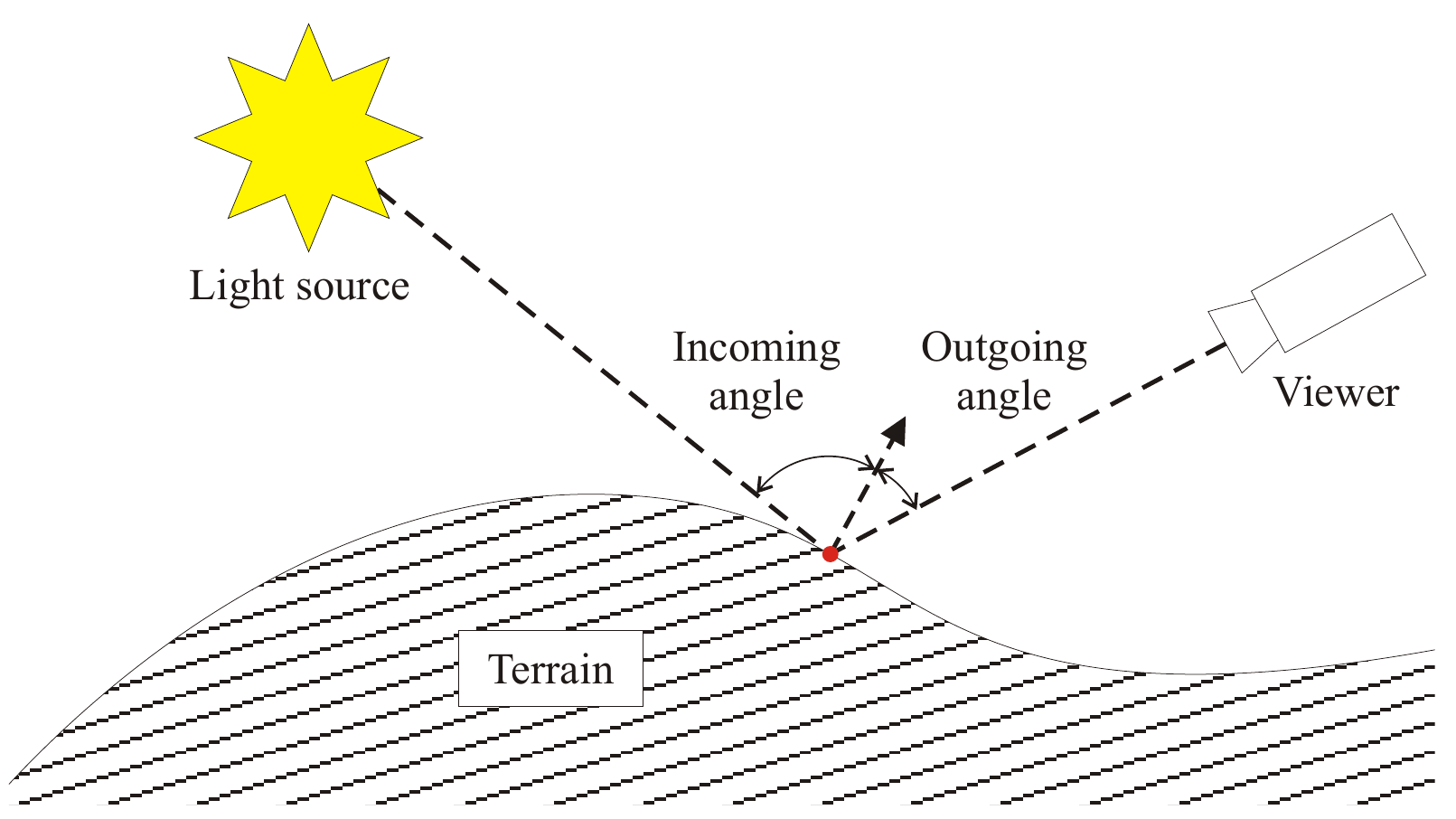}
    \vspace{-0.65cm}
    \caption{Hapke's model relates the reflectance to the incidence angles of the light source and observer/viewer shown in this figure, given the material's single scattering albedo and photometric parameters~\cite{heylen2014review}.}
    \label{fig:bidirectional_reflectance_fcn}
\end{figure}

Although a constant scaling model is intuitive and simple, a more rigorous conclusion can be achieved by analyzing the dependence of radiative transfer models with the topography of the scene.
To this end, one could resort to the model developed by Hapke~\cite{Hapke1981,HapkeBook1993}, which describes the bidirectional reflectance (i.e., the reflectance as a function of the incidence angles of the light source and observer/viewer depicted in Fig.~\ref{fig:bidirectional_reflectance_fcn}) as a function of the single scattering albedo and of photometric parameters of the material~\cite{heylen2014review}.

Hapke's model suggests a more complex relationship between the endmember signatures and the topography.
In this context, the mixture of materials is assumed to happen at the macroscopic level, allowing for the consideration of the LMM in the albedo domain, where Hapke's model acts separately on each endmember. %
Besides the dependency on the spectral signature with photometric parameters, which shall be discussed in the next section, the dependence on the single scattering albedo\footnote{i.e. the ratio between reflected and received radiation, as a function of the viewing angle.} indicates that changes in incident angles can affect each material in a pixel differently from the others, since the behavior of the reflectance as a function of the angle is different for each material.
This indicates that each endmember/material in a pixel can be differently affected by topographic effects. Furthermore, the nontrivial relationship between geometry and the spectral signatures leads to a more complex variation than single scaling for each endmember for high albedo materials~\cite{drumetz2016blindUnmixingELMM,drumetz2019ELMM_from_Hapke}.
Besides, even small topographic variations can significantly affect the ground reflectance. For instance, in~\cite{combal2002effectSmallTopographicVariatVeget} experimental studies found that even small slopes (of less than 10 degrees) originating from irregularities in tree canopy can lead to appreciable (enough to influence the results of subsequent tasks) changes in the measured reflectance of vegetation spectra.

\subsection{\textbf{Intrinsic spectral variability}}

Another important source of spectral variability is the intrinsic variation pertaining the definition of a material, which is also called \emph{intrinsic variability}.
The characterization of this type of variability has been prominently studied in the area of vegetation monitoring, where it poses a huge challenge to the ability to identify tree species from spectral measurements~\cite{cochrane2000variabilityVegetationClassification,zhang2006variabilityInfluenceClassfTropicalTree}, and also to the characterization of soil and mineral spectra.
Vegetation spectral signature can change due to many factors, including micro-climates, soil characteristics, precipitation, presence of heavy metals and drought, foliage age and colonization by leaf pathogens~\cite{cochrane2000variabilityVegetationClassification}. The spectral signature of soil is also heavily affected by variations in its composition and moisture content~\cite{baumgardner1986reflectancePropSoil}.
Furthermore, intrinsic spectral variability is also common in mineral spectra due to variations in the grain size distribution and the presence of variable amounts of impurities~\cite{crowley1986VariabReflectanceRocksImpurities,clark1999spectroscopyMineralsManual}. 
Moreover, it also depends on what level of detail is adopted to represent a given material (e.g., a \emph{tree} endmember may possibly be split into \emph{trunk} and \emph{leaf} endmembers), which is generally application dependent~\cite{franke2009hierarchicalMESMA}.
Although imposing a large impact on the endmember spectral signatures, the dependence of intrinsic spectral variability on physico-chemical parameters, which are usually unknown, makes it very \mbox{hard to tackle.}

\begin{figure}
    \centering
    \begin{minipage}{.32\linewidth}
        \centering
        \includegraphics[width=\linewidth]{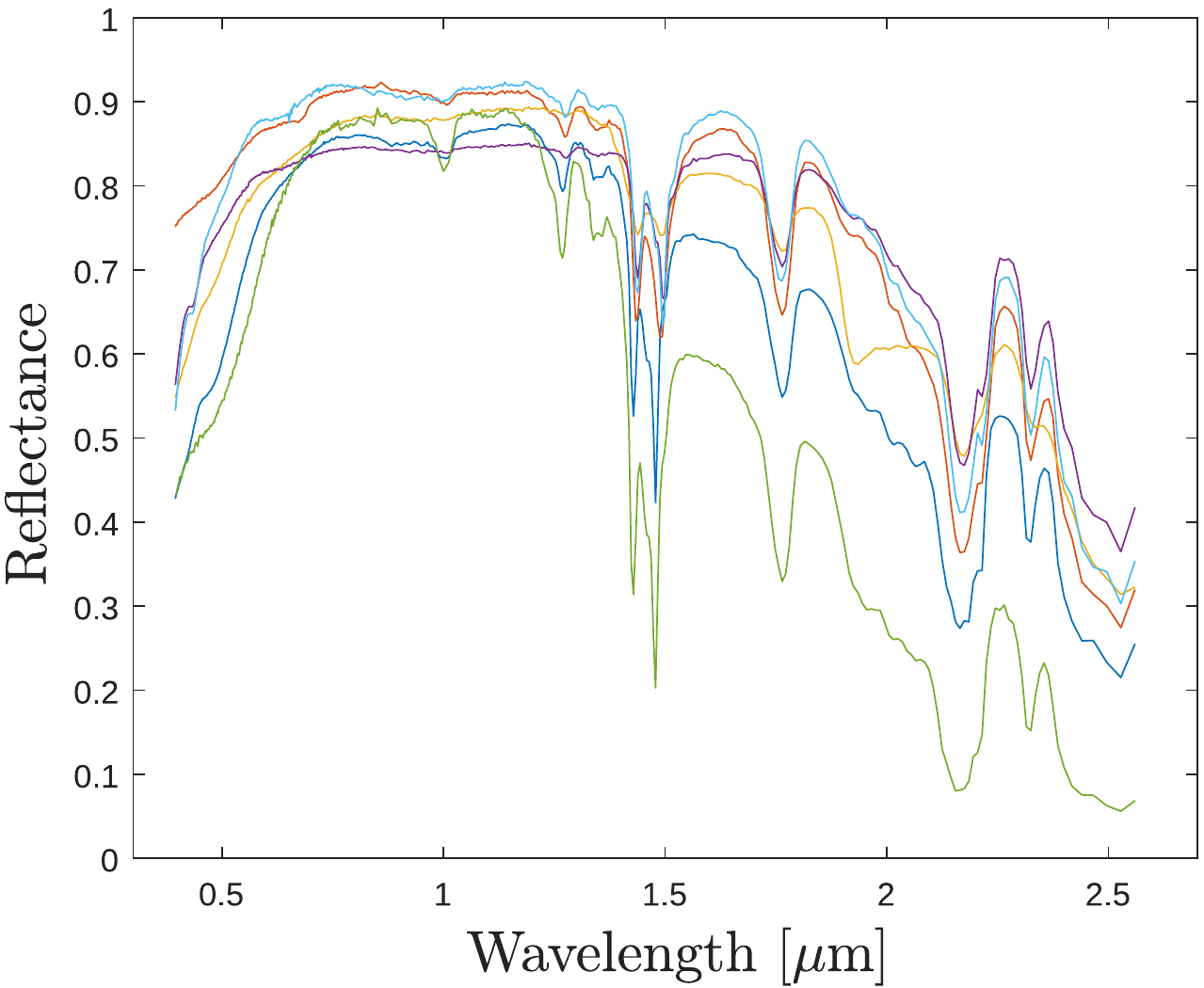} \\ (a)
    \end{minipage}
    \begin{minipage}{.32\linewidth}
        \centering
        \includegraphics[width=\linewidth]{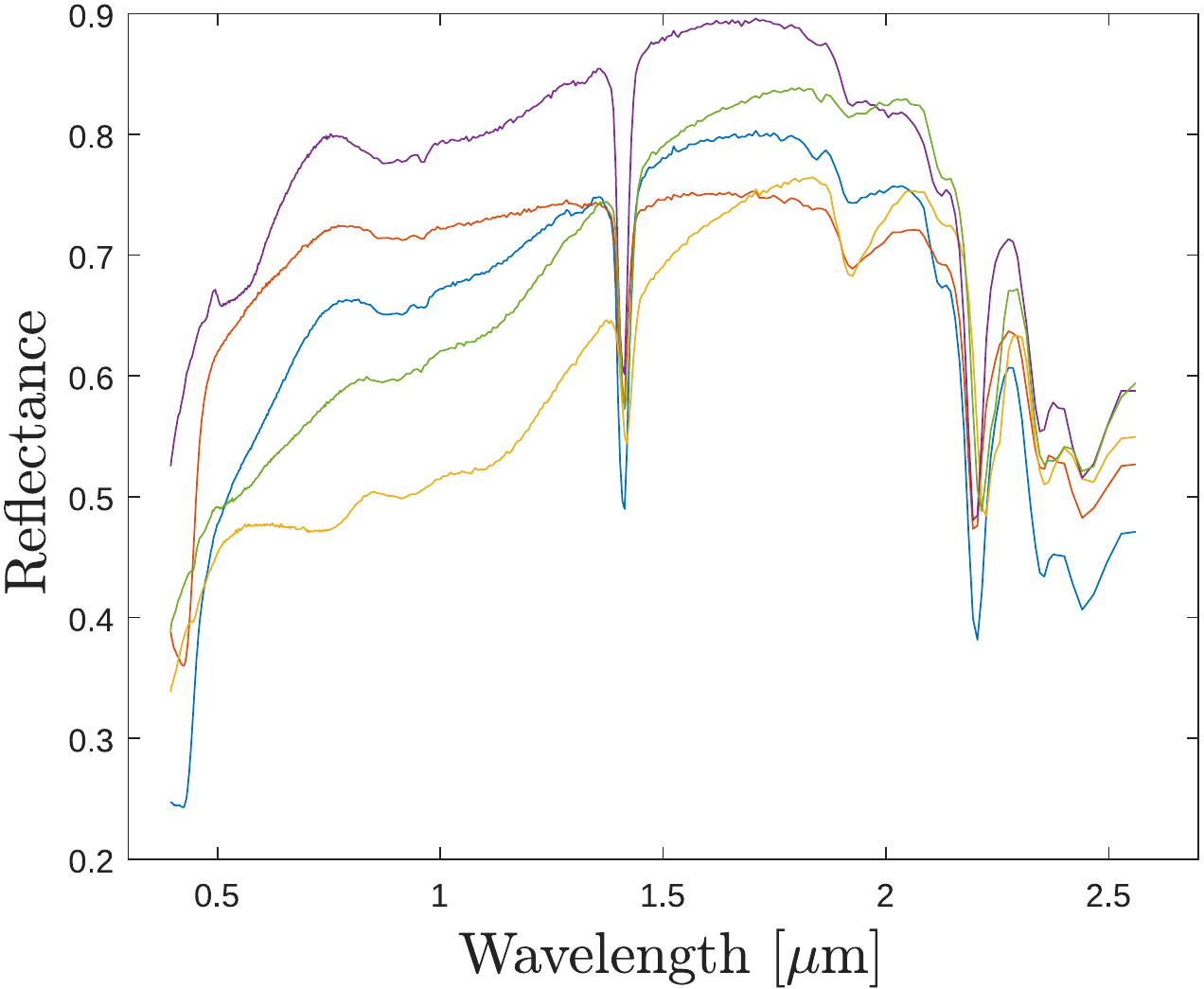} \\ (b)
    \end{minipage}
    \begin{minipage}{.32\linewidth}
        \centering
        \includegraphics[width=\linewidth]{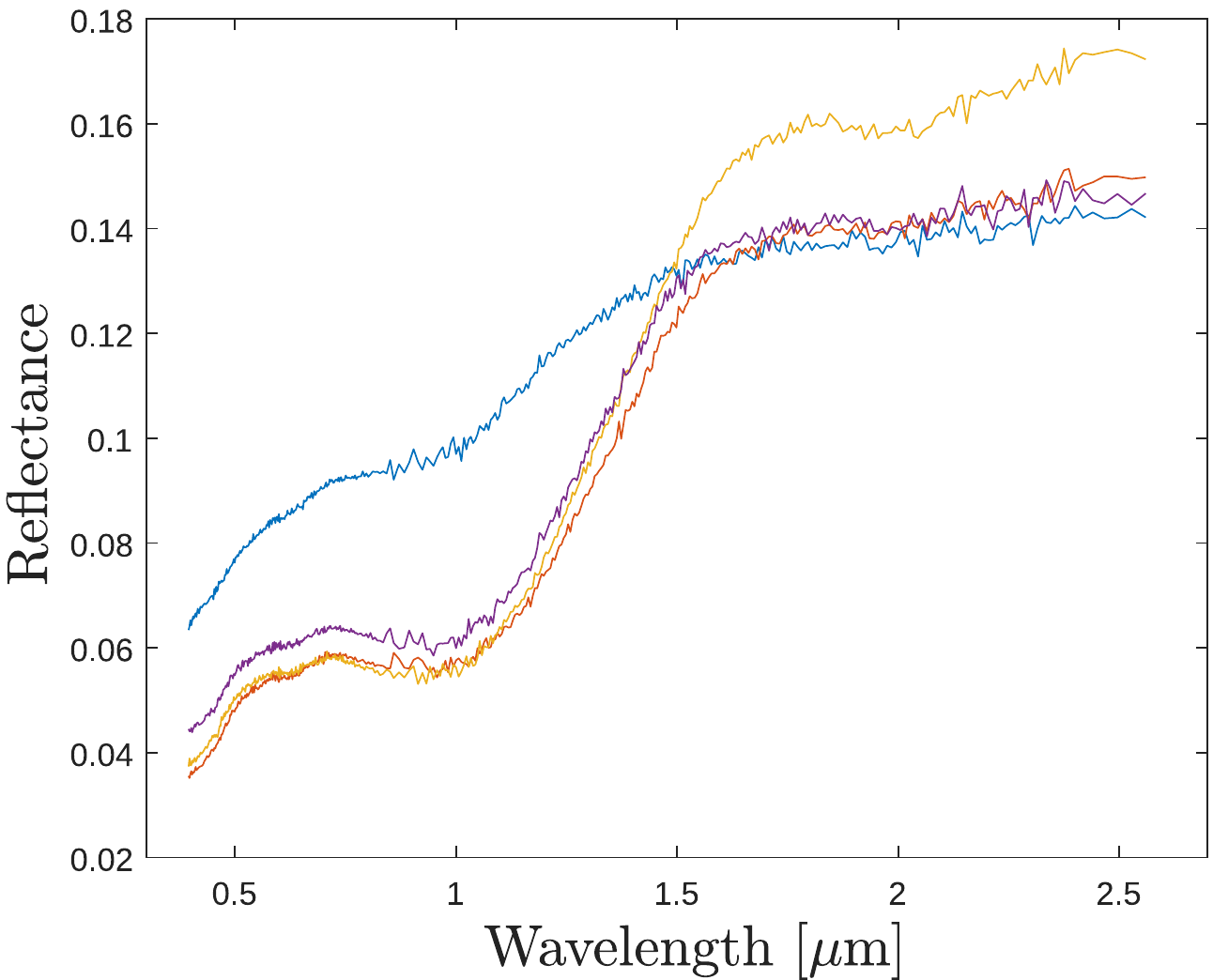} \\ (d)
    \end{minipage}
    \caption{Samples of variation of spectra from the USGS library. (a) Alunite. (b) Muscovite. (c) Pyrite.}
    \label{fig:sampleSpectraVarUSGS}
\end{figure}

One characteristic consistently observed in experimental studies is the smoothness of the observed spectra (i.e., the reflectance varies slowly between spectral bands). This behavior can be taken into account when designing SU algorithms.
Moreover, unlike spectral changes caused by illumination and topography effects, intrinsic spectral variability frequently presents a considerable dependence of the variability amplitude with the spectral wavelength. 
For instance, the signatures of different instances of minerals in the USGS library depicted in Fig.~\ref{fig:sampleSpectraVarUSGS} show complex dependence between the reflectance variation and wavelength.
The samples from alunite and muscovite show a variability that is far from uniform across the spectrum. Moreover, different instances from pyrite display complex variation, which is not consistent across all samples, occurring independently in different regions of the spectra.
This behavior has been verified in similar experimental studies in other works, and poses a significant challenge for differentiating mineral classes based on their spectral signatures~\cite{price1994HowUniqueSignatures}.

These characteristics are even more prominent and well known in the spectral variation of vegetation reflectance, which shows significant dependency on the wavelength and behaves very differently in visible, near-IR, and short-wave-IR ranges~\cite{asner1998biophysicalVariabilityCanopy}.
This means that a simple scaling of a reference spectral signature is usually not sufficient to account for variations within tree species~\cite{cochrane2000variabilityVegetationClassification}.
Extensive experimental studies support this claim. In~\cite{cochrane2000variabilityVegetationClassification} the author found that the variation of spectral reflectance in the visible and near-infrared regions can occur independently when measuring tropical forest canopy in Brazil.
Similar inhomogeneity in spectral variation was also observed in other studies with tropical tree species~\cite{ferreira2013variabilityTropicalLeaf} and also in many distinctive environments, including conifer~\cite{gong1997coniferVariability} and boreal tree species~\cite{lukevs2013variabilityLeaves}.
Similar non-uniform variation trends are also consistently observed in seasonal changes as indicated by many experiments, including in salt marshes~\cite{gao2006multiSeasonalSaltMarshChina}, semi-arid environments~\cite{schmidt2000variabilityVegetationArid} and boreal tree species~\cite{mottus2014seasonalSpectraBirchLeaves}.
Furthermore, nonuniform spectral variations have also been observed in samples from mineral, soil and rock spectra~\cite{price1994HowUniqueSignatures}.

Numerous works model the spectral signature of materials as a function of photometric or chemical properties of the medium, being based on either radiative transfer modelling or in empirical approaches.
A well known example is Hapke's model, which describes the spectra of a surface composed of particles as a function of parameters such as surface roughness and density and size of the particles~\mbox{\cite{Hapke1981,HapkeBook1993}}.

Another prominent line of work models the spectral characteristics of vegetation and soil samples as a function of biophysical parameters~\cite{jacquemoud2001leafOpticalPropertiesReview}. Models of this kind have been applied for the estimation of leaf biochemistry from the observed spectra.
An important example consists of the characterization of leaf reflectance spectra as a function of leaf biophysical parameters~\cite{jacquemoud2001leafOpticalPropertiesReview}, for which a wide variety of models have been used, ranging from a simple description of leaf scattering and absorption properties to complex models which perform a detailed description of the plant cells' shape, size, position, and biochemical content~\cite{jacquemoud2001leafOpticalPropertiesReview}.
Some instances of those models include the characterization of the spectra of broadleaf vegetation as a function of leaf mesophyll structure, pigment and water concentration~\cite{jacquemoud1990PROSPECTmodelLeaf} or as a function of leaf angular profiles~\cite{verhoef1984reflectanceGeometrySAILmodel}, and of pine needles as a function of cellulose, lignin and water content~\cite{dawson1998LIBERTYmodelBiochemicalPineNeedles}.
Other works model soil reflectance spectra as functions of moisture conditions~\cite{lobell2002moistureEffectsOnReflectance,somers2009modelSoilSpectraMoisture,sadeghi2015modelSoilSpectraMoisturePhysical}, and snow albedo as a function of snow grain sizes and liquid equivalent depth~\cite{wiscombe1980modelSnowAlbedoRTF}.

\begin{figure}[t]
    \centering
    \begin{minipage}{.32\linewidth}
        \centering
        \includegraphics[width=\linewidth]{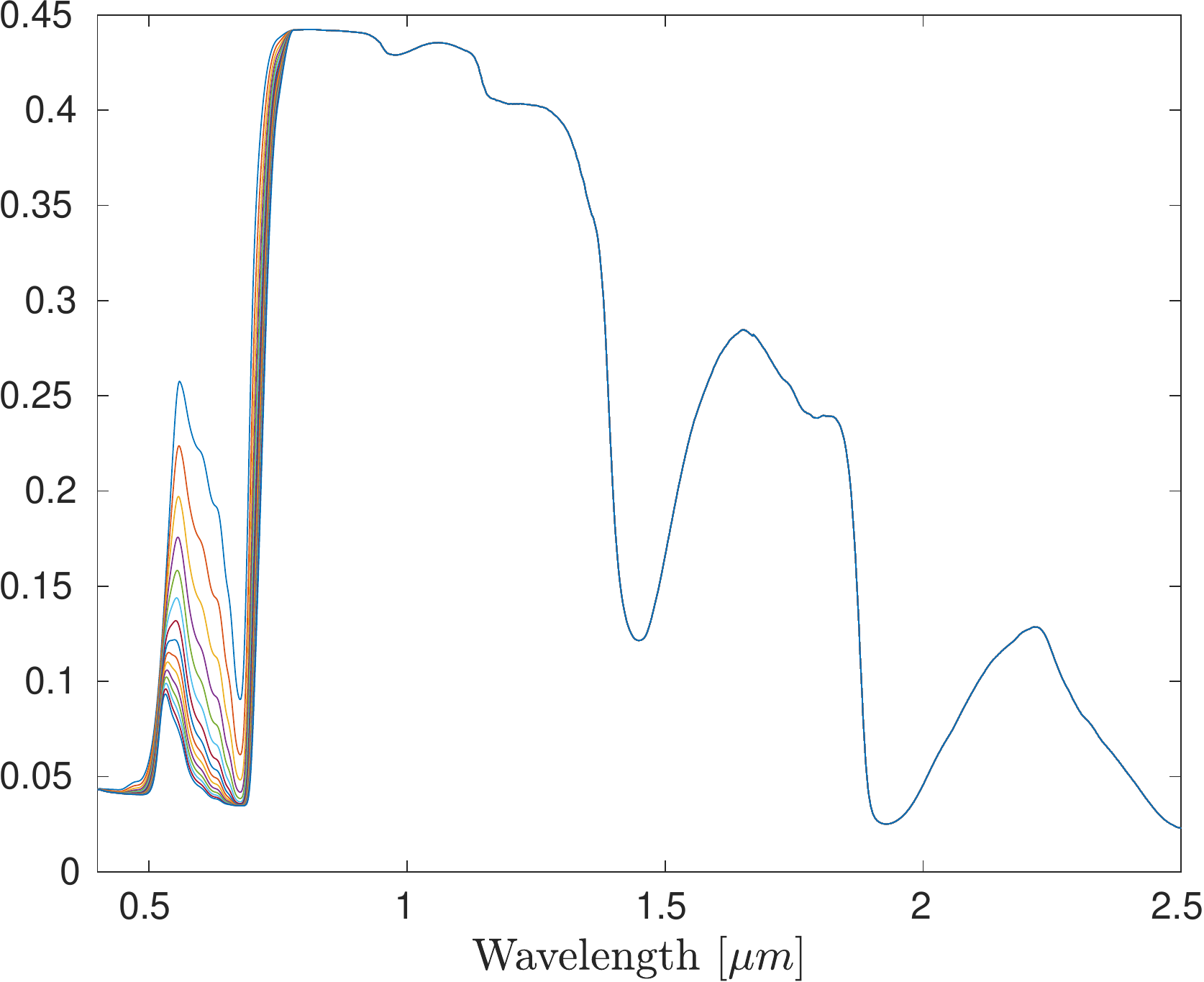} \\ (a)
    \end{minipage}
    \begin{minipage}{.32\linewidth}
        \centering
        \includegraphics[width=\linewidth]{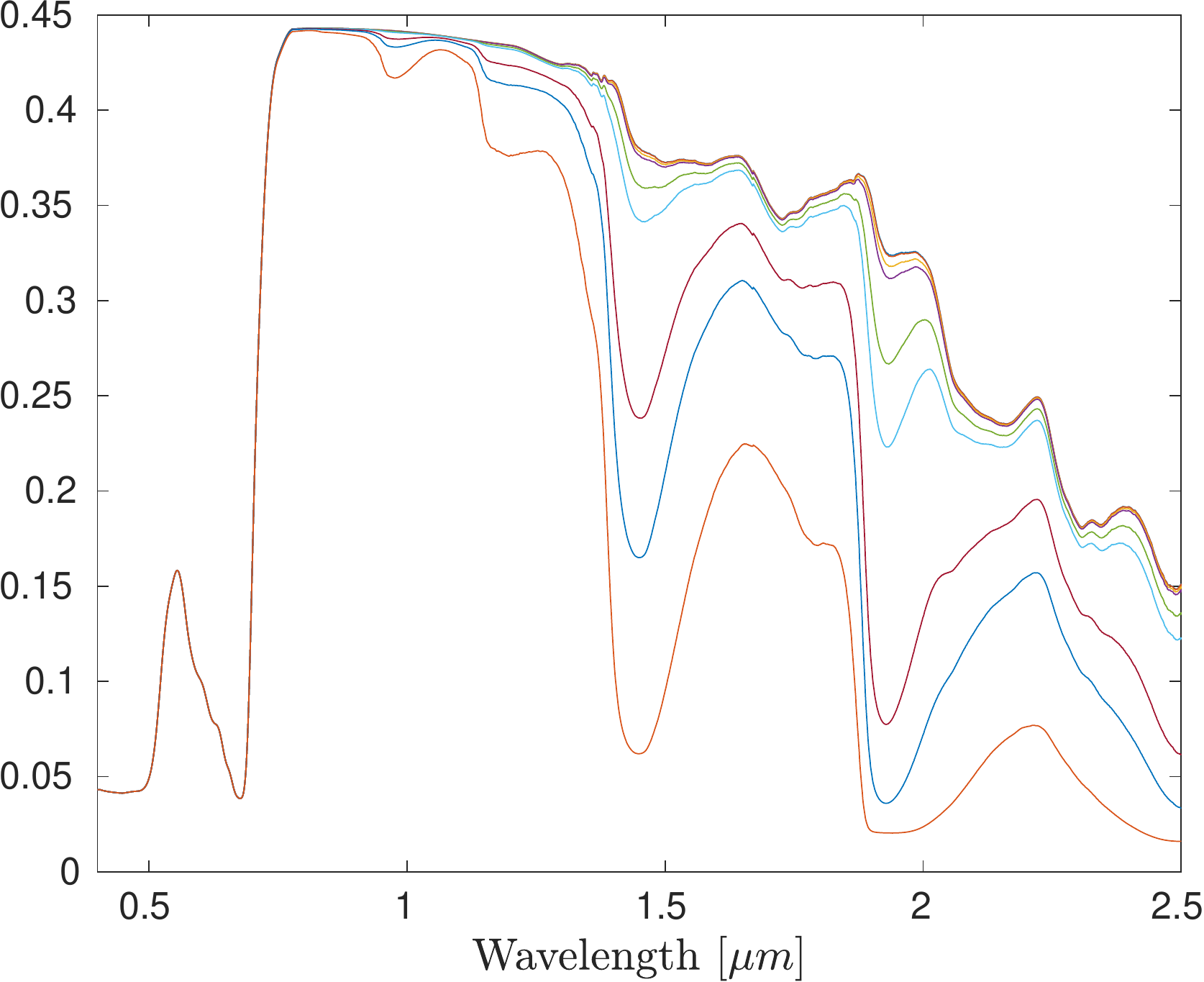} \\ (b)
    \end{minipage}
    \begin{minipage}{.32\linewidth}
        \centering
        \includegraphics[width=\linewidth]{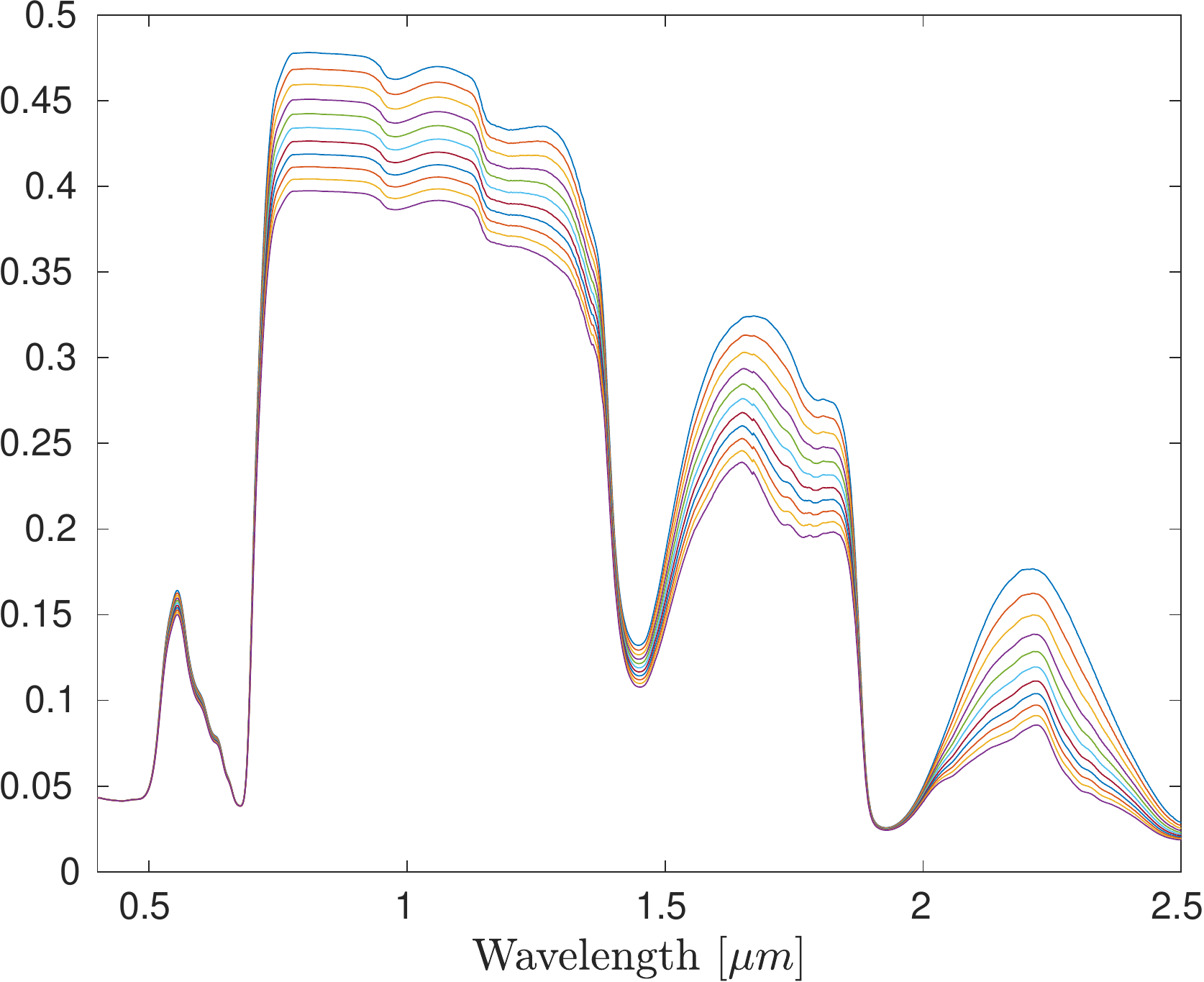} \\ (c)
    \end{minipage}
    \caption{Reflectance spectra for vegetation generated with the \mbox{PROSTECT-D} model~\cite{feret2017prospect_d} for varying degrees of (a) chlorophyll content, (b) equivalent water thickness, and (c) dry matter content.}
    \label{fig:vegetationSpectraPROSPECTD}
\end{figure}

As an illustrative example, we generated spectral signatures of vegetation spectra using the \mbox{PROSPECT-D} model~\cite{feret2017prospect_d} as a function of varying degrees of chlorophyll content, equivalent water thickness and dry matter content. The resulting signatures, depicted in Fig.~\ref{fig:vegetationSpectraPROSPECTD}, show that intrinsic spectral variability can present complex patterns and non-uniformity, as it is often concentrated in specific regions of the spectrum.

Through their analytical characterization of EM spectra, these kinds of models confine spectral variability to lie on a low-dimensional manifold. This constitutes important information that can be leveraged to alleviate/reduce the severe ill-posedness of unsupervised SU problems accounting for spectral variability.

Another important characteristic is that endmembers affected by intrinsic spectral variability usually display significant spatial correlation~\cite{webster1989spatialCorrelationEndmembers}.
For instance, many experimental geostatistical works evaluating the spatial distribution and variability of the physico-chemical properties of the soil (e.g., sand and clay concentration, electrical conductivity, pH, compaction and available elements such as nitrogen, phosphorus and potassium) have reported significant spatial correlation/smoothness in these properties.
Reports include measurements performed in Rhodes grass crop terrain~\cite{tola2017soilPhysicochemicalVariability}, calcareous soils~\cite{najafian2012spatialChemicalCalcareousSoils}, rice fields~\cite{wei2009spatialVariabilitySoilRiceField} and tobacco plantations~\cite{hou2010spatialSoilVariabilityTobacco}.
Besides directly impacting the spectral signature of the soil, these characteristics have been widely acknowledged to directly influence vegetation growth (e.g., they show strong correlation with crop productivity~\cite{tola2017soilPhysicochemicalVariability}), and hence their spectral signature~\cite{asner1998biophysicalVariabilityCanopy,tola2017soilPhysicochemicalVariability}. Therefore, spatial correlation in the variability is expected both in soil/terrain and in vegetation signatures. 
A similar behavior has also been observed in mineral spectra in the presence of spatially correlated grain size distributions and impurity concentrations~\cite{crowley1986VariabReflectanceRocksImpurities,clark1999spectroscopyMineralsManual}.
This implies that the variability tends to be small in small spatial neighborhoods, even though it may be large across a large scene. This fact can be leveraged to design SU algorithms since it supplies information that can be used to reduce the severe ill-posedness of the problem.

\begin{figure}
    \centering
    \hspace{-3.5ex}
    \begin{minipage}{.32\linewidth}
        \centering
        \includegraphics[width=1.17\linewidth]{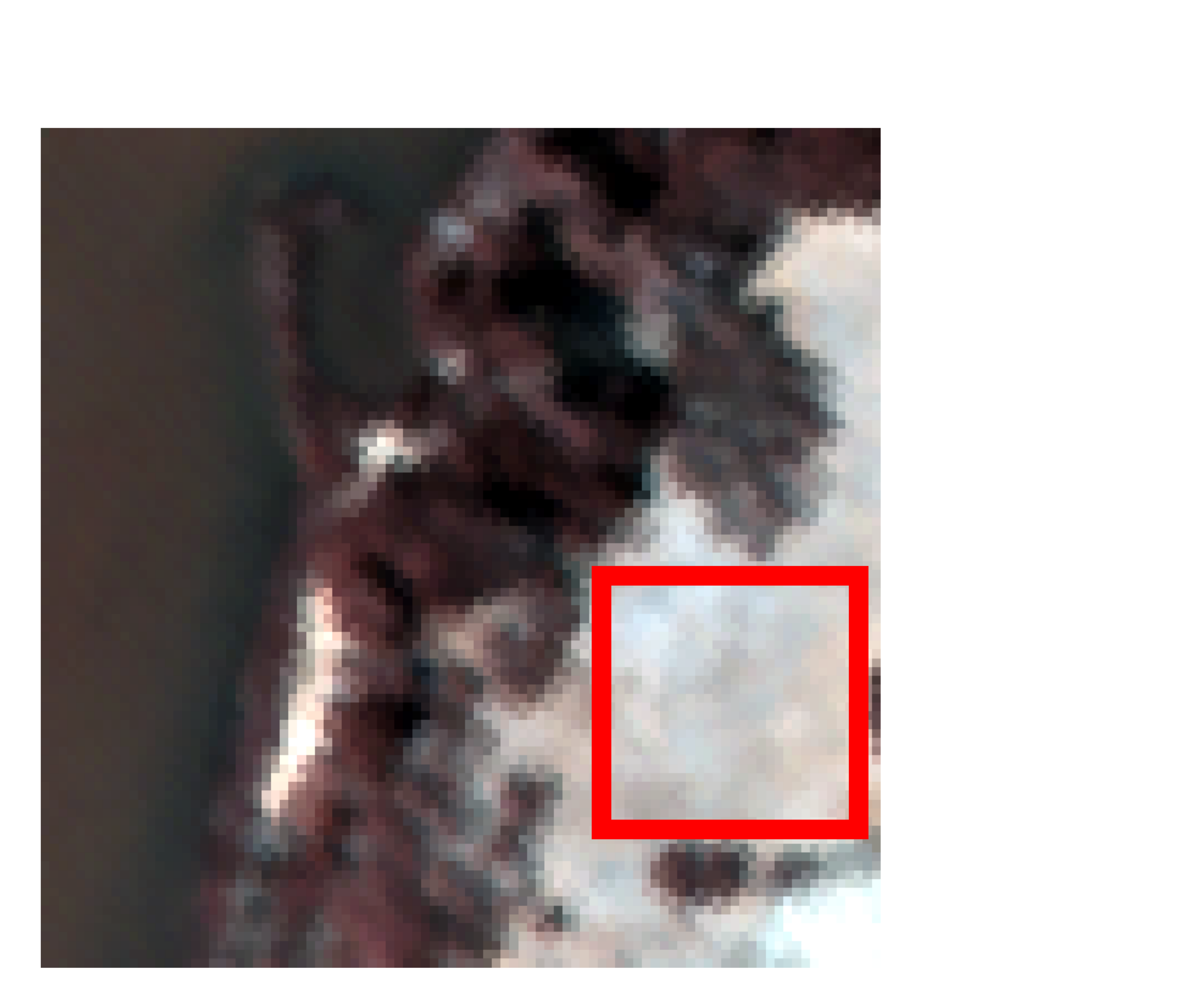} \\ (a)
    \end{minipage}
    \hspace{-3ex}
    \begin{minipage}{.32\linewidth}
        \centering
        \includegraphics[width=1.17\linewidth]{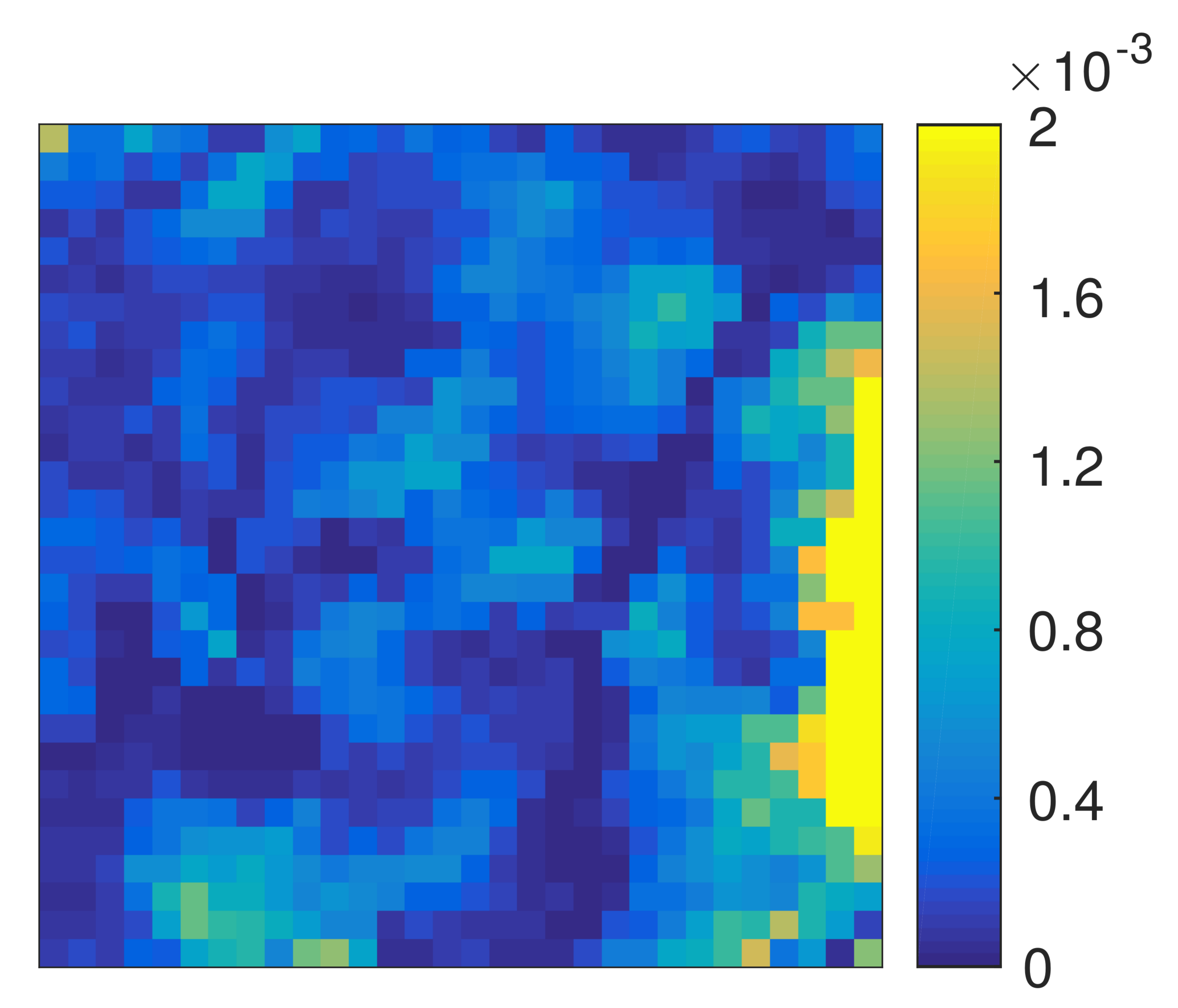} \\ (b)
    \end{minipage}
    \hspace{0.3ex}
    \begin{minipage}{.32\linewidth}
        \centering
        \includegraphics[width=1.17\linewidth]{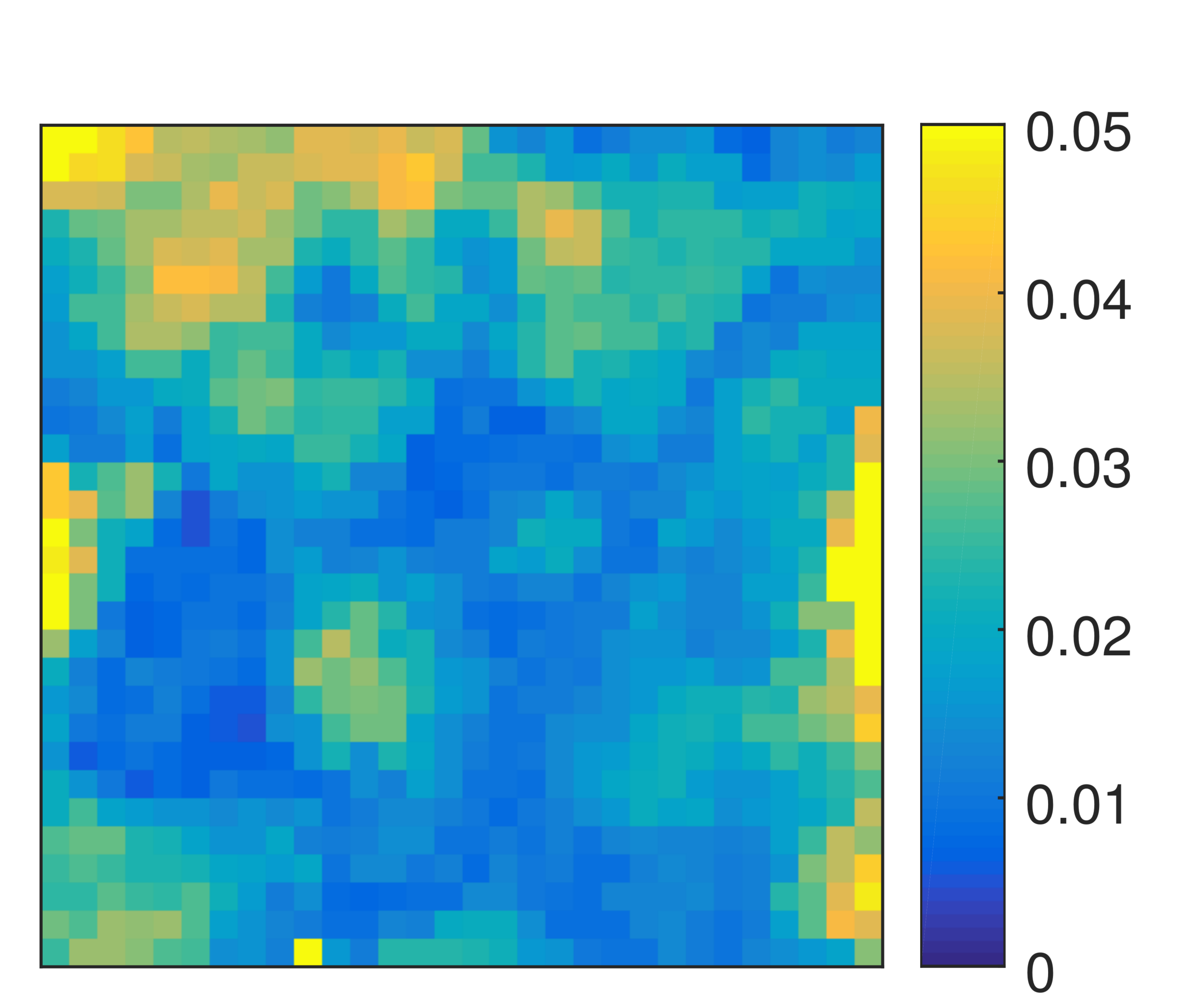} \\ (c)
    \end{minipage}
    \caption{Spatial behavior of endmember variability. (a) Soil subregion of the Samson image (highlighted by a red square). (b) Euclidean distance and (c) spectral angle between each pixel and the average spectra of the region.}
    \label{fig:samson_soil_ex_variability}
\end{figure}

To illustrate this effect, we performed an experiment by measuring the spectral variability in a homogeneous region (composed by mostly pure pixels) of soil in the Samson image, depicted in Fig.~\ref{fig:samson_soil_ex_variability}-(a).
We then computed the Euclidean\footnote{The Euclidean distance between $\bx$ and $\by$ is computed as $\sqrt{\frac{1}{N} \sum_{i=1}^N (\bx_i-\by_i)^2}$.} distance and the spectral angle between each soil pixel and the average spectra of all pixels in the subregion, which was used as a reference material signature. The results are depicted in Figs.~\ref{fig:samson_soil_ex_variability}-(b) and~\ref{fig:samson_soil_ex_variability}-(c), where it can be seen that the variability shows strong spatial correlation, as observed both in the Euclidean distance and spectral angle.

\section{Unmixing Methods That Use Spectral Libraries}
\label{sec:overall_methods_libs}

\enhancedcoloredtbox{Methods that use Spectral Libraries:}{
\begin{itemize}
    \item[$+$] The methods are usually conceptually simple and easy to interpret
    \item[$-$] The quality of the SU results depends strongly on the spectral library
\end{itemize}}

One of the main approaches to address spectral variability in SU is to consider large libraries of spectra acquired \textit{a priori}. These libraries contain different instances of each material in a scene, and the unmixing problem becomes generally equivalent to finding which of these signatures can best represent each pixel in the scene. Different algorithms have been proposed for this task, which we review in the sequel.

The spectral libraries used by these methods are sometimes called \emph{bundles}, and should in principle account for all possible variations of each material. Mathematically, they are represented as
\begin{align} \label{eq:EMs_as_sets_model}
    \mathcal{M}_p {}={} \big\{ \widetilde{\bm}_{p,1},\ldots,\widetilde{\bm}_{p,M_p}
     \big\}
    , \,\,\, p=1,\ldots,P \,,
\end{align}
where~$\mathcal{M}_p$ is a library/bundle containing $M_p$ reference spectral signatures~$\widetilde{\bm}_{p,i}\in\amsmathbb{R}^L$ of the $p^{\rm th}$ material, and $P$ is the number of materials in the scene. The spectral signature of each material in the $n^{\rm th}$ pixel $\by_n$ of a hyperspectral image is then represented as an unknown element $\bm_{n,p}\in\mathcal{M}_p$ belonging to this bundle.

Those sets can be readily used to constrain the endmember matrices of the LMM for the $N$ pixels to belong to a new set $\bM_n\in\mathcal{M}$, with $n = 1, \dots, N$, where
\begin{align} \label{eq:EM_libs_structured}
    \mathcal{M} = \Big\{[\bm_{1},\ldots,\bm_{P}], \bm_{p}\in\mathcal{M}_p ,\,p=1,\ldots,P\Big\}
    \,,
\end{align}
is the set of all possible endmember matrices, with $\prod_{p=1}^PM_p$ elements.
This definition assumes that only one signature from each library~$\mathcal{M}_p$, $p=1,\ldots,P$ is present in each pixel. However, other representations of the EM signatures as, e.g., sparse or convex combinations of the elements in $\mathcal{M}_p$ can also be considered in order to obtain more flexibility (see, e.g.,~\cite{yuan2015NMFusingSpectralLibrary,uezato2018SU_variabilityAdaptiveBundlesDoubleSparse,drumetz2019SU_bundlesGroupSparsityMixedNorms}). Such strategies will be discussed in Sections~\ref{sec:MESMA_and_variants} and~\ref{sec:sparse_SU}.

Different methods have been proposed to solve the SU problem using spectral libraries. These can be roughly divided into four groups of formulations: MESMA, Sparse SU, machine learning, and spectral transformations.
The MESMA algorithm and its variants formulate SU as a computationally demanding optimization problem, and achieve good quality.
Sparse SU formulations use mathematical relaxations to the MESMA problem that are computationally easier to solve.
Machine learning algorithms provide more flexible ways to do SU but also at a large computational complexity.
Spectral transformations are empirically-oriented techniques that can be used to improve methods from the first three categories.

\cred{Although all these families of methods use spectral libraries to address spectral variability in SU, the reasoning underlying each of them can be quite different, leading to varying degrees of required user supervision, computational complexity, and abundance estimation quality, as illustrated in the diagram of Fig.~\ref{fig:decision_tree}. Moreover, additional prior knowledge can be considered in different ways, including, e.g., the design of principled neural network architectures and the manual specification of the robustness of particular spectral bands to variability. We review each family of approaches in the following.}

\subsection{\textbf{MESMA and variants for small spectral libraries}} \label{sec:MESMA_and_variants}

\enhancedcoloredtbox{MESMA and Variants:}{
\begin{itemize}
    \item[$+$] Generally provide good SU results
    \item[$+$] Are easy to setup (few or no parameters)
    \item[$-$] Have a very high computational complexity
    \item[$-$] The results depend strongly on the quality of the spectral library available
\end{itemize}}

The Multiple Endmember Spectral Mixture Analysis (MESMA) algorithm~\cite{roberts1998originalMESMA} and its variants (sometimes also referred to as \textit{iterative mixture analysis cycles}) are among the most widely used algorithms for this task.
These methods allow the endmember signatures to vary on a per-pixel basis while following the model in~\eqref{eq:EMs_as_sets_model}.
The unmixing problem is solved by searching for the endmember and abundance combinations that result in the smallest reconstruction error (RE) for each observed pixel, i.e.,
\begin{align} \label{eq:opt_prob_MESMA_i}
\begin{split}
    & \underset{\ba_n,\bM_n}{\arg\min} \,\,\, 
    \|\by_n - \bM_n\ba_n \|^2
    \\
    & \text{subject to} \quad \bM_n \in \mathcal{M}, \,
    \ba_n \geq0, \, \cb{1}^\top\ba_n = 1 \,.
\end{split}
\end{align}
The endmember matrices $\bM_n$ constructed by taking spectra from the bundles are sometimes called \textit{endmember models}.

The MESMA algorithm has been employed in a wide variety of situations, including natural, urban and extra-terrestrial environments~\cite[p.1607]{somers2011variabilityReview}, and in single and multi-date scenarios~\cite{lippitt2018multidateMESMAshrublands}.
However, even though MESMA is very amenable to parallelization~\cite{bernabe2015parallelMESMA}, it consists of a combinatorial optimization problem whose associated computational cost can become very high.
More specifically, its computational cost scales as the product of the sizes of the individual libraries, as it consists of solving an FCLS problem $\prod_{p=1}^P M_p$ times~\cite{heylen2016alternatingAngleMinimization}. This can make the complexity of unmixing unrealistic for large library sizes.
Furthermore, the problem~\eqref{eq:opt_prob_MESMA_i} can become ill-posed when there are many endmembers in the bundles, since different material combinations can lead to very similar reconstruction errors. In order to circumvent these limitations, several modifications to the original MESMA algorithm have been proposed. %

Many variants of MESMA aim to provide computationally efficient approximate solutions to~\eqref{eq:opt_prob_MESMA_i}.
Early simplifications consist of an early stop of the exhaustive search optimization procedure~\eqref{eq:opt_prob_MESMA_i} by selecting the first EM model that presents a reconstruction error that is both below a threshold and well distributed across spectral bands~\cite{roberts1998originalMESMA}.
Another approach proposed is to solve~\eqref{eq:opt_prob_MESMA_i} approximately by performing unconstrained least squares with every possible endmember model, and then select the solution that yields positive abundances and the smallest reconstruction error~\cite{combe2008MELSUMunmixingMARS}.

Although these simple modifications successfully reduce the computational complexity of MESMA, the approximations involved can also negatively impact the abundance reconstruction results~\cite{dennison2004comparisonErrorMetricsEndmemberSelection}, what imposes practical limitations on the selection of the thresholds and tolerances.
This motivated the consideration of more elaborated strategies to provide more significant reductions of its complexity without impacting the unmixing performance.

An alternative approach to MESMA attempted to lessen the computational complexity by solving an angle minimization problem with respect to each library separately~\cite{heylen2016alternatingAngleMinimization,tits2015alternatingAngleMinimization2EM}. Although not guaranteed to converge to the optimal solution of~\eqref{eq:opt_prob_MESMA_i}, this strategy performed similarly to MESMA on practical experiments, and scales linearly with the library sizes, leading to computational improvements for large numbers of signatures~$M_p$, $p=1,\ldots,P$, in the EM bundles.
Another work considered a mixed integer linear program (MILP) reformulation of the MESMA problem. This approach allows for a more efficient computation of an exact solution to~\eqref{eq:opt_prob_MESMA_i} for small to medium scale problems~\cite{mhenni2018MESMA_MILP}.

A simple approach which is largely employed to reduce the computational complexity of MESMA is to perform a careful pruning of the spectral libraries $\mathcal{M}_p$, $p=1,\ldots,P$. This process attempts to remove redundant or irrelevant spectra from the libraries before unmixing. These approaches will be described in detail in Section~\ref{sec:sub_lib_reduction}.

Besides reducing its complexity, other approaches modify MESMA with the purpose of improving its accuracy.
For instance, an early practice attempts to alleviate the ill-posedness of MESMA by prioritizing models with a smaller number of endmembers, for otherwise comparable reconstruction errors~\cite{Song2005variabilityWeightedProbability,roberts2003countBasedEndmemberSelection}. This avoids increasing the complexity of the model for marginal gains. 
When consideration of material nuances is important, it may be important to allow multiple signatures of the same broader endmember class in the model. This was the case in~\cite{tan2014modifiedMultipleMESMA}, where effects such as different vegetation species in a single pixel were of interest.
Spatial information has also been considered with MESMA by using segmentation algorithms to divide the image into different homogeneous objects, which are then unmixed individually using a library also constructed from \mbox{object-based spectra~\cite{zhang2014objectBasedMESMA,zhang2016multiscaleObjectBasedMESMA}}.

A different formulation attempted to increase the flexibility of MESMA by allowing the endmembers of each pixel to be represented as a sparse, non-negative combination of the signatures contained in the library for their respective material class~\cite{yuan2015NMFusingSpectralLibrary,uezato2018SU_variabilityAdaptiveBundlesDoubleSparse}. Under this model, SU was then formulated as a non-convex optimization problem with different sparsity constraints, including both $L_{1/2}$~\cite{yuan2015NMFusingSpectralLibrary} and $L_0$-norm-based penalties~\cite{uezato2018SU_variabilityAdaptiveBundlesDoubleSparse}. This problem was solved using a multiplicative update rule in~\cite{yuan2015NMFusingSpectralLibrary}, and using the proximal alternating linearized minimization method in~\cite{uezato2018SU_variabilityAdaptiveBundlesDoubleSparse}.

Another set of approaches related to MESMA is referred to as \textit{fuzzy unmixing}. These methods consider a measure of uncertainty or indeterminacy in the estimated abundances by computing quantities such as average, maximum and minimum cover fractions.
One of the first approaches of this kind used linear programming methods to determine 
maximum and minimum fractional abundances for each material using spectral libraries extracted from the observed image~\cite{bateson2000endmemberBundlesFuzzyU}.
Another approach attempted to determine the abundance indeterminacy (i.e., its fuzzy membership amount for each value of abundance fractions) by evaluating how close synthetically mixed spectra with all possible endmember combinations were to the observed pixel spectra~$\by_n$~\cite{petrou1999fuzzyUnmixingTwoEndmembers}. This procedure, however, required the discretization of the abundance values and its computational complexity does not scale well with the number~$P$ of endmembers classes.

Other approaches performed linear SU with a large number of endmember models selected at random from the library. Afterwards, measures of uncertainty in the estimated fractional abundances such as maximum, minimum and average cover fractions were computed from these results, providing a more detailed characterization of the abundances~\cite{asner2000biogeophysicalAutoSWIR,asner2002spectralUnmixingAutoMCU,asner2003scaleBiophysicalAutoMCU}.
A similar work proposed to compute the final abundance fractions as a weighted sum of the abundances obtained from SU with each possible combination of signatures drawn from the library $\mathcal{M}$~\cite{Song2005variabilityWeightedProbability}. The weights corresponded to the probability of each EM model being actually present in the scene, which was supposed to be known \textit{a priori}.

\subsection{\textbf{Sparse unmixing}} \label{sec:sparse_SU}

\enhancedcoloredtbox{Sparse Unmixing:}{
\begin{itemize}
    \item[$+$] Generally is very computationally efficient (especially compared to MESMA)
    \item[$-$] SU results might not be as accurate as MESMA
    \item[$-$] Can be harder to interpret (e.g., it might select multiple signatures of the same material to represent a given pixel)
    \item[$-$] SU results are sensitive to the selection of the regularization coefficients
\end{itemize}}

An alternative approach to perform spectral unmixing with spectral libraries is to formulate the SU as a sparse regression problem, where we want to select a small number of spectral signatures from the library which can best represent each observed pixel according to the LMM.

Most sparse unmixing methods are based on an \emph{unstructured} library, which can be derived from~\eqref{eq:EMs_as_sets_model} by concatenating all the signatures in a single matrix~$\Mlib$, defined as:
\begin{align} \label{eq:EM_libs_unstructured}
    \Mlib {}={} \big[\widetilde{\bm}_{1,1}, \ldots,\widetilde{\bm}_{p,k},\widetilde{\bm}_{p,k+1},\ldots,\widetilde{\bm}_{P,M_P}
    \big]
    \,.
\end{align}

Using the spectral library defined in~\eqref{eq:EM_libs_unstructured}, the sparse unmixing problem can be formulated as the optimization problem~\mbox{\cite{bioucas2010ADMM_sunsal,iordache2011sunsal}}:
\begin{align} \label{eq:sparse_unmixing_opt_i}
\begin{split}
    & \underset{\ba_n\geq0}{\arg\min}
    \,\,\, \|\by_n-\Mlib\,\ba_n\|^2
    \\
    & \text{subject to } \|\ba_n\|_0 \leq P, \,\,\, 
    \cb{1}^\top \ba_n = 1 \,,
\end{split}
\end{align}
where $\|\cdot\|_0$ is the $L_0$ pseudo-norm, which counts the number of non-zero elements in a vector.
Different strategies have been proposed to solve the sparse SU problem using the $L_0$ pseudo-norm using, for instance, greedy (e.g., matching pursuit or forward-backward) algorithms~\cite{shi2014greedyMatchingPursuit_L0normSU1,tang2014greedyMatchingPursuit_L0normSU2}, Lagrangian function (regularized) formulations~\cite{shi2018ADMM_L0normSU}, or multi-objective optimization procedures that consider the reconstruction error and the sparsity of the solution jointly~\cite{xu2017multiObjective_L0normSU_genetic1,xu2018multiObjective_L0normSU_genetic2,xu2019multiObjective_L0normSU_genetic3}.
Note that~\eqref{eq:sparse_unmixing_opt_i} would be equivalent to MESMA if we added an additional linear structuring constraint to enforce the occurrence of only a single nonzero abundance per material class~\cite{mhenni2018MESMA_MILP}.

The optimization problem~\eqref{eq:sparse_unmixing_opt_i} is, however, non-convex and generally NP-hard to solve. It is therefore common to relax the $L_0$ pseudo-norm constraint into its convex surrogate, leading to the following optimization problem~\cite{bioucas2010ADMM_sunsal}:
\begin{align} \label{eq:sparse_unmixing_opt_ii}
\begin{split}
    & \underset{\ba_n\geq0}{\arg\min} \,\,\, 
    \|\by_n-\Mlib\,\ba_n\|^2 + \lambda \|\ba_n\|_1 \,,
\end{split}
\end{align}
where $\|\cdot\|_1$ is the $L_1$ norm and the parameter $\lambda$ controls the level of sparsity of the estimated abundances.
The sum-to-one constraint is not used in~\eqref{eq:sparse_unmixing_opt_ii} due to its incompatibility with the $L_1$ norm~\cite{bioucas2010ADMM_sunsal}.
Although problem~\eqref{eq:sparse_unmixing_opt_ii} is non-smooth, it is convex and can be solved very efficiently. Besides, it produced good experimental performance. This motivated a great deal of interest in sparse unmixing methods, resulting in a number of works proposing improvements such as the use of alternative sparsity promoting penalties~\cite{iordache2014collaborativeSparseRegressionUnmixing,qian2011unmixing_L12_NMF} or different means of spatial regularization~\cite{iordache2012sunsal_TV,borsoi2018superpixels1_sparseU}.
Sparse unmixing methods would merit a more comprehensive review, which is beyond the scope of this paper. Thus, in the following we restrict ourselves to modifications of the sparse SU framework specifically aimed at dealing with spectral variability or with structured libraries.

In~\cite{iordache2011unmixingGroupLasso}, $L_{2,1}$-norm based group sparsity constraints have been used to favor the selection of abundance vectors containing many entire material classes with zero proportions.
A later formulation considered a fractional group $(p,q)$-norm sparsity constraint as a generalization of the approaches based on the $L_{2,1}$-norm~\cite{drumetz2019SU_bundlesGroupSparsityMixedNorms}. The~$(p,q)$-norm penalty permits a better control of the sparsity within each group of variables, as well as the addition of the sum-to-one constraint. However, this comes at the expense of making the optimization problem non-convex.

Another sparse SU formulation~\cite{fu2016sparseUnmixingLibraryMismatches} proposed to explicitly represent mismatches between the library spectra and the hyperspectral image caused by different acquisition conditions. In this case, the spectral signatures of the library are also estimated in the SU process. However, they are constrained to be within a given Euclidean distance of a corresponding element of the library known \textit{a priori}. This allows the estimated signatures to vary arbitrarily within Euclidean balls centered at the library elements to compensate for spectral mismatches.

A different approach~\cite{berman2017comparisonSparseUnmixingMining} proposed to modify the LMM for unmixing mineral spectra in mining applications by including an additional term representing the mixture of the ``background'' spectrum of the endmembers. This background spectrum was defined as the low-frequency part of the spectral signatures, and was estimated \textit{a priori} from the library as a parametric function of smooth splines. The performance of an $L_1$-norm based sparse SU framework under this model was reported to be similar to MESMA, albeit at a much smaller computational cost.

\subsection{\textbf{Machine Learning Algorithms}}
\label{sec:machineLearning_SU}

\enhancedcoloredtbox{Machine Learning Algorithms:}{
\begin{itemize}
    \item[$+$] Very flexible approaches, in principle can deal with any effect that is represented in the training data
    \item[$-$] Most methods either have a large computational complexity or do not have a clear physical motivation
    \item[$-$] The SU quality depends on the \cmag{representativeness} of the training data (which is usually generated using a spectral library)
    \item[$-$] Generally do not return an EM spectra for each pixel
\end{itemize}}

Some works propose to address spectral variability using machine learning methods by formulating SU as a supervised regression problem. The objective is to learn transformations mapping the observed (mixed) pixel to the abundance fractions~\cite{guilfoyle2001comparativeUnmixingNeuralNetworksRBF,baraldi2001comparisonNeuroFuzzyUnmixing,okujeni2014comparisonMLregressionUrbanCover,bovolo2010fuzzyTrainingUnmixing,licciardi2011SU_NN_dimReduction} using a supervised training procedure. Mixed pixels with known proportions are employed as training data for algorithms such as neural networks, random forests or Support Vector Machines (SVMs).
These techniques can be straightforwardly adapted to address spectral variability by considering multiple spectral signatures for each endmember when generating the synthetic training dataset. This has been done either by directly applying regression methods~\cite{okujeni2013learningSVRunmixingUrban} or by converting SU into a classification problem by quantizing the solution space of abundance values and using a one-against-all strategy~\cite{mianji2011unmixingVariabilitySVMclassification}.
Another work modified the SVM cost function to directly minimize the unmixing reconstruction error during the training process~\cite{wang2013unmixingSVMregressionResidueConstr}.

Usually, these approaches result in extremely large training sets for large spectral libraries. Thus, even though some strategies such as bootstrap aggregation have been employed to speed-up the training process~\cite{okujeni2017ensembleLearningUnmixingSVM,rosentreter2017unmixingComparisonSVMlearning}, the computational cost is still very high.
Although methodologies to discard irrelevant (regarding the impact on performance) subsets of the training data~\cite{plaza2009smalTrainingSetsNNunmixing,plaza2010selectingTrainingSamplesNNunmixing} could in principle be applied to accelerate training, recent works have instead focused on modifying the algorithms to reduce their complexity.

One of the main reasons for this large complexity is that the training data must jointly describe spectral variations due to changes in both the abundances and in the endmembers. Recent works have tried to address this issue by using only pure pixels from a spectral library as training data. 
One such approach, which received considerable attention, consists of extended SVMs. Extended SVMs employ hybrid soft-hard classification or regression to address spectral variability.
It is assumed that the spectral space is separable by hyperplanes delimiting two complementary regions containing only pure and only mixed pixels, respectively~\cite{wang2009unmixingExtendedSVMpurePixels}.
The extended SVM is then trained to find a soft-hard classifier containing both 1) a hard classification rule consisting of the hyperplanes delimiting the regions in which the pixels are considered pure, and 2) a soft classification rule which determines the abundances of the pixels considered to be mixed.

Different forms of the extended SVM have been considered, using either a single~\cite{wang2009unmixingExtendedSVMpurePixels} or multiple kernels~\cite{gu2013unmixingExtendedSVMpurePixelsMultiKernel}, considering the abundance indeterminacy by computing the maximum and minimum proportion values similarly to the fuzzy SU procedures~\cite{li2015unmixingExtendedSVMgeometricAnalysis}, or using Fisher discriminant analysis to reduce the within-class spectral signature variability in the spectral library before training~\cite{li2016unmixingSVMandFisher}.
Although hybrid soft-hard classification methods can be fast to train, they lack a clear physical interpretation of the results since they have no direct relation to the physical mixing model. Moreover, the influence of spectral variability on the regions of the spectral space containing mixed pixels is limited since it only comes from the marginal hyperplanes that separate the pure from the mixed pixels regions~\cite{wang2009unmixingExtendedSVMpurePixels}.

A related strategy that also uses only pure pixels in the training process consists of modeling the latent function from the mixed pixel spectra to the abundance maps in a probabilistic framework as a multi-task Gaussian Process~\cite{uezato2016unmixingGaussianProcessVariability}. In this case, the abundance means and covariance matrices are obtained through the posterior distribution of the abundances conditioned on the training set (i.e., the spectral library) and on the mixed pixels.
This strategy was also extended to consider spatial correlation in a two step process by using the Gaussian Process results from~\cite{uezato2016unmixingGaussianProcessVariability} as input to the abundances prior information in a maximum a posteriori estimation problem~\cite{uezato2016GaussianProcessVariabilitySpatialInfo}.
Although this strategy has a strong statistical motivation, the introduction of additional constraints (e.g., abundance non-negativity and sum-to-unity) is not straightforward and results in high computational complexity.

\begin{color}{red}
Another work proposed to mitigate the influence of spectral variability by first processing the image using a geodesic SU method~\cite{Heylen2011}, before applying Gaussian process regression to estimate the final abundances~\cite{koirala2020geodesicSupervisedSUvariability}. Although possibly inaccurate, the preliminary abundances estimated by the geodesic SU algorithm are not affected by endmember variations caused by differences in illumination and acquisition conditions. The Gaussian process regression then learns to map the inaccurate initial abundances to the desired ones. Despite increasing the robustness to spectral variability, geodesic SU can introduce significant distortions in the abundances for complex data manifolds, which may not be trivial to compensate.

Note that other machine learning techniques have also been recently employed to perform SU without directly addressing spectral variability. This includes the use of convolutional neural networks~\cite{zhang2018unmixingDeepCNNs}, the consideration of neural networks well adapted to learn from fewer samples~\cite{zeng2020attentionScatteringNN_SU_fewSamples}, and the use of autoencoders to perform unsupervised SU by identifying the latent codes with the fractional abundances, and the decoder with the mixture model~\cite{palsson2018autoencoderUnmixing_IEEEaccess,su2019deepAutoencoderUnmixing,su2018autoencodersUnmixing,qu2018udas_autoencoderUnmixing}.
Other works also considered specific neural network architectures inspired by unfolding iterative optimization algorithms~\cite{qian2020modelInspiredNNs_SU}, or used Hopfield neural networks to optimize the SU cost functions more efficiently~\cite{li2016hopfieldNN_bilinearSU}.
Machine learning methods have also been recently applied in different experimental settings, such as in unmixing spectrally similar vegetation types~\cite{cooper2020ML_SU_mappingSimilarVegetation} and urban surfaces~\cite{mitraka2016nonlinearSU_urban_ML}, and when using training data collected at multiple locations~\cite{okujeni2018ML_SU_multi_site_libraries}.

Given the success that recent machine learning methods are achieving at different problems, particularly in the area of remote sensing~\cite{zhang2016deepLearningRemoteSensingReview,zhu2017zhang2016deepLearningRemoteSensingReview}, such techniques may bring important advances \cdgreen{if used to address} the EM variability problem in SU in the future.
\end{color}

\subsection{\textbf{Spectral transformations}}
\label{sec:spectral_transformations}

\enhancedcoloredtbox{Spectral Transformations:}{
\begin{itemize}
    \item[$+$] Can be seen as a ``pre-processing'' strategy that can be used jointly with other library-based SU methods
    \item[$+$] Are conceptually simple and of low-complexity
    \item[$-$] Many of the methods are empirical and require a significant degree of expert knowledge about the underlying application
    \item[$-$] The performance of the less-supervised methods depends strongly on the representativeness/quality of the library
\end{itemize}}

An approach frequently used to mitigate the effect of spectral variability in library-based SU consists of selecting a subspace of the spectral space that is minimally influenced by the variability of the endmembers to be prioritized in the unmixing process.
This idea was first introduced to improve the classification of materials under varying atmospheric illumination conditions~\cite{healey1999hyperAtmosphericConditions,lau2004atmosphericComparissonHymap}.

The majority of these methods are based on affine transformations of the observed pixels defined as
\begin{align} \label{eq:band_transformation_variability_general}
    \bW_{\!\lambda} \by_n + \bb_n = \bW_{\!\lambda} \bM_n \ba_n + \bW_{\!\lambda} \be_n + \bb_n \,,
\end{align}
where the matrix $\bW_{\!\lambda}$ and the affine term $\bb_n$ are determined to minimize the effects of endmember variability in the subsequent SU process.
Besides modifying the observed pixel spectrum $\by_n$, this transformation is also applied to the elements of the spectral library, yielding: 
\begin{align} \label{eq:band_transformation_variability_library}
    \bW_{\!\lambda}\mathcal{M}_p +\bb_n \triangleq
    \big\{ \bW_{\!\lambda}\bm + \bb_n: \bm \in \mathcal{M}_p \big\} \,,
\end{align}
for $p=1,\ldots,P$. Different particular cases of this model have been considered in the literature, most notably with $\bW_{\!\lambda}$ being a diagonal matrix with positive real (band weighting) or binary (band selection) elements.
Note that although traditional dimensionality reduction (e.g., PCA) or band selection methods used to compress the hyperspectral image could be implemented using this transformation with $\bb_n=\cb0$, the direct application of compression techniques does not necessarily improve the robustness to spectral variability~\cite{li2004waveletFeaturesUnmixing}.

Spectral transformation approaches can be generally divided into two major groups: those defined \textit{a priori} based on expert knowledge by the user, and those constructed automatically using information in a spectral library. We will review each case in the following.

\subsubsection{\textbf{User-defined spectral transformations}}

The first user-defined spectral transformations were proposed to normalize the effects of illumination and brightness variations, or to emphasize useful spectral features.
These approaches include subtracting the reflectance value of a selected (specific) spectral band from all remaining bands~\cite{asner2000biogeophysicalAutoSWIR}, subtracting from each endmember its mean value in the spectral dimension to reduce the variability due to differences in brightness~\cite{wu2004normalizedUnmixingUrbanETM}, or normalizing/dividing the reflectance value at each wavelength by the corresponding value of the convex hull of the spectral signatures~\cite{youngentob2011eucalyptusMESMAcontinuumRemoval}.
Other examples also include using the first or second derivatives~\cite{zhang2004SecondDerivativeUnmixing,debba2006firstDerivativeUnmixing}, or the wavelet transform of the spectral signatures~\cite{li2004waveletFeaturesUnmixing} \mbox{for SU.}

A later spectrum-based approach that has become very popular for solving this problem consists of using band selection methods. These methods basically work by performing SU using only selected wavelength intervals in which there is little spectral variability between different spectral signatures of the same material~\cite{asner2000biogeophysicalAutoSWIR,somers2011variabilityReview}. Although many of these approaches rely on expert knowledge about the specific underlying application, they are simple and easily interpretable and also help in reducing the computational cost of the SU problem.
Examples of band selection methods defined \textit{a priori} by the user include the selection of the the SWIR2 spectral region (2100--2400~nm) for unmixing of soil and vegetation in arid and semi-arid environments~\cite{asner2000biogeophysicalAutoSWIR}, and the combination of various spectral regions such as visible, NIR and SWIR for other applications~\cite{asner2002spectralUnmixingAutoMCU}.

\subsubsection{\textbf{Library-based spectral transformations}}

Spectral transformations proposed more recently leverage information contained in the spectral library to compute the terms~$\bW_{\!\lambda}$ and~$\bb_n$ of the affine transformation. This circumvents one of the main downsides of the previous approaches by making the process automated instead of delegating the choice to the user.
These techniques can be further divided in three groups, namely band selection, band weighting, and more general spectral transformations.

\paragraph{\textbf{Band selection}}

Band selection methods proposed more recently seek to identify the robust spectral regions based on the samples in the spectral library. Different strategies have been proposed.

One of the first approaches is based on the analysis of the spectral residuals obtained by performing a preliminary unmixing of the image using the LMM with an average EM matrix~\cite{miao2006yellowStarthistleBandSelection}. Only the spectral bands with minimal residual variance are then used for SU, based on the empirical observation that they correspond to more robust spectral zones.

Another method, called stable zone unmixing, proposes to select spectral bands that are robust to spectral variability by minimizing an instability index defined as the ratio between the intra-class and the inter-class endmember variances (computed based on a spectral library)~\cite{somers2010stableZoneUnmixing}.
This method was later extended in order to minimize both the instability index and the correlation between signatures of different endmember classes at the same time, aiming to improve the numerical conditioning of the SU problem~\cite{somers2013uncorrelatedBandSelectionInstabilityIndex,ghaffari2017optimalBandSelectionCorrelationVariability,tane2018evaluatingEndmemberAndBandSelection}.

The work~\cite{somers2014pixelDependentBandSelectionVariability} proposed to improve the separability between classes by employing the stable zone unmixing framework to select an individual set of spectral bands for each possible subset of endmember/material classes that could be tested with MESMA when considering endmember models with fewer then $P$ signatures in the SU process.

\paragraph{\textbf{Band weighting}}

Band weighting methods are more flexible techniques which allow one to prioritize the spectral bands in the unmixing process according to their reliability or significance using a continuous weight term.
This is usually done by weighting the reconstruction error of each band in the SU cost function.
Different approaches have been proposed to compute the weight to be applied to each band. For instance, a weighting strategy based on two terms was proposed in~\cite{somers2009weightedUnmixingVariabilityAgricultural}. One term normalizes the energy of the reflectance spectra to equalize the contributions to low- and high-reflectance bands, and another term accounts for the robustness of each band to spectral variability using its instability index (i.e., ratio between intra-class and inter-class endmember variance).
This approach was later applied to monitor both the level of defoliation in Eucalyptus plantations~\cite{somers2010weightedMESMAeucalyptus} and invasive plant species using multi-temporal data~\cite{somers2013invasiveHawaiiMultiTemporalBandWeighting}. It was also later extended in~\cite{somers2009magnitudeShapeWeightingWeedsCitrus} to consider SU integrating both reflectance and derivative spectra.
Band weights based on the instability index were also used to prioritize the more stable spectral bands when designing spectral filters robust to spectral variability, which are low-complexity alternatives that approximate the solution of the SU problem as a direct application of a single linear transformation~\cite{krippner2018opticalFilterBandSelectionVariability}.

\paragraph{\textbf{General spectral transformations}}

Another group of approaches proposed to use more flexible linear transformations to better mitigate the effect of spectral variability. These techniques consist of variations of the Fisher Discriminant Analysis (FDA), which is widely used for pattern classification. FDA aims to find a transformation of the data to obtain a feature space with the best separability between different classes~\cite{Bishop:2006:PRM:1162264}. In the context of SU, this amounts to minimizing the variance of the signatures of each material while also maximizing the distance between the mean values of the different endmember classes~\cite{chang2006weightedUnmixingAndFisher}. Mathematically, this is formulated as 
\begin{equation} \label{eq:variability_transformations_fisher}
    \bW_{\!\lambda} {}={} \mathop{\arg\min}_{\bW} \,\, \frac{\bW^\top \bS_{\rm within} \bW}{\bW^\top \bS_{\rm between} \bW} \,,
\end{equation}
where $\bS_{\rm within}$ is the weighted sum of the within-class covariance matrices, and $\bS_{\rm between}$ is the covariance matrix of the mean endmember spectra.

The first approaches applied FDA to SU directly by either using spectral libraries known \textit{a priori}~\cite{chang2006weightedUnmixingAndFisher} or constructed using pure pixels extracted from the observed hyperspectral image~\cite{jin2010fisherDiscriminantUnmixing}. Another work also considered the augmentation of the spectral library with pure pixels extracted from the image to improve the discrimination among spectrally similar vegetation species~\cite{bue2015MESMA_FisherDiscriminant_inSceneSpectra}.
Later approaches considered other variations, such as the iterative addition of more column vectors to $\bW_{\!\lambda}$ using a Gram–Schmidt orthonormalization procedure to increase the dimensionality of its output space for multispectral images with a small number of bands~\cite{liu2017orthogonalFisherTransformationSemiarid}.
Another work proposed to make the spectral signatures of different endmembers orthogonal to each other, and the spectral signatures of the same endmember all unitary and collinear to improve the numerical conditioning of the SU problem~\cite{jafari2016endmemberOrthogonalTransformationVariability}.
The FDA was also successfully used to improve the performance of MESMA when unmixing urban surfaces (containing vegetation, soil, water and manmade materials) using image-extracted \mbox{spectral libraries~\cite{xu2019mappingImperviousSurfaceMESMA_FDA}.}

In contrast with its improved flexibility, the FDA has as a downside its dependence on a good estimation of the covariance matrices to be used in~\eqref{eq:variability_transformations_fisher}. Thus, the FDA may not perform well if the amount of samples in the libraries is not statistically representative~\cite{du2007modifiedFisherClassification}.

\section{Unmixing Methods That Estimate the Endmembers from the Image}
\label{sec:overall_methods_blind}

In more recent years, a large number of works proposed to address spectral variability in SU without relying on prior knowledge about spectral libraries. Different strategies have been proposed to this end, which we divide into four groups.
Local unmixing methods are both computationally and conceptually simple but require significant user supervision.
Parametric endmember models provide more flexibility to represent EM spectral variability but make the SU problem harder to solve.
EM-model-free methods address spectral variability by using different modifications to the SU cost function.
Bayesian methods use statistical representations for the endmembers, which leads to a smaller amount of user supervision at the price of a high computational complexity.

\cred{All these families of methods are able to estimate both the EMs and the abundances directly from the image. However, as seen in Section~\ref{sec:overall_methods_libs}, the reasoning underlying each of them is quite different, which leads to different \cdgreen{levels of required} user supervision, computational complexity, and abundance estimation quality, as illustrated in the diagram of Fig.~\ref{fig:decision_tree}. Moreover, prior knowledge used in the design of the algorithms is an important ingredient to guarantee their good performance, and includes, e.g., the spectral and spatial correlation of endmember signatures and their statistical properties. We review each family of approaches in the following.}

\subsection{\textbf{Local unmixing methods}}
\label{sec:local_SU}

\enhancedcoloredtbox{Local Unmixing:}{
\begin{itemize}
    \item[$+$] Conceptually simple and physically motivated
    \item[$+$] Computationally efficient
    \item[$-$] Usually requires a significant amount of user supervision
    \item[$-$] The selection of the local image regions has a significant impact on the results
    \item[$-$] Local EM extraction can be difficult
    \item[$-$] Grouping the local estimates into global results is also challenging
\end{itemize}}

A conceptually simple and efficient method to deal with spectral variability is to perform both endmember extraction and spectral unmixing locally for small, non-overlapping regions of the hyperspectral image.
This approach, called \textit{local unmixing}, assumes the endmember signatures to be constant in each region of the image, benefiting from the knowledge that spectral variability is often negligible in small regions.
The basic framework of local unmixing can be summarized into the following steps:
\begin{enumerate}
    \item Divide the observed image into a set of regions;
    
    \item Estimate the number of spectral signatures and extract the endmembers in each region;
    
    \item Perform SU with the local endmember signatures;
    
    \item Combine all the local SU results into global sets of endmember signatures and global abundance maps using, e.g., clustering procedures. 
\end{enumerate}

Although local unmixing methods proposed up to date share similar overall methodologies, there are important differences in the way the hyperspectral image is partitioned (e.g., using simple square tiles or more advanced image segmentation) and how the endmembers are extracted from each region. This can have a significant impact on the results.

The first approaches for local unmixing required complete user supervision. For instance, the variable MESMA (VMESMA) algorithm proposed in~\cite{garcia2005VMESMA} used manual image segmentation to divide the image into local regions. SU was then performed iteratively, updating the segmentation maps and manually including additional endmembers in the process until a satisfying result was obtained.
Later approaches attempted to reduce the need for user supervision in the process.
For instance, endmember extraction and SU were performed individually in local (square) image tiles in~\cite{canham2011spatiallyAdaptiveUnmixing,goenaga2013unmixingTimeSeriesPuertoRico}.
Afterwards, the locally extracted endmembers and abundance maps were then merged into the global endmember sets and abundance maps using clustering algorithms.

Image segmentation methods were later used to provide more flexibility when dividing the hyperspectral image into local regions. For instance, in~\cite{li2015segmentationLocalUnmixing} manual endmember extraction and spectral unmixing (using the FCLS algorithm) were performed individually in each image region defined by a segmentation algorithm.
Another work considered a superpixel decomposition of the image aided by external map metadata in order to compute a more accurate segmentation~\cite{sun2017superpixelLocalUnmixing}.
A more sophisticated method was proposed in~\cite{drumetz2014localUnmixingBinaryPartitionTree,veganzones2014hyperspectralSegmentationBPT} by using a binary partition tree to divide the image into different regions from a coarse to a fine spatial scale. Local unmixing was then performed at the scale of the partition tree yielding the smallest \mbox{reconstruction errors.}

Besides the choice of the segmentation procedure, endmember extraction is also a challenging part of local unmixing and has a great impact on the performance of these algorithms.
A spatially adaptive unmixing method was proposed in~\cite{deng2013spatiallyAdaptiveUnmixing} to estimate the distribution of different surfaces in urban environments. Endmember spectra for each pixel were synthesized as a weighted average of pure pixels extracted in a spatial neighborhood specified by the user, with weights given as a function of their distance to each mixed pixel at hand. 
A similar approach used as endmembers the mean values of pure pixels extracted within each (square) image region, which were identified using a classification strategy~\cite{wu2014spatiallyConstrainedMESMAurban}.
These approaches can positively weight pure pixels that are spatially close to each pixel being unmixed. This idea was also explored in other works such as in~\cite{cao2015localUnmixingMultipleCriteriaMESMA}, which performed SU using a variant of the MESMA algorithm, or in~\cite{mei2015localUnmixingSpatialEMselection}, which used only the spatially closest pure pixels to process each \mbox{mixed pixel.}

Other local unmixing approaches considered hierarchical segmentation approaches in which the hyperspectral image was divided into two spatial scales, a coarse one where unmixing was performed with MESMA, and a fine spatial scale in which the spectral libraries were extracted using either the spectral signatures of small and homogeneous objects~\cite{deng2016multipleNeighborhoodwiseLocalMESMA} or \textit{a priori} knowledge about the abundances obtained from external high resolution classification maps~\cite{deng2015segmentationLocalMESMAcoarseResolution}.

An important issue of local unmixing algorithms is the determination of the number of endmembers contained in each local image region. While in most experimental works this was performed empirically or even manually, it is desirable to have automated methodologies to estimate the number of local endmembers and their spectral signatures. 
This usually involves the estimation of the intrinsic dimensionality of the local subset of the hyperspectral image~\cite{robin2015estimationIntrinsicDimensinalityComparisson}. However, the performance of intrinsic dimensionality estimators is often negatively impacted when the size of the data set is small~\cite{drumetz2016localIntrinsicDimensionality}. This strongly limits the characteristics (i.e., size) of the subsets or segmentation procedures that are selected for unsupervised local unmixing.
Collaborative sparse regression approaches~\cite{iordache2014collaborativeSparseRegressionUnmixing} were proposed to deal with the shortcomings of intrinsic dimensionality estimation by avoiding the selection of repeated or mixed signatures during unmixing~\cite{drumetz2017localUnmixingCollaborativeSpartityAlgorithmic}. The sparsity level was selected using a Bayesian information criterion in order to obtain a good compromise between small reconstruction errors and a small number of \mbox{selected signatures.}

A different line of work attempted to relax the assumption of connectedness of the local spatial regions, performing SU in different subsets of the hyperspectral image which are not necessarily spatially adjacent.
For instance, the piecewise convex model proposed in~\cite{zare2013piecewiseConvexNMF} considered a set of different endmember matrices, all estimated from the entire image. Each pixel was then assigned to one of these EM matrices using a (fuzzy) membership function, which was estimated along with the other variables in a non-convex matrix factorization problem.
Other works extended this approach by considering cluster validity indices~\cite{anderson2012piecewiseConvexNMFparameters} or sparsity promoting priors~\cite{zare2012bootstrappingPCEparameterEstimation} to estimate parameters such as the number of EM matrices and the number of material classes in each segment, or using spatial constraints to encourage neighboring pixels to have similar membership values~\cite{zare2010spatialRegularizationPiecewiseSmooth}.

A similar work considered the estimation of multiple EM matrices in a non-negative matrix factorization framework by using abundance sparsity constraints instead of employing (fuzzy) membership functions, while also penalizing the mutual coherence between the signatures of different material classes to improve inter class separability~\cite{castrodad2011learningSpectralLibrariesSpCoding}.
A related strategy considered a self-dictionary model where the multiple EM signatures are selected directly as the hyperspectral image pixels that can best reconstruct most of the remaining pixels in the scene as a sparse linear combination~\cite{iordache2014selfDictionaryLibExtraction}.
Another approach with even more flexibility considered an individual EM matrix for each image pixel in a non-negative matrix factorization formulation~\cite{revel2018inertiaConstrainedPixelbypixelNMF}. A regularization term penalizing the trace of the covariance matrix of the estimated spectral signatures for each class was also considered to reduce the ill-posedness of the estimation problem.

\subsection{\textbf{Parametric models}}
\label{sec:parametric_mdl_SU}

\enhancedcoloredtbox{Parametric Endmember Models:}{
\begin{itemize}
    \item[$+$] The SU algorithms are computationally efficient
    \item[$+$] Very flexible and physically motivated models to represent any kind of variability
    \item[$+$] Easy to incorporate prior information
    \item[$-$] Determining a good EM model might require some degree of expert knowledge
    \item[$-$] Require significant user supervision for tuning free model parameters
    \item[$-$] Estimating the parameters of the EM models (along with the abundances) can be challenging due to the presence of non-convex optimization problems and sensitivity to parameter choice or initialization
\end{itemize}}

A flexible and physically reasonable way to address spectral variability consists of employing parametric models to represent the endmember spectra. These strategies allow for great freedom to incorporate constraints and information from the underlying application.
They are generally based on representing the EM spectra as
\begin{align} \label{eq:general_parametric_model}
    \bM_n {}={} f(\bM_0,\btheta_n) \,,
\end{align}
where $f(\cdot)$ is a function of an average or reference EM matrix $\bM_0$ and of a vector of parameters $\btheta_n$. The number of parameters in $\btheta_n$ is usually small, which allows one to confine the endmember spectra to a low-dimensional manifold.
The SU problem is then formulated as the recovery of the abundances and of the parameters $\btheta_n$ for all pixels of \mbox{the image.}

The model in~\eqref{eq:general_parametric_model} can be defined either based on the underlying physics describing material spectra as a function of numerous geometric and photometric parameters, such as Hapke's~\cite{Hapke1981,HapkeBook1993} or Shkuratov's~\cite{shkuratov1999modelNonlinearUnmixng} for packed particles and the PROSTECT or PROSAIL models for vegetation~\cite{jacquemoud1990PROSPECTmodelLeaf,jacquemoud2001leafOpticalPropertiesReview}. However, the model~\eqref{eq:general_parametric_model} can also be inspired by physics but chosen in order to allow for more flexibility and mathematical tractability.
We will review these approaches \mbox{in the following.}

\subsubsection{\textbf{Physics-based methods}} \label{sec:parametricMDL_physicsBasedMdls}

The first SU approaches using parametric models aimed to obtain fractional abundances from intimate mineral mixtures by inverting the Hapke model~\cite{johnson1992hapkeModelUnmixingInversion}.
With perfect knowledge about the viewing geometry, the scattering properties of the different materials and the single scattering albedo of the EMs, the SU problem using Hapke's model becomes linear in the albedo domain~\cite{heylen2014linearMixtureIntimateModels}.
However, since these variables are hardly available in practice, many works attempted to invert Hapke and related models blindly. This inversion is mathematically and computationally very difficult in general and requires hyperspectral images acquired at multiple viewing geometries~\cite{johnson1992hapkeModelUnmixingInversion}.
Thus, subsequent works proposed simplifications of the scattering characteristics of the materials in the model~\eqref{eq:general_parametric_model} to improve its mathematical tractability~\cite{mustard1989hapkeModelUnmixingApproximations}.
These methods have been successfully applied to estimate abundance maps at different scenes, including the Cuprite mining district at Nevada~\cite{shipman1987hapkeModelCuprite} \mbox{and the Moon~\cite{mustard1998nonlinearUnmixingMoonCrater,dhingra2011nonlinearUnmixingMoonHapke}.}

This approach has also been applied to SU of vegetation mixtures based on the inversion of radiative transfer models. The first works simplified the problem by assuming external knowledge of biophysical parameters.
For instance, a model for mixtures of vegetation, shadowed and illuminated soil was proposed for SU by approximating plant geometry with spatially distributed cylinders containing layers of leaves~\cite{gilabert2000geometricModelMixtureVegetationUnmixing}. Although spectral variability was allowed by means of changes in biophysical parameters, these were assumed to be known \textit{a priori} to solve the SU problem.
Another approach considered SU of soil and vegetation using a simplified mixing model as a function of NDVI values instead of the full spectral signatures~\cite{song2017estimatingNDVImultianglePhysicalModel}. In this case, a physical model was used to represent the variability of the NDVI values as a function of parameters such as the viewing geometry, leaf density and clumping effect. However, the NDVI ``endmembers'' for each pixel had to be estimated before SU by using multi-angled observations and assuming prior knowledge of the leaf biophysical parameters. 
A later approach for SU of soil and vegetation mixtures proposed to estimate the biophysical parameters blindly from the hyperspectral image using the PROSAIL model for vegetation spectra~\cite{li2018PROSAILmodelRTFvegetationVariabilityUnmixing}.
The SU problem was formulated as the recovery of both the abundances and the two parameters of the PROSAIL model, and solved using an alternating optimization procedure. Note that the other parameters of the PROSAIL model had to be fixed \textit{a priori}.

Although these models carry a strong physical motivation, their use in SU leads to computationally intensive and mathematically challenging (i.e., non-convex, significantly ill-posed~\cite{cannon2017monteCarloUnmixingHapke}) problems.
This occurs because physics models were originally devised as forward models that accurately describe the reflectance spectra based on a set of parameters, and were not originally designed to be inverted, which limits their use for SU in practical problems~\cite{Dobigeon-2014-ID322}.

\subsubsection{\textbf{Physically motivated and non-physics-based methods}}

The low mathematical tractability of physics-based models has motivated recent studies leading to more flexible or parsimonious models that are only inspired by the underlying physics.
Although these models are not as precise as those presented in Section~\ref{sec:parametricMDL_physicsBasedMdls} when representing physical phenomena underlying spectral variability, they allow for more efficient SU algorithms estimating the involved parameters $\btheta_n$ from the observed image.
Moreover, although models inspired by physics can be ill-posed, the EM spectra are often confined to a low-dimensional manifold since they only depend on a small number of physico-chemical variables. This property can be exploited to design parsimonious models with possible constraints and reduce the ill-posedness of the SU problem.

Several parametric models have been recently proposed with these objectives.
One of the resulting SU algorithms is the scaled constrained least squares method~\cite{Nascimento-2005-ID325-doesICAplaysArole}, which attempts to represent uniform illumination variations in each pixel by introducing an additional scaling factor $\psi_n\in\amsmathbb{R}_+$ in the EM matrices as
\begin{align} \label{eq:scls_model}
    \bM_n = \psi_n \bM_0 \,.
\end{align}
SU can be performed using model~\eqref{eq:scls_model} by solving a simple non-negative least squares problem, which is convex and computationally efficient. However, this model lacks capability to represent more complex spectral variability that have been observed in practical scenes, motivating the search for more flexible models.

An extended version of the LMM (ELMM) was later proposed in~\cite{veganzones2014newELMM,drumetz2016blindUnmixingELMM} by allowing each endmember in a pixel to be individually scaled by a constant factor, resulting in the following representation for the EM matrices:
\begin{align} \label{eq:elmm_model}
    \bM_n = \bM_0 \, \diag(\bpsi_n)  \,,
\end{align}
where vector $\bpsi_n\in\amsmathbb{R}^P_+$ contains the scaling factors for each of the $P$ materials.
The ELMM can represent more complex variability originated from variations in both illumination and topography, which can affect each material in the hyperspectral image differently. Furthermore, the ELMM can be obtained from successive physical approximations of the Hapke model for small-albedo materials~\cite{drumetz2019ELMM_from_Hapke}.
Based on an estimate of $\bM_0$ obtained from the observed image, SU under the ELMM was formulated as a non-convex matrix factorization problem in which the model~\eqref{eq:elmm_model} was enforced by means of an additive penalty in the cost function~\cite{drumetz2016blindUnmixingELMM}. A regularization promoting spatial homogeneity of the scaling factors $\bpsi_n$ was also considered to reduce the ill-posedness of the SU problem~\cite{drumetz2016blindUnmixingELMM}.
The ELMM has also shown good performance for multitemporal data~\cite{henrot2016dynamical} and has been used to facilitate the interpretation of local unmixing results~\cite{tochon2016localUnmixingELMMscalingFactors}. Moreover, the ELMM can be derived from a Taylor series expansion of a general nonlinear mixture model~\cite{drumetz2017relationshipsBilinearELMM}, what introduces SU with spectral variability (viewed as a locally linear SU problem) as a direct way to address the general nonlinear SU problem. This shows that some mixture models originally devised to represent spectral variability (such as the ELMM) can achieve good performance in nonlinear SU.

Despite its physical motivation, the ELMM model lacks flexibility to represent more complex spectral variability, e.g., affecting the spectra non-uniformly.
To address this limitation, the generalized LMM (GLMM) was later proposed in~\cite{imbiriba2018glmm} by introducing an individual scaling factor for each band, leading to the following EM model
\begin{align} \label{eq:glmm_model}
    \bM_n = \bPsi_n \circ \bM_0 \,,
\end{align}
where the matrix $\bPsi_n\in\amsmathbb{R}^{L\times P}$ contains the scaling factors for each element of $\bM_0$ and~$\circ$ denotes the Hadamard (element-wise) matrix product.
Note that the amount of spectral variability brought by the GLMM is proportional to the amplitude of the reference spectra $\bM_0$ in each band.
However, the larger number of parameters makes the SU problem resulting from~\eqref{eq:glmm_model} more ill-posed with challenging estimation problems.
This motivated the development of a tensor interpolation framework to estimate the matrices $\bPsi_n$ from training hyperspectral data obtained based on prior knowledge about the positions of pure pixels in the hyperspectral image~\cite{borsoi2019tensorInterpolationICASSP}. However, the performance of the method proposed in~\cite{borsoi2019tensorInterpolationICASSP} depends strongly on the amount of pure pixels available in the image.
The GLMM has also been successfully used in multitemporal SU~\cite{borsoi2020multitemporalUKalmanEM} and to represent spectral variability when fusing hyperspectral with multispectral images acquired at different \mbox{time instants~\cite{Borsoi_2018_Fusion}.}

Note that the performance of unmixing methods based on the ELMM and GLMM depends strongly on the quality of the reference EM matrix~$\bM_0$, which must be estimated from the observed image. In order to reduce the dependence of the ELMM on~$\bM_0$, the authors of~\cite{drumetz2019EMsDirectionalDataU_journal} proposed to estimate~$\bM_0$ jointly with the remaining variables during SU.
Each column of~$\bM_0$ was also constrained to have a unit norm in order to obtain EMs as directional data in the spectral space. Moreover,~$\bM_0$ was initialized using a simple cosine-based k-means clustering of the observed data-cube, which improved the robustness of the method to the presence of \mbox{shadowed pixels.}

A different EM model was proposed in~\cite{Thouvenin_IEEE_TSP_2016_PLMM} by considering an additive term to the mean EM matrix, resulting in the following EM representation
\begin{align} \label{eq:plmm_model}
    \bM_n = \bM_0 + \bd\bM_n \,,
\end{align}
where the matrix $\bd\bM_n\in\amsmathbb{R}^{L\times P}$ is an additive perturbation representing spectral variability.
In this case, both the reference EM matrix $\bM_0$ and the pixel-dependent additive perturbation terms $\bd\bM_n$ were estimated blindly from the hyperspectral image.
However, this model has a large number of parameters. Thus, to mitigate the ill-posedness of the SU problem, a regularization term consisting of the Frobenius norm of $\bd\bM_n$, $n=1,\ldots,N$ was included in the unmixing cost function.
Besides the simplicity and mathematical tractability, the use of an additive perturbation in~\eqref{eq:plmm_model} also makes the problem amenable to an interesting interpretation when only a single additive perturbation matrix is considered for all image pixels. In this case, the SU problem becomes equivalent to a total least squares problem with constraints~\cite{arablouei2018unmixingPerturbedEndmembersCRLB}.
Furthermore, the Perturbed Linear Mixing Model (PLMM) has also been considered for robust SU using an outlier-insensitive reconstruction error metric with an $L_p$-quasi norm~\cite{syu2019outlierInsensitivePLMM} 
and for multitemporal and distributed SU~\cite{sigurdsson2017sparseDistU,thouvenin2016online}.

One difficulty of parametric EM models is to construct functions $f(\cdot)$ that are parsimonious but still flexible enough to represent complex spectral variations. To circumvent this issue, a deep generative EM model was proposed in~\cite{borsoi2019deepGun} based on the hypothesis that the EMs lie on low-dimensional manifolds. Instead of fixing the EM model \textit{a priori}, variational autoencoders with neural networks were used to learn the parametric function $f(\cdot)$ in~\eqref{eq:general_parametric_model} using pure pixels extracted from the observed hyperspectral image. SU was then formulated as the recovery of the abundances and of the representations of the EMs in the learned manifold, which can be of very small dimension. Despite making SU more well-posed, the resulting cost functions are non-convex and can be difficult to optimize.

A different work proposed to exploit the spatial correlation of the endmembers and abundance maps by proposing a general multiscale mixing model addressing EM variability~\cite{Borsoi_multiscaleVar_2018}. The SU problem was solved using a multiscale representation of the mixing model, which allowed for the use of any parametric EM models as in~\eqref{eq:general_parametric_model}. This resulted in improved results when compared to standard spatial regularization strategies. Although the formulation was algebraically involved, an approximate algorithm with small complexity was derived under some simplifying assumptions.

\subsection{\textbf{EM-model-free methods}}
\label{sec:EM_mdl_free_SU}

\enhancedcoloredtbox{EM-model-free unmixing:}{
\begin{itemize}
    \item[$+$] Algorithms are usually computationally efficient
    \item[$+$] Involves different strategies with a wide range of model complexity or user supervision
    \item[$+$] Methods usually make few or restrained assumptions about the endmember models (unlike Bayesian or parametric models)
    \item[$-$] Some approaches have a more limited modeling capability
\end{itemize}}

Some methods have proposed to mitigate the effects of spectral variability blindly without assuming any specific model to represent the endmember signatures.
One simple approach consists of using a metric depending on the reconstruction error in the SU cost function in order to improve the robustness of SU to endmember variability. It can be motivated by the fact that the commonly used Euclidean distance is very sensitive to variations in the amplitude of the pixel spectra, being thus significantly influenced by illumination variations~\cite{chen2009unmixingShapeMetric}.
This motivated the consideration of the  spectral angle mapper, spectral correlation and spectral information divergence metrics due to their insensitiveness to scaling variations~\cite{chen2009unmixingShapeMetric,tits2012shapeBasedUnmixingVegetation}.

The downside of this approach lies in the nonlinear and possibly non-convex nature of the resulting SU optimization problem, which becomes harder to solve. An efficient strategy based on the projected gradient descent algorithm was proposed to optimize the SU cost function when using the spectral angle mapper metric~\cite{kizel2017projectedGradientAngleCostFunction}.
Although conceptually simple, these approaches focus on specific effects such as brightness variations and it is not clear how they can be generalized to address more complex spectral variability.

More recent SU methods consider more general models in order to deal with complex intrinsic variability effects.
For instance, an additive residual term in the LMM~\eqref{eq:LMM} was considered in~\cite{halimi2017fastNonparametricVariability} in order to account for spectral variability and other unmodelled effects. This term was represented as the product of two matrices. The first matrix corresponded to the first columns of the discrete cosine transform, forcing the additive terms to be spectrally smooth. The second matrix was defined for the pixel-dependent coefficients, which were forced to be spatially sparse and were estimated by solving a convex optimization problem.

A similar approach included ideas from physically motivated parametric models by considering the LMM with a constant scaling factor for each pixel to account for rough illumination variations and an additional non-parametric additive term to account for other types of spectral variability~\cite{hong2019augmentedLMMvariability}.
This additive term was defined as the product between an approximately orthonormal basis matrix having low coherence with the endmember signatures, and a coefficient matrix representing the variability contribution to each pixel.
However, these constraints make the resulting optimization \mbox{problem non-convex.}

A different idea was to estimate a subspace projection of the observed hyperspectral image that minimizes the effect of spectral variability in SU~\cite{hong2018SULoRA_lowRankEnbeddingUnmixingVar}. This strategy allows SU to be performed by minimizing the reconstruction error in the projected space. This subspace is forced to be of low-dimension by penalizing the nuclear norm of the projection operator in the cost function, which is estimated jointly with the abundances during SU.

A more recent method considered the multidimensional representation of the pixel-dependent EM matrices and abundance vectors by employing mathematical tools from tensor decomposition~\cite{imbiriba2018ULTRA_V}.
By assuming that the endmembers and abundance tensors are approximately low-rank, the SU problem was formulated as a non-convex non-negative tensor factorization problem.
This led to a parsimonious model without the need for explicit parametric representations of the endmembers that are tied to specific applications.

\subsection{\textbf{Bayesian Methods}}
\label{sec:Bayesian_SU}

\enhancedcoloredtbox{Bayesian Methods:}{
\begin{itemize}
    \item[$+$] Benefit from well-developed statistical estimation tools to derive the SU methods
    \item[$+$] Can have a very low degree of user supervision once the statistical distributions are selected
    \item[$-$] Can use unrealistic distributions (e.g., isotropic Gaussians) to represent the EMs for mathematical tractability
    \item[$-$] Generally do not return the specific spectral signatures at each image pixel
    \item[$-$] Suffer from a very high computational cost
    \item[$-$] Hyperparameters may need to be set by the user, and specifying hyperprior distributions for hierarchical Bayesian models may not be trivial
\end{itemize}}

Another set of methods considers endmembers to be random vectors, following multivariate statistical distributions, i.e.,
\begin{align}
    \bm_{n,p}\sim \mathfrak{D}(\btheta_{n,p}) \,,
\end{align}
where $\btheta_{n,p}$ encodes parameters of a distribution $\mathfrak{D}$. The spectral signatures actually present in each pixel are realizations of this random vector, and SU is then formulated as the problem of finding a statistical estimator for the abundances and for the endmembers.

These approaches depend on the statistical distribution $\mathfrak{D}$ employed to represent EM spectra, on the amount of user supervision that is required and on the computational algorithm used to solve the problem. Some methods require the parameters of the distribution $\btheta_{n,p}$ to be set \textit{a priori}, which might be difficult in the absence of a large spectral library. Other works reduce user supervision by employing hierarchical Bayesian methods to estimate $\btheta_{n,p}$ jointly with the remaining parameters at the cost of a higher computational cost~\cite{moussaoui2006bayesianGammaEMdist,dobigeon2009jointBayesianEMunmixing}.
The different Bayesian methods addressing spectral variability can be classified according to the statistical distribution used to represent the EMs: a Gaussian distribution, which provides mathematical tractability or more complex distributions providing a more physically reasonable representation. We will discuss both cases in the following.

\subsubsection{\textbf{The Normal Compositional Model}}

The first statistical model that has been considered to represent endmember spectra was a multivariate Gaussian distribution, in the so-called Normal Compositional Model (NCM), given by
\begin{align}
    \bm_{p,n} \sim \mathcal{N}(\btheta_{n,p}) \,,
\end{align}
where $\mathfrak{D}\equiv\mathcal{N}$ and $\btheta_{n,p}=\{\operatorname{mean},\operatorname{covariance}\}$ contains the mean vector and covariance matrix for the $p^{\rm th}$ endmember of the $n^{\rm th}$ pixel.
The NCM has been widely used due to its mathematical tractability~\cite{stein2003earlyNCM,eismann2004otherNCMearly}.
The first works employing the NCM for SU considered expectation-maximization strategies in which the abundances, the mean endmember values and their covariance matrices were estimated iteratively~\cite{stein2003earlyNCM}. However, due to the non-convexity of the estimation problem, the direct application of expectation maximization approaches is unable to decide whether variations observed in the mixed pixel spectra $\by_n$ are due to different abundances or to the endmember variability. This might result in the endmembers absorbing all variation in the observed scene with nearly constant abundances~\cite{eismann2004otherNCMearly}.
Some approaches proposed to address this problem by considering the use of diagonal covariance matrices and empirical strategies to estimate the endmember data more easily from the observed mixed pixels. For instance, endmember means and covariances were estimated both \textit{a priori} using pure pixels selected from the hyperspectral image~\cite{liu2009bayesianNCMestimationSelfOrgMaps}, and iteratively based on large regions of observed pixels with homogeneous abundances (obtained from the segmentation of estimates of the abundances available \textit{a priori})~\cite{gao2016regionBasedCovarianceEstNCM}.

Other works attempted to improve different aspects of this method, by using a particle swarm optimization algorithm to solve the (usually intractable) integrals involved in the estimation of the abundances in the ``expectation'' step of the algorithm~\cite{zhang2014bayesianNCMparticleSwarmOpt}, or by incorporating \textit{a priori} information in the form of additional constraints penalizing the nuclear norm of the abundances in groups of pixels determined through image segmentation methods (in order to promote spatial homogeneity)~\cite{ma2019unmixingGMM_LowRank_generalizedEM}.

Despite these advances, the susceptibility of expectation-maximization-based methods to converge to poor local minima of the non-convex cost function prevented their large-scale applicability for this problem. Instead, most recent approaches rely on more robust (although costly) techniques based on Markov chain Monte Carlo methods to sample the posterior distribution. 
Although the works that adopt this approach share the same underlying idea, they differ significantly in the way in which the endmembers and abundances are represented and in the amount of user supervision that is required.
For instance, different strategies have been proposed to represent the mean and covariance matrices of the endmembers in the NCM.
One of the first approaches considered the endmember mean values to be known \textit{a priori} and their covariance matrices to be multiples of the identity matrix~\cite{eches2010bayesianNCM}, while employing conjugate distributions to make the estimation of the parameters easier.
Later works attempted to add more flexibility by considering, for instance, a single full covariance matrix shared by all endmembers~\cite{kazianka2011bayesianNCMfullCovariance} or a positive definite matrix defined \textit{a priori} and multiplied by EM-dependent scaling parameters~\cite{kazianka2012hyperpriorAnalysisNCM}.
Diagonal covariance matrices were employed in~\cite{halimi2015unsupervisedBayesianUnmixing}, which also considered the estimation of the EM mean values in a hierarchical Bayesian framework, using hyperpriors to estimate the distribution parameters directly from the observed hyperspectral image.
The Bayesian framework has also been used in~\cite{eches2010NCMestimateNumberEM} to estimate the number of EMs in the scene blindly using a uniform discrete prior.

Other works attempted to address physically motivated particular cases of the general NCM. This includes the consideration of statistical dependence between different EMs to represent spectral variability that may affect all materials in the scene equally (e.g., atmospheric effects)~\cite{jafari2017independentBaseVectorsNCM}, and the explicit representation of the higher correlation between adjacent spectral bands to introduce spectral smoothness to the signatures, leading to a well-posed model that is also fast to compute~\cite{puladas2018sumProductUnmixingNCM}.

An alternative approach which has been used to simplify the unmixing process associated with the NCM is to estimate the endmember means and covariance matrices \textit{a priori} based on spectral libraries extracted from the observed image. This has been performed considering libraries obtained both using pure pixel-based endmember bundle extraction methods~\cite{zhuang2015NCMcovarianceEstimationBundles} and on multiple endmember matrices estimated by a piecewise convex blind SU algorithm~\cite{zare2012bootstrappingPCEparameterEstimation}. However, these methods suffer from the limitations of image-based EM bundle extraction techniques, which will be discussed in detail in Section~\ref{sec:constructing_libs}.

Other works also considered a piecewise convex model which uses a set of different Gaussian distributions to model the endmembers. Afterwards, during SU each image pixel is assigned to one of these distributions using a membership function represented by a Dirichlet random variable.
The unmixing problem under this model was solved by considering both an alternating optimization method in a maximum \textit{a posteriori} framework~\cite{zare2010piecewiseConvexPCE} and a Markov chain Monte Carlo sampling approach providing an estimate of the posterior \mbox{distribution of interest~\cite{zare2013bayesianPiecewiseConvex}.}

Although the Dirichlet prior distribution is frequently used to represent the abundances, many works have considered variations which incorporate useful information from the underlying practical problem. 
Examples include the enforcement of abundance sparsity using a sparse Dirichlet prior~\cite{amiri2018sparsityBayesianUnmixingNCM}, or the encouragement of spatial homogeneity by dividing the abundance maps into a finite number of classes sharing the same Dirichlet distribution parameters. This division has been performed either blindly by means of a classification prior using the Potts model~\cite{halimi2015unsupervisedBayesianUnmixing} or through an \textit{a priori} segmentation of the hyperspectral image in a latent Dirichlet allocation framework~\cite{zou2017variabilityUlatentDirichletSuperpixels}.

More recently, the NCM has also been applied to problems other than that of linear unmixing or spectral variability.
For instance, the NCM has been considered to represent the uncertainties in EM estimation instead of the intrinsic variability of the material classes, which changes the problem by introducing statistical dependence between the different image pixels~\cite{zhou2016spatialCompositionalModel}.
Other works also applied the NCM to problems such as nonlinear SU with a bilinear mixing model \cite{luo2018bayesianNCMbilinearMM}, for the linear unmixing of sediment grain size distribution (where the EMs represent the grain sizes of constituent materials) to study transport and deposition of sediments~\cite{yu2016bayesianNCMunmixingSedimentGrainSize}, or to represent the variability of the endmembers across multiple images in multitemporal SU, using additional spatially sparse terms accounting for potential abrupt spectral changes between the different images~\cite{thouvenin2018hierarchicalBU}.

\subsubsection{\textbf{Other Endmember Distributions}}

Despite its popularity, the NCM does not have a strong physical motivation, which led to the consideration of more accurate distributions to represent the EMs.
For instance, a Beta distribution was considered in~\cite{du2014spatialBetaCompositional} in order to constrain reflectance values to physically meaningful ranges and to allow for possible skewness in the distribution. Unfortunately, a direct solution to the SU problem cannot be obtained. Thus, a piecewise constant model was assumed for the abundances, which allowed the parameters of the distribution to be estimated using a combination of a clustering algorithm and a variant of the method of moments. %

A Gaussian mixture model has also been considered in~\cite{zhou2018variabilityGaussianMixtureModel} in order to allow for possibly multi-modal EM distributions. The SU problem was solved as a maximum a posteriori estimation problem using a generalized expectation-maximization approach. However, since learning the parameters of Gaussian mixture models can be difficult, they were estimated before performing SU based on spectral libraries assumed to be known \textit{a priori}.

Another approach proposed to represent EM spectra as a sum of an average spectral signature known \textit{a priori} and a spatially and spectrally smooth function representing EM variability to provide a model that is physically more reasonable~\cite{halimi2016unmixingVariabilityNonlinearityMismodeling}. Bilinear mixing models were also considered along with an additive residual term to account for \cmag{mismodelling} effects or outliers.

A different approach has been proposed which does not make an explicit assumption about the distribution of EM spectra and instead only relies on some of their statistics.
This is the case of~\cite{bosdogianni1997unmixingVariabilityMomentsFitting}, which formulates the SU problem similarly to the method of moments by trying to find the abundance values which match the mean and covariances obtained through the LMM to those of the observed mixed pixels. A similar work applied the same idea using transformed statistics constructed from the ratio between the means and covariances of the pixels and endmembers in different spectral bands~\cite{faraklioti2001unmixingVariabIlluminationMomentsFitting}. This strategy increases the robustness of the method since band ratios are invariant to illumination variations. However, similarly to~\cite{du2014spatialBetaCompositional}, a piecewise constant abundance model is used to estimate the covariance matrix of the observed pixels. Moreover, the covariance matrices of the EMs are assumed to be known \textit{a priori}.

\section{Spectral Libraries}
\label{sec:overall_concerning_libs}

\begin{figure*}[!t]
    \centering
    \includegraphics[width=0.9\linewidth]{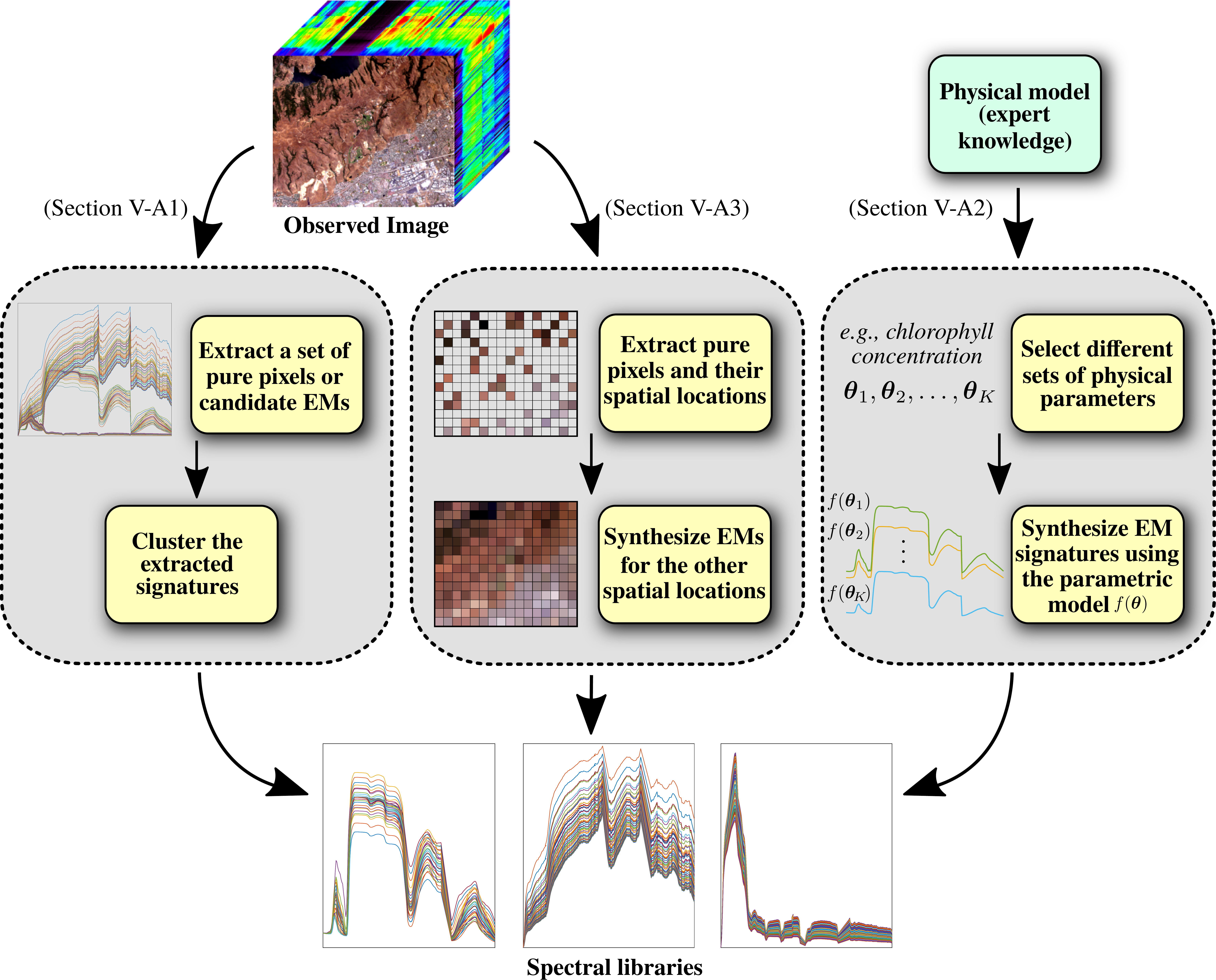}
    \caption{Illustrative diagram depicting existing approaches to generate spectral libraries: image based library generation using endmember extraction (left, discussed in Section~\ref{sec:constructing_libs_imageBased}), spatial interpolation of pure pixels extracted from the image at known locations (center, discussed in Section~\ref{sec:constructing_libs_interp}), and the generation of synthetic signatures from physics-based models (right, discussed in Section~\ref{sec:constructing_libs_physicsMdl}).}
    \label{fig:diagram_library_extraction}
\end{figure*}

A large number of SU techniques discussed in Section~\ref{sec:overall_methods_libs} address spectral variability by using spectral libraries or bundles known \textit{a priori}. The performance of these methods is often heavily impacted by how well the libraries can represent the endmembers actually present in the scene. Moreover, in many practical situations it is either very costly or even impossible to obtain laboratory or \textit{in situ} measurements of endmember spectra.
Another problem with many methods presented in Section~\ref{sec:overall_methods_libs} (like MESMA) is that their computational complexity increases very quickly with the library size, which can make the problem intractable for large libraries.

Thus, the problems of removing redundant or irrelevant spectra before SU and, especially, of extracting spectral libraries directly from observed hyperspectral images are of central importance in order to allow the techniques discussed in Section~\ref{sec:overall_methods_libs} to be widely applicable. Fortunately, several techniques have been proposed to address both of these problems, which we will discuss in detail in \mbox{this section.}

\subsection{\textbf{How to construct spectral libraries?}}
\label{sec:constructing_libs}

Many library-based SU works assume that spectral libraries are manually obtained from \textit{in situ} or through controlled laboratory measurements~\cite{roberts1998originalMESMA,somers2016endmemberLibrariesChapter}, which may be complicated in practical applications.
Moreover, existing libraries may have been acquired at conditions which do not reflect those actually observed in the scene, which introduces errors in the SU process~\cite{iordache2011sunsal,tompkins1997endmemberSelection,roberts1998originalMESMA}. Even the spatial resolution at which the hyperspectral image is acquired was found to have a considerable impact on the results of SU with MESMA in urban environments when the library was fixed \textit{a priori}~\cite{wetherley2017multiscaleLibraryEvaluationMESMA}.

Traditional endmember extraction algorithms (EEAs), on the other hand, typically consider only a single spectral signature per material and are thus unable to appropriately address spectral variability~\cite{plaza2004comparisonEMextraction,veganzones2008endmemberExtractionShortReview}.
These shortcomings make the construction of spectral libraries one of the main challenges of library-based SU methods~\cite{somers2016endmemberLibrariesChapter}. A simple and reliable method that has been employed to construct spectral libraries in practice depends on expert knowledge to manually select pure pixels of each material from the hyperspectral image~\cite{quintano2013spectralLibraryPPIandManual,wetherley2017multiscaleLibraryEvaluationMESMA}. However, there has been a growing interest in developing methods that can reduce the amount of user supervision and automatically extract libraries directly from observed hyperspectral images. Three main general lines of research can be identified in this direction:
\begin{itemize}
    \item[a)] extract multiple pure pixels from the observed hyperspectral image to generate a candidate library, and then cluster the extracted signatures into their respective material classes;
    \item[b)] generate libraries using radiative transfer models that represent endmember variability mathematically;
    \item[c)] extract pure pixels while keeping information about their spatial locations, and apply an interpolation algorithm to generate endmember signatures for each image pixel.
\end{itemize}
The diagram in Fig.~\ref{fig:diagram_library_extraction} gives an illustrative overview of the key ideas underlying each of these approaches, which are reviewed in the following.

\subsubsection{\textbf{Image based library construction}}
\label{sec:constructing_libs_imageBased}

\hfill
\enhancedcoloredtbox{Image-based Library Extraction:}{
\begin{itemize}
    \item[$+$] Allows spectral libraries to be extracted with signatures that are at the same conditions of the image pixels
    \item[$+$] Can benefit from expert knowledge to reliably identify pure pixels in the image
    \item[$-$] Depends strongly on the presence of pure pixels
    \item[$-$] The observed image should not be too small
    \item[$-$] Mixed pixels may be included in the library by mistake
    \item[$-$] Clustering the extracted signatures into their correct material classes is challenging
\end{itemize}}

The simplest approaches for the construction of image-based spectral libraries are completely supervised. Image pixels are included in the library either based on their correlation to some initial endmembers manually selected as the extreme points of the PCA of the observed image~\cite{bateson1996manualEndmemberSelection,bateson2000endmemberBundlesFuzzyU}, or simply by manually screening a large number of pure pixels extracted from the image using expert knowledge about the spectral characteristic of the materials in the scene~\cite{quintano2013spectralLibraryPPIandManual}. Pure pixels were also extracted from multiple hyperspectral images of the same scene acquired at different spatial resolutions to increase the diversity of the resulting spectral library in urban environments~\cite{wetherley2017multiscaleLibraryEvaluationMESMA}.
Other work used only partially labeled data in order to reduce the amount of domain knowledge that is required~\cite{meerdink2019bundleExtractionPartiallyLabelled}.
Recent strategies attempted to automate this process by extending EEAs for the extraction of multiple signatures of each material in the observed image.
The first work in this direction proposed to apply traditional EEAs to random subsets of pixels that are sampled from the hyperspectral image (with or without replacement)~\cite{somers2012automatedBundlesRansomSampling}. Different sets of EM signatures are generated using this method. All the extracted signatures are then grouped into different sets corresponding to the material classes by using a clustering algorithm (e.g., k-means). 
The size of the image subsets, however, must be selected with great care in order for EEAs to work satisfactorily~\cite{drumetz2016localIntrinsicDimensionality}, and the clustering step can be challenging.

Later works proposed different strategies for the extraction or selection of multiple pure pixels or endmember candidates from the observed image.
One simple iterative strategy consists of including in the library all pixels that are within a given spectral distance of some reference EMs~\cite{gao2015findingEndmemberClassesIterative}. This process is performed iteratively, with the reference EMs initialized using a standard EEA and then updated as the mean values of the library signatures at the previous iteration. Besides being very simple, this procedure does not require the library signatures to be clustered afterwards.
A related strategy worked in a reverse way, by iteratively removing pure pixels from a large initial set of candidate signatures in order to obtain the final spectral library~\cite{xu2016bundleExtractionReconstructionError}. A pixel candidate is removed if it can be represented with small error as a convex combination of the remaining signatures in the library. %
A clustering procedure is then performed to group the selected spectra into EM classes.

Recent works have proposed more involved empirical approaches to differentiate between spectrally similar materials when extracting or clustering the EM signatures, or to remove mixed pixels from the constructed library.
For instance, in~\cite{andreou2016bundlesExtractionVariabilityMultiscaleBand} EM extraction was performed multiple times for different subsets of the spectral bands constructed at multiple spectral scales and intervals. These signatures were afterwards clustered into EM classes based on a metric constructed from features derived from applying clustering algorithms individually to the spectral scales and intervals used previously.

A related strategy considered both the extraction and clustering of the library signatures based on subsets of the wavelet transform coefficients of the reflectance spectra that are robust to spectral variability~\cite{uezato2016novelBundlesExtractionClustering}. These subsets were selected based on how much their empirical statistical distribution deviates from an uncorrelated Gaussian distribution. %
The hyperspectral image was also partitioned into spatial segments using a hierarchical clustering algorithm, and only one signature for each spatial segment was considered to be included \mbox{in the library.}

Another strategy proposed to extract spectral signatures as image pixels which can best represent all other pixels in the observed image as a sparse linear combination~\cite{yin2019bundleExtractionSpecFeaturesLunar}. Afterwards, these signatures were grouped into material classes using spectral features derived from the slopes of a piecewise linear approximation of each signature. %

Auxiliary libraries available \textit{a priori} have also been used to aid in the extraction of image-based spectral libraries in~\cite{xu2019mappingImperviousSurfaceMESMA_FDA}. The $k$-nearest neighbor algorithm was first used to classify the image pixels in the different material classes, using library spectra known \textit{a priori} as training data. This led to a set of candidate EMs for each material class. Based on the classification results, the image-extracted library was then defined as the average spectra of those candidate EMs of each class that were contained in a spectral neighborhood of each of the training samples (from its corresponding class)~\cite{xu2019mappingImperviousSurfaceMESMA_FDA}.

Another group of approaches makes use of the empirical observation that pure pixels are more likely to be contained in spatially homogeneous regions.
Spectral libraries can be constructed either by restricting EM candidates to be contained in sufficiently homogeneous regions~\cite{xu2015bundlesExtractionSpatialSpectral,hua2019bundleExtractionSuperpixelPPI}, by applying an image over-segmentation strategy before pure pixel extraction~\cite{xu2018regionalClusteringPreprocessingSLIC}, or by considering EM candidates as the average of homogeneous regions obtained from a coarse spatial scale selected from a multiscale image decomposition~\cite{torres2014multiscaleEMextraction}. These strategies should be applied with care to avoid the inclusion of pixels extracted from mixed, homogeneous regions into \mbox{the library.}

Some alternatives tried to build spectral libraries by using different forms of matrix factorization of the hyperspectral image.
For instance, spectral libraries for each material class are constructed in~\cite{castrodad2011learningSpectralLibrariesSpCoding} by learning sparse representations of sets of pure pixels of each material, which are extracted from the observed image. More precisely, dictionary learning is applied to the pure pixels of each material, from which the resulting basis matrices are used to construct the spectral library.
Another approach proposed to extract the spectral library using the results of an SU procedure using a matrix factorization approach which does not accounts for spectral variability~\cite{zhao2017unmixingArchetypalAnalysisEMextraction}. However, besides depending on the results of another SU algorithm, there is no guarantee that the selected signatures are pure pixels.

\subsubsection{\textbf{Generating spectral libraries from physics models}}
\label{sec:constructing_libs_physicsMdl}

\hfill
\enhancedcoloredtbox{Physics-based Library Synthesis:}{
\begin{itemize}
    \item[$+$] Can generate libraries independently of the observed image
    \item[$+$] Can represent a wide range of spectral variability if more complex models are employed
    \item[$-$] Depends on the availability of an accurate physical model for the spectra of the EMs
\end{itemize}}

An alternative approach to generate spectral libraries which does not depend on the observed hyperspectral image is to employ a physics-based (i.e., radiative transfer) model describing the reflectance of the EMs as a function of physico-chemical parameters. This allows us to generate different instances of the material spectra to constitute a synthetic library by sampling the free parameters of the model. Examples of such models include the PROSPECT model~\cite{jacquemoud1990PROSPECTmodelLeaf} for vegetation or Hapke's~\cite{Hapke1981} and Shkuratov's models~\cite{shkuratov1999model} for packed \mbox{particle spectra.}

Different models inspired by physics have been employed to generate or augment spectral spectral libraries for SU in many applications. These applications include models for canopy as a function of its height and canopy radius~\cite{peddle1999canopyModelLibraryBorealRTF}, fire temperature radiance as a function of view and solar geometry and atmospheric conditions~\cite{dennison2006wildfireTemperatureMESMA_RTF} and soil reflectance as a function of moisture content~\cite{somers2009modelSoilSpectraMoisture}.
This strategy has also been applied to generate training data for SU of binary mixtures of vegetation and impervious materials using machine learning algorithms~\cite{yang2017fractionalVegetationCoverML_RTF,jia2016fractionalVegetationCoverML_RTF,verger2011fractionalVegetationCoverML_RTF,peddle1999canopyModelLibraryClassificationRTF}.
Note, however, that directly sampling all parameters of complex models such as PROSPECT might lead to a very large number of signatures. This has motivated strategies to sample the parametric models more efficiently or to remove redundant spectra from the generated library~\cite{tits2012clusteringMESMAlibrary_RTF}.

In spite of their advantages, a significant drawback of these methods is the requirement of accurate knowledge of the physical process governing the observation of the reflectance of the materials by the sensor. %
A different approach attempted to circumvent this issue by proposing a data augmentation strategy, where one wishes to synthesize additional signatures to be included in small, pre-existing libraries~\cite{borsoi2019EMlibManInterpVAE}. The spectral signatures in the library are used as training data in order to learn the statistical distribution of the EMs using deep generative models such as variational autoencoders and deep neural networks. This allows one to sample new signatures from the learned distributions to augment the existing library.

\subsubsection{\textbf{Spatial interpolation of endmember signatures}}
\label{sec:constructing_libs_interp}

\hfill
\enhancedcoloredtbox{Spatial Endmember Interpolation:}{
\begin{itemize}
    \item[$+$] Uses the hypothesis of spatially correlated EM signatures
    \item[$-$] Needs knowledge of the spatial position of pure pixels in the scene
    \item[$-$] The amount of pure pixels available can strongly affect the performance of the methods
\end{itemize}}

A number of approaches based on the assumption that EMs are spatially correlated proposed to synthesize pixel-dependent EM signatures based on a set of pure pixels at known spatial locations using interpolation techniques.
Many of these works aim to perform SU of vegetation and soil mixtures by using vegetation indices (i.e., spectral features given by ratios of band differences, such as the NDVI) \textit{in lieu} of \mbox{traditional endmembers.}

For instance, the spatial interpolation of vegetation and soil NDVIs based on linear regression has been considered for SU of coarse resolution images, where the training samples for the EMs were obtained using using classification maps from complementary, high resolution images available \textit{a priori}~\cite{maselli2001estimatingSpatiallyVariableEndmembers}.
A similar strategy considered the use of spatially weighted kriging employing as training samples pure pixels which were either manually extracted from the scene~\cite{johnson2012spatialInterpolationEndmembers} or obtained by randomly sampling the vertices of the simplex obtained by a low-dimensional projection of the hyperspectral image~\cite{li2016geostatisticalEndmemberInterpolation}. This strategy allowed one to weight the contribution of the training samples according to their spatial distance to each interpolated signature.

Other works also considered the spatial interpolation of actual spectral signatures instead of just vegetation and soil indices using spatially weighted linear regression or kriging.
This has been performed considering training data obtained both from complementary high resolution classification maps~\cite{zhang2015neighbourhoodSpecificEndmemberSignatureGeneration}, or from pure pixels extracted from the image inside sub-regions appropriately selected with the aid of a classification algorithm~\cite{li2017classificationAndEndmemberInterpolation}.

\subsection{\textbf{Library pruning techniques}}
\label{sec:pruning_libs}

One significant problem with many SU methods based on spectral libraries such as MESMA is that their computational complexity increases quickly with the size of the spectral library. 
Furthermore, databases containing laboratory acquired spectra often contains hundreds of different materials. Using a library of this size can actually decrease the performance of SU since the problem becomes more and more ill-posed.

One solution to this problem consists of removing redundant or irrelevant signatures from large spectral libraries before the SU process. These approaches, also called library pruning, have been largely applied in order to reduce the complexity and improve the accuracy of both MESMA~\cite{dennison2003libraryPruningMESMAcost} and sparse unmixing algorithms~\cite{iordache2012dictionaryPruningSparseUnmixingCost}.
There are three main groups of library pruning techniques. Library reduction techniques just remove redundant signatures to improve the computation time. Endmember selection techniques identify which materials are present in each hyperspectral image pixel to remove absent EM classes from the library before SU. Same-class library pruning attempt to identify and remove signatures which are acquired at different conditions from those of the observed image. These approaches will be reviewed in detail in the following.

\subsubsection{\textbf{Library reduction techniques}}
\label{sec:sub_lib_reduction}

\hfill
\enhancedcoloredtbox{Spectral Library Reduction:}{
\begin{itemize}
    \item[$+$] Very simple strategies that do not depend on the observed hyperspectral image
    \item[$-$] Only reduces computational complexity, but does not improve the quality of the SU results
\end{itemize}}

Library reduction techniques attempt to remove redundant spectral signatures from the library regardless of the observed hyperspectral image, which tends to improve the computational complexity of SU but not necessarily its quality. 
A common idea is to find a small set of signatures which can best represent the remaining spectra of the same EM class in some sense~\cite{roth2012libraryPruningComparisoncost} such as the squared error~\cite{dennison2003libraryPruningMESMAcost}, the average spectral angle~\cite{dennison2004comparisonErrorMetricsEndmemberSelection} or the count-based EM selection metric, where one counts the number of signatures one candidate can represent with error below a threshold~\cite{roberts2003countBasedEndmemberSelection}.
An alternative method also divided the library signatures into groups according to their Euclidean norm, selecting one signature from each group to explicitly account for brightness variations~\cite{xu2015endmemberSelectionNormPartitioning}.

\subsubsection{\textbf{Endmember selection methods}}
\label{sec:lib_pruning_EM_selection}

\hfill
\enhancedcoloredtbox{Endmember Selection:}{
\begin{itemize}
    \item[$+$] Remove only entire material classes from the library for each pixel, and is also effective for variability-free SU
    \item[$+$] Leverage information from the observed hyperspectral image
    \item[$+$] Can improve the SU quality and reduce the computational complexity
    \item[$-$] Usually depend on some sort of classification procedure
    \item[$-$] Rely on the observation that usually only a few materials are contained in each pixel
\end{itemize}}

Endmember selection techniques attempt to identify which EM classes are present in each pixel using information such as classification maps~\cite{garcia2005VMESMA,degerickx2017libraryPruningUrbanVariability} to remove entire absent materials from the library and improve the unmixing results~\cite{garcia2005VMESMA,degerickx2017libraryPruningUrbanVariability}. This relies on the observation that hyperspectral image pixels usually contain only a small number of materials, and has also been applied to SU without considering spectral variability~\mbox{\cite{rogge2006iterativeEndmemberRemoval,bian2017endmemberSelectionIterativeDistanceToPixels}}.

The simplest EM selection methods use classification algorithms to select the EM classes present in mixed pixels~\cite{roessner2001classificationUrbanEndmemberClassIdentification,deng2016classBasedMESMA}.
Another work employed a block sparse unmixing algorithm as a preprocessing step in order to remove material classes with low abundances values from the library for each image pixel before applying the MESMA algorithm to obtain the final SU results~\cite{chen2016endmemberSelectionSparseMESMA}.
A more elaborate approach proposed to semantically organize subsets of material classes in a hierarchical tree, starting from a rough (e.g. pervious and impervious) up to a fine differentiation between the endmembers (e.g., different vegetation species)~\cite{franke2009hierarchicalMESMA}. Afterwards, SU is performed at each level of the tree, using the abundance results in the previous, coarser level to constrain which EMs can be selected at the current one (i.e., a pixel containing only a pervious EM in the coarse scale cannot have concrete EM in the finer one).

Some recent approaches have also proposed to use external, complementary data in order to aid in identifying which materials are present in each pixel.
For instance, in~\cite{liu2013urbanMESMAstratifiedClassification} the hyperspectral image was divided into rural and urban subsets by using external data of road network density, which allowed for the use of a separate set of EM classes for each of the subsets.
Another work proposed to use additional LIDAR data to remove material classes from the library of each pixel based on its height distribution (e.g., a ``tree'' or ``building'' endmember can be removed from a pixel that has low height)~\cite{degerickx2019libraryPruningWithLIDAR}.

\subsubsection{\textbf{Pruning libraries within the same class}}
\label{sec:lib_pruning_same_class}

\hfill
\enhancedcoloredtbox{Same-Class Endmember Pruning:}{
\begin{itemize}
    \item[$+$] Remove spectral signatures from each material class that are not representative of the observed hyperspectral image
    \item[$+$] Can improve the SU quality and reduce the computational cost even for libraries with few materials classes
    \item[$-$] Identify which signatures in the spectral library do not share the same acquisition conditions with the observed image is generally difficult 
\end{itemize}}

Recent approaches proposed to remove signatures from the library that have been acquired at conditions different from those of the hyperspectral image, keeping only signatures that are representative of the observed image. However, measuring the representativeness of the EM signatures is a difficult task.
A simple approach proposed to remove signatures that have a large spectral angle and spectral $L_1$ distance relative to the observed pixels~\cite{fan2014endmemberSelectionMESMAprunning}. However, this strategy might discard relevant signatures in the presence of many mixed pixels.
Another work proposed to compare only pure pixels extracted from the image with the library spectra in the \mbox{wavelet domain~\cite{singh2017transformationsSelectingEMlibraryThermalIR}.}

A different approach proposed to remove library elements that have large distances to a small set of the leading eigenvectors of the observed hyperspectral image, and are thus unlikely to be present therein~\cite{iordache2014musicLibraryPrunning}. This strategy eliminates the direct need for pure pixels in the scene. It has also been successfully applied for plant production system monitoring~\cite{iordache2014dynamicPlantMonitoringLibraryPruning}, and was later extended to consider a brightness normalization pre-processing step and other strategies from Section~\ref{sec:sub_lib_reduction} to additionally remove redundant spectra~\cite{degerickx2017libraryPruningUrbanVariability}.

Another work also proposed to perform library pruning iteratively in a sparse unmixing formulation by removing signatures corresponding to low abundance values during the SU process~\cite{zhang2018iterativeLibraryPruningSparseUnmixing}. However, this process depends directly on the accuracy of the SU process at the first iterations.

\section{Experimental Evaluation}
\label{sec:experimental}

This section presents a brief discussion about the experimental evaluation of the unmixing algorithms when spectral variability is considered. We first discuss the generation of synthetic data in detail. Afterwards, some existing software packages that can be useful for practitioners are presented. Finally, an illustrative, tutorial-style simulation example is presented in order to demonstrate the use of a few of the SU techniques reviewed in the paper, after selection using the decision tree in Fig.~\ref{fig:decision_tree}.

\subsection{\textbf{Generating Synthetic Data}}

One challenge in the evaluation of unmixing methods is the lack of reliable ground truth data for the abundances of real hyperspectral images. The difficulty in collecting ground truth data is even more pronounced when endmember variability is considered. 
Thus, being able to generate realistic synthetic data (for which the true abundances are available) turns out to be important to allow a quantitative evaluation of SU algorithms.

More precisely, the generation of synthetic data can be roughly divided into three steps:
\begin{enumerate}
    \item generating synthetic abundances;
    \item generating endmember signatures for each pixel in the image;
    \item applying the mixing model of choice (in our case, the LMM) to generate the mixed image pixels.
\end{enumerate}
We discuss each of these steps in the following.

\subsubsection{\textbf{Generating synthetic abundance data}}

The generation of synthetic abundance maps can be performed in different ways. A simple strategy is to sample the abundance values randomly from a Dirichlet distribution. This approach allows one to control the amount of pure pixels in the image, and can be useful when performing Monte Carlo simulations in which large amounts of data must be generated. 
Another approach consists in introducing spatial contextual information (i.e., pixels that are close in space tend to have similar abundance values) into the generated abundances to generate more realistic data. Such data can be generated using, for instance, piecewise smooth images sampled from a Gaussian random field~\cite{kozintsev1999computations}. This approach is able to generate images containing smooth regions, sharp transitions and fine details whose spatial composition and regularity characteristics can be controlled by the user~\cite{kozintsev1999computations}.
One software tool that can be used to generate abundance maps according to Gaussian random fields is the Hyperspectral Imagery Synthesis tool for Matlab (available for download \href{http://www.ehu.eus/ccwintco/index.php?title=Hyperspectral_Imagery_Synthesis_tools_for_MATLAB}{here}).
Another way to obtain realistic synthetic abundance maps is to consider abundances obtained by applying an existing spectral unmixing algorithm on a real hyperspectral image~\cite{hao2016semiRealisticHSIsimulation}. The resulting abundance maps will have a realistic spatial distribution, and can be used as ground truth to generate new synthetic datasets.

\subsubsection{\textbf{Generating synthetic endmember variability}}
\label{sec:example_synth_EM_var}

Generating realistic endmember variability data is not a simple task since, as explained before, the spectral signatures of the materials present a complex dependence of different physico-chemical and environmental parameters.
Fortunately, very accurate radiative transfer models have been developed for many applications. Such models describe the physical processes governing, e.g., vegetation spectra~\cite{jacquemoud2001leafOpticalPropertiesReview}, mineral interactions~\cite{Hapke1981,shkuratov1999model} and atmospheric effects~\cite{berk2004modtran5}.

Well-calibrated radiative transfer models can be used to generate realistic simulated image scenes that allow one to simultaneously study nonlinear mixtures and endmember variability effects.
Experimental studies have found that the data simulated using such models show a very strong agreement with reference ground truth data collected under the same circumstances using ground-based spectral measurement set-ups~\cite{somers2013validatingRTFsOrchards,somers2009nonlinearMixtureOrchards}. This approach has been already used to evaluate nonlinear unmixing models in~\cite{dobigeon2014comparisonNonlinearMixturesRTF}.
Thus, well-calibrated radiative transfer models can be used to generate realistic simulated hyperspectral data that allow us to develop, optimize, test and compare different SU techniques considering endmember variability.

Although complex ray-tracing simulations can be considered (e.g.,~\cite{jacquemoud2001leafOpticalPropertiesReview,Hapke1981,shkuratov1999model,berk2004modtran5,tits2012rayTracingSimulNonlinearVarEvaluation}), here we present some simplified models for illustrative purposes, which describe variability present in vegetation spectra and caused by different viewing geometries.

The first model we consider is the PROSPECT-D~\cite{feret2017prospect_d}, which represents vegetation leaf spectra as a function of, e.g., the chlorophyll and dry matter content, and of the equivalent water thickness. PROSPECT-D and other related models for vegetation spectra can be downloaded \href{http://teledetection.ipgp.jussieu.fr/prosail/}{here}, for different software platforms (including Matlab and Python).

We also consider a simplification of Hapke's model~\cite{Hapke1981} by assuming a Lambertian (isotropic) scattering and a densely packed medium. This simplified model describes variations in the reflectance spectra of a material $y_{\rm sensor}$ (at each wavelength) as a function of the viewing geometry~\cite{heylen2014review,drumetz2019ELMM_from_Hapke}:
\begin{align}
    y_{\rm sensor}
    = \frac{\omega}{(1+2\mu_1\sqrt{1-\omega})(1+2\mu_2\sqrt{1-\omega})} \,,
    \label{eq:approximate_hapke_model}
\end{align}
where $\omega$ is the single scattering albedo of the material, and $\mu_1$ (resp.~$\mu_2$) is the cosine of the angle between the incoming (resp.~outgoing) radiation and the normal to the surface. This model allows us to generate different EM spectra by varying the values of $\mu_1$ and $\mu_2$. While~\eqref{eq:approximate_hapke_model} is approximately linear for small albedo values, important nonlinearities occur for large albedo values~\cite{drumetz2019ELMM_from_Hapke}.

We also consider variability introduced by errors occurring in a simple atmospheric compensation model, where the reflectance of each pixel at each wavelength is obtained by dividing the corresponding pixel's radiance by the radiance observed at a perfectly reflective calibration panel~\cite[Section~IV-A1]{uezato2016novelBundlesExtractionClustering},\cite{ramakrishnan2015simpleIlluminationCompensationModel}.
Assuming full visibility and that the adjacency effect is negligible, this model is given by:
\begin{align}
    y_{\rm sensor} = y_{\rm s} 
    \frac{E_{\rm sun-gr} \mu_1 + E_{\rm sky}}{E_{\rm sun-gr} \mu_2 + E_{\rm sky}}
    \,,
\end{align}
where $y_{\rm s}$ and $y_{\rm sensor}$ denote the reflectance at the ground and at the sensor, $E_{\rm sun-gr}$ denotes the solar radiance observed at the ground level, and $E_{\rm sky}$ denotes the skylight. Parameters $\mu_1$ and $\mu_2$ are the cosines of the angles between the surface normal and the direction of the sun at each pixel and at the calibration panel, respectively. By fixing $\mu_2$, $E_{\rm sun-gr}$ and $E_{\rm sky}$ \textit{a priori}, $\mu_1$ can be varied to simulate spectral signatures at different viewing geometries.

\subsubsection{\textbf{Generating the mixed pixels}}

Finally, each pixel can be generated according to the spectral variability accommodating LMM described in~\eqref{eq:model_variab_general}, with the endmember for each pixel ($\bM_n$ columns) sampled randomly from the set of synthetically generated signatures\footnote{More complex models can also be employed to consider the spatial relationship between the endmembers by, e.g., generating the viewing geometry according to a digital terrain model as in~\cite{drumetz2016blindUnmixingELMM}.}. Additive noise can also be introduced to obtain a desired signal-to-noise ratio.

\subsection{\textbf{Available Software Resources}}

Several software packages are available to perform SU with spectral variability. Classical techniques such as MESMA and some of its alternatives (including library pruning and transformation methods) can be found in the VIPER tools software package \cite{roberts2019VIPER_tools_v2}, which is available as plug-ins for well-established software such as ENVI (\href{https://sites.google.com/site/ucsbviperlab/viper-tools}{here}) and QGIS (\href{https://pypi.org/project/mesma/}{here}).
An implementation of the MESMA algorithm is also available in~R in the RStoolbox (\href{https://github.com/bleutner/RStoolbox}{here}).

Algorithms that were developed more recently, on the other hand, are usually only available as standalone prototypes implemented in Matlab or Python. A list of software packages for some of the papers reviewed in this work (most of which found at the authors' websites) is contained in Table~\ref{tab:listOfCodes}. Also, the OpenRemoteSensing website, which aims to share and disseminate codes and papers, also has an increasing number of SU methods (\href{https://openremotesensing.net/kb/codes/spectral-unmixing/}{here}), some of which considering spectral variability.

\begin{table}[th]
\centering
\renewcommand{\arraystretch}{1.3}
\caption{List of computational codes containing implementations of some of the works reviewed in this paper (provided by the respective authors)}
\begin{tabular}{|ccc|}
\hline
Method & Link & Language \\
\hline
\multicolumn{3}{|c|}{Methods that use spectral libraries} \\
\hline

MESMA~\cite{roberts1998originalMESMA}, AAM~\cite{heylen2016alternatingAngleMinimization}  & \href{https://sites.google.com/site/robheylenresearch/code/AAM.zip?attredirects=0&d=1}{link} & Matlab \\

SUnSAL~\cite{iordache2011sunsal}, SUnSAL-TV~\cite{iordache2012sunsal_TV} & \href{http://www.lx.it.pt/~bioucas/publications.html}{link} & Matlab\\

Sparse SU with mixed norms~\cite{drumetz2019SU_bundlesGroupSparsityMixedNorms} & \href{https://openremotesensing.net/knowledgebase/hyperspectral-image-unmixing-with-endmember-bundles-and-group-sparsity-inducing-mixed-norms/}{link} & Matlab\\

\hline
\multicolumn{3}{|c|}{Bayesian methods} \\
\hline

BCM~\cite{du2014spatialBetaCompositional} & \href{https://github.com/GatorSense/BetaCompositionalModel}{link} & Matlab \\

NCM-E (NCM by Eches \textit{et al.})~\cite{eches2010bayesianNCM} & \href{http://olivier.eches.free.fr/research.html}{link} & Matlab \\

UsGNCM~\cite{halimi2015unsupervisedBayesianUnmixing} &  \href{https://sites.google.com/site/abderrahimhalimi/publications}{link} & Matlab \\

Bayesian OU~\cite{thouvenin2018hierarchicalBU} & \href{https://pthouvenin.github.io/robust-unmixing-plmm/}{link} & Matlab\\

PCOMMEND~\cite{zare2010piecewiseConvexPCE} & \href{https://github.com/GatorSense/PCOMMEND}{link} & Matlab\\

GMM~\cite{zhou2018variabilityGaussianMixtureModel} & \href{https://github.com/zhouyuanzxcv/Hyperspectral}{link} & Matlab\\

\hline
\multicolumn{3}{|c|}{Parametric EM models} \\
\hline

ELMM~\cite{drumetz2016blindUnmixingELMM} & \href{https://openremotesensing.net/knowledgebase/spectral-variability-and-extended-linear-mixing-model/}{link} & Matlab\\

PLMM~\cite{Thouvenin_IEEE_TSP_2016_PLMM} & \href{https://pthouvenin.github.io/unmixing-plmm/}{link} & Matlab\\

GLMM~\cite{imbiriba2018glmm} & \href{https://github.com/talesimbiriba/GLMM}{link} & Matlab\\

DeepGUn~\cite{borsoi2019deepGun} & \href{https://github.com/ricardoborsoi/Unmixing_with_Deep_Generative_Models}{link} & Matlab\\

MUA-SV~\cite{Borsoi_multiscaleVar_2018} & \href{https://github.com/ricardoborsoi/DataDependentSUvarRelease}{link} & Matlab\\

OU~\cite{thouvenin2016online} & \href{https://pthouvenin.github.io/online-unmixing-plmm/}{link} & Matlab \\

\hline
\multicolumn{3}{|c|}{EM-model-free methods} \\
\hline

RUSAL~\cite{halimi2017fastNonparametricVariability} & \href{https://sites.google.com/site/abderrahimhalimi/publications}{link} & Matlab \\

SULoRa~\cite{hong2018SULoRA_lowRankEnbeddingUnmixingVar} & \href{https://sites.google.com/view/danfeng-hong/data-code}{link} & Matlab \\

ALMM~\cite{hong2019augmentedLMMvariability} &  \href{https://openremotesensing.net/knowledgebase/an-augmented-linear-mixing-model-to-address-spectral-variability-for-hyperspectral-unmixing/}{link} & Matlab\\

ULTRA-V~\cite{imbiriba2018ULTRA_V} & \href{https://github.com/talesimbiriba/ULTRA-V}{link} & Matlab \\

\hline
\end{tabular}
\label{tab:listOfCodes}
\end{table}

\begin{figure}
    \centering
    \includegraphics[width=\linewidth]{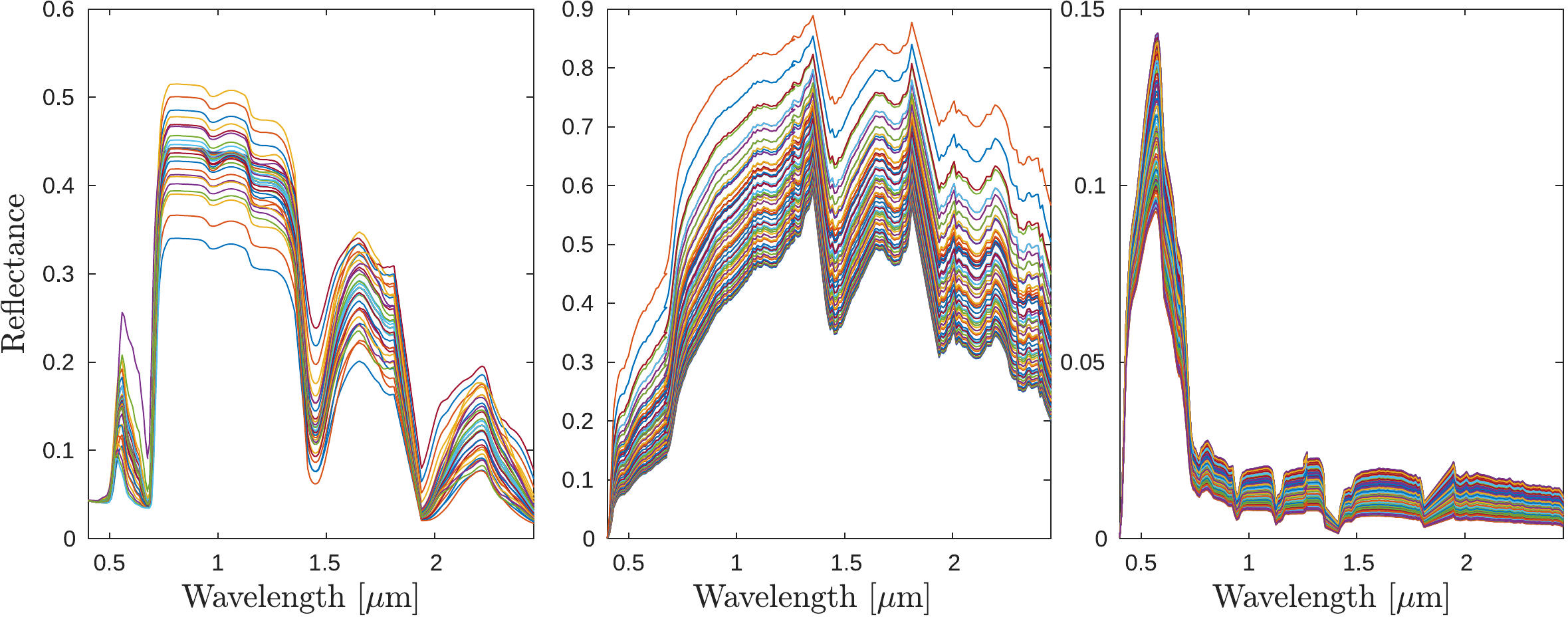}
    \caption{Generated spectral signatures used in the synthetic hyperspectral image, vegetation (left), dirt (center) and water (right).}
    \label{fig:example_synth_signatures}
\end{figure}

\subsection{\textbf{Experimental setup and results}}
\label{sec:exp_example_setup}

We now present a simulation to illustrate the application of some of the algorithms reviewed in this work. Note that this simulation is merely illustrative, and not a comprehensive performance evaluation. We generated a synthetic hyperspectral image containing vegetation, dirt and water as constituent materials. Spatially correlated abundances with $50\times50$ pixels were first sampled from a Gaussian random field. Then, we followed the procedure described in Section~\ref{sec:example_synth_EM_var} to generate different endmember spectra for each material in the scene. The PROSPECT-D model was used to generate vegetation spectra, while the simplified Hapke and atmospheric models were used to generate dirt and water spectra, respectively, at different viewing geometries. The generated synthetic signatures, containing $L=198$ bands, can be seen in Fig.~\ref{fig:example_synth_signatures}. The endmembers contained in each pixel were sampled randomly from this set of synthesized signatures, and the pixel spectra were then generated following the LMM with variability in~\eqref{eq:model_variab_general}, with white Gaussian noise added to the image to obtain a signal-to-noise ratio of 30~dB.

To evaluate the SU results, we considered as quantitative quality measures the Root Mean Squared Error (RMSE) and the Spectral Angle Mapper (SAM). The RMSE between two generic variables $\bX$ and $\,\widehat{\!\bX}$ is defined as
\begin{align}
    \operatorname{RMSE}_{\bX} = \sqrt{\frac{1}{N_{\bX}}\|\bX-\,\widehat{\!\bX}\|_F^2} \,,
\end{align}
where $\|\cdot\|_F$ is the Frobenius norm and $N_{\bX}$ denotes the number of elements in $\bX$. We used the RMSE to evaluate the estimated abundances $\,\widehat{\!\bA}$, the reconstructed images $\widehat{\bY}$ and the estimated, pixel dependent endmembers $\widehat{\bM}_n$ (for the cases when this estimate was available).
The SAM was also used to evaluate the estimated endmembers as:
\begin{align}
    \operatorname{SAM}_{\bM} = \frac{1}{LPN} \sum_{n=1}^N \sum_{p=1}^P \arccos \bigg( \frac{\bm_{p,n}^\top \widehat{\bm}_{p,n}}{\|\bm_{p,n}\| \|\widehat{\bm}_{p,n}\|}\bigg)
    \,,
\end{align}
where $N$ is the number of pixels and $P$ the number of materials in the hyperspectral image.

We also evaluated the complexity of the algorithms through their execution times, measured in an Intel Core~I7 processor with 4.2~GHz and 16~Gb of RAM. Finally, in order to increase the reliability of the results, we executed the simulation for ten independent Monte Carlo realizations and report the average values for all metrics.

\subsubsection{\textbf{Algorithm selection and setup}}

For illustrative purposes, we considered the recovery of the abundance maps following four different paths in the decision tree of Fig.~\ref{fig:decision_tree} (selected according to the algorithm implementations available in Table~\ref{tab:listOfCodes}).
\begin{enumerate}
    \item Small spectral libraries extracted directly from the image, no expert knowledge available:
    \begin{itemize}
        \item[a)] less user supervision: MESMA and variants~\cite{roberts1998originalMESMA};
        \item[b)] less computational cost: Sparse unmixing (fractional sparse SU~\cite{drumetz2019SU_bundlesGroupSparsityMixedNorms});
    \end{itemize}
    
    \item Spectral libraries not available \textit{a priori}:
    \begin{itemize}
        \item[a)] less user supervision necessary: Bayesian methods (NCM-E~\cite{eches2010bayesianNCM}, BCM~\cite{du2014spatialBetaCompositional,du2019codesForTheBetaCompositionalModel})
        \item[b)] less computational cost: Parametric models (ELMM~\cite{drumetz2016blindUnmixingELMM}, DeepGUn~\cite{borsoi2019deepGun}); EM-model-free methods (RUSAL~\cite{halimi2017fastNonparametricVariability})
    \end{itemize}
\end{enumerate}

We additionally considered the FCLS solution as a baseline, using a single set of endmembers extracted from the image using the Vertex Component Analysis (VCA) algorithm~\cite{Nascimento2005}. The EMs extracted by VCA were also used as initialization or as reference/mean signatures for some of the algorithms (ELMM, DeepGUn, RUSAL, NCM-E).
For MESMA and sparse SU, the spectral libraries were extracted from the observed image, as will be described in greater detail in the next subsection. The spectral libraries were also used to estimate the parameters of the beta distribution in the BCM.
The regularization/tuning parameters of the algorithms (fractional sparse SU, ELMM, DeepGUn, RUSAL) were manually adjusted to maximize the abundance reconstruction performance measured in an independent dataset generated following the same specifications as in the beginning of Section~\ref{sec:exp_example_setup}.

\begin{figure}
    \centering
    \includegraphics[width=\linewidth]{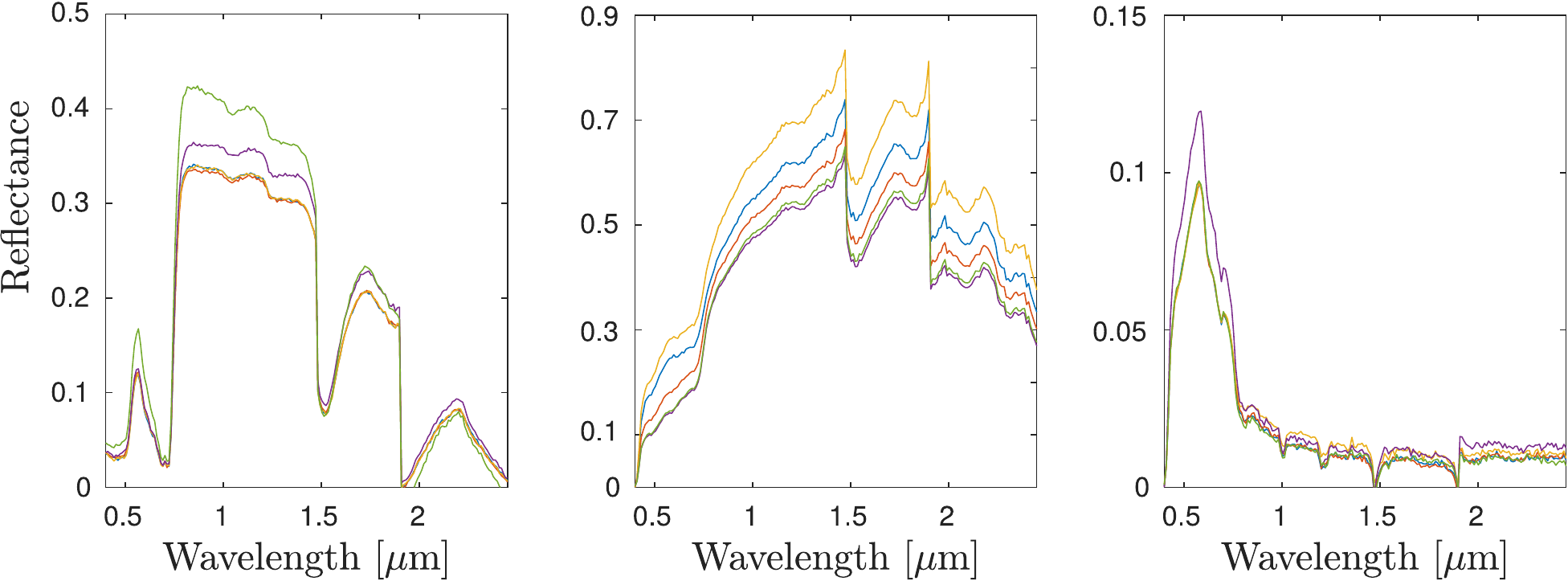}
    \caption{Endmember bundles extracted by batch VCA~\cite{somers2012automatedBundlesRansomSampling}, for vegetation (left), dirt (center) and water (right).}
    \label{fig:extracted_signatures}
\end{figure}

\subsubsection{\textbf{Library extraction}}

In order to demonstrate the use of library-based SU methods in practical settings, the spectral libraries used by MESMA and Fractional sparse SU were extracted directly from the observed image. We used the method described in~\cite{somers2012automatedBundlesRansomSampling}, which consists of performing endmember extraction (in this case, using the VCA algorithm) in subsets of pixels randomly sampled from the image. We extracted five sets of endmembers, using subsets of 500 pixels each (sampled with replacement). 

\begin{figure}
    \centering
    \includegraphics[width=\linewidth]{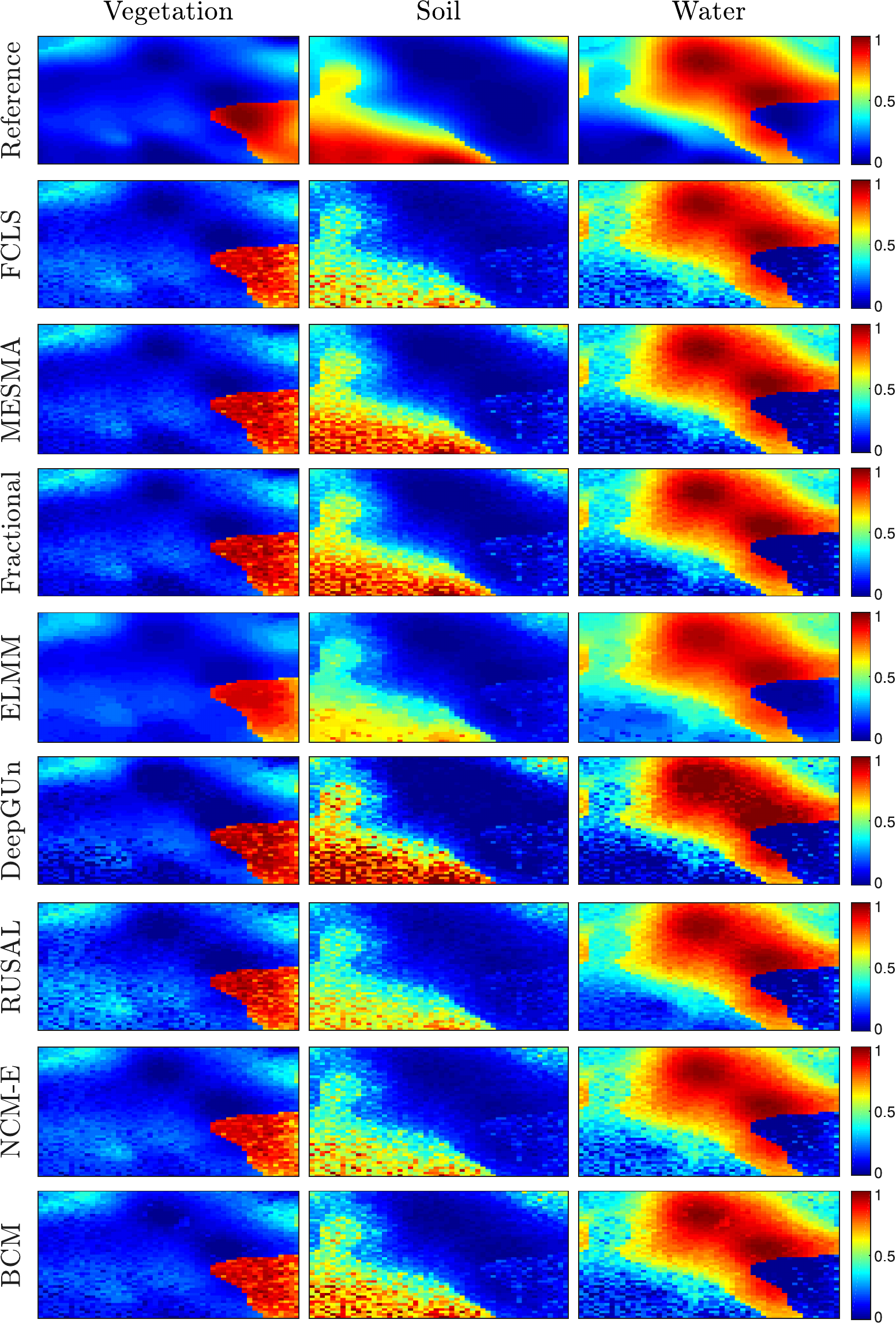}
    \caption{Abundance maps estimated by the algorithms (values are mapped to colors ranging from blue ($a=0$) to red ($a=1$).}
    \label{fig:example_estim_abundances}
\end{figure}

The library was kept small in order to prevent the inclusion of redundant signatures and also to reduce the probability of selecting mixed pixels by mistake. As a byproduct, this also keeps the computational complexity of methods such as MESMA very low, while providing good experimental results in this example. The estimated signatures can be seen in Fig.~\ref{fig:extracted_signatures}. Although the spectral variability in Fig.~\ref{fig:extracted_signatures} is less accentuated than that of the true endmembers in Fig.~\ref{fig:example_synth_signatures}, the estimated signatures are good representatives of the materials in the scene. The good performance of the library extraction method can be explained by the presence of multiple pure pixels in the synthetically generated abundance maps, which can be seen in the first row of Fig.~\ref{fig:example_estim_abundances}.

\begin{table}%
\small
\caption{Quantitative simulation results (RMSE results are multiplied by $10^4$).}
\vspace{-0.2cm}
\centering
\renewcommand{\arraystretch}{1.2}
\setlength{\tabcolsep}{3.3pt}
\begin{tabular}{lcccccc}
\midrule
&	$\operatorname{RMSE}_{\bA}$	&	$\operatorname{RMSE}_{\bM}$	&	$\operatorname{SAM}_{\bM}$	&	$\operatorname{RMSE}_{\bY}$	&	Time [s] \\
\midrule
FCLS	&	9.899	&	--	&	--	&	0.239	&	0.37	\\
MESMA	&	6.083	&	0.504	&	0.234	&	0.159	&	4.90	\\
Fractional	&	5.993	&	0.525	&	0.232	&	0.159	&	3.41	\\
ELMM	&	8.695	&	0.697	&	0.560	&	0.127	&	28.84	\\
DeepGUn	&	7.203	&	0.447	&	0.395	&	0.324	&	80.42	\\
RUSAL	&	9.509	&	--	&	--	&	0.108	&	1.05	\\
NCM-E	&	9.897	&	--	&	--	&	0.239	&	2482.85	\\
BCM	&	8.105	&	--	&	--	&	0.472	&	468.69	\\
\midrule		
\end{tabular}
\label{tab:results_synthData}
\end{table}

\begin{figure}
    \centering
    \includegraphics[width=\linewidth]{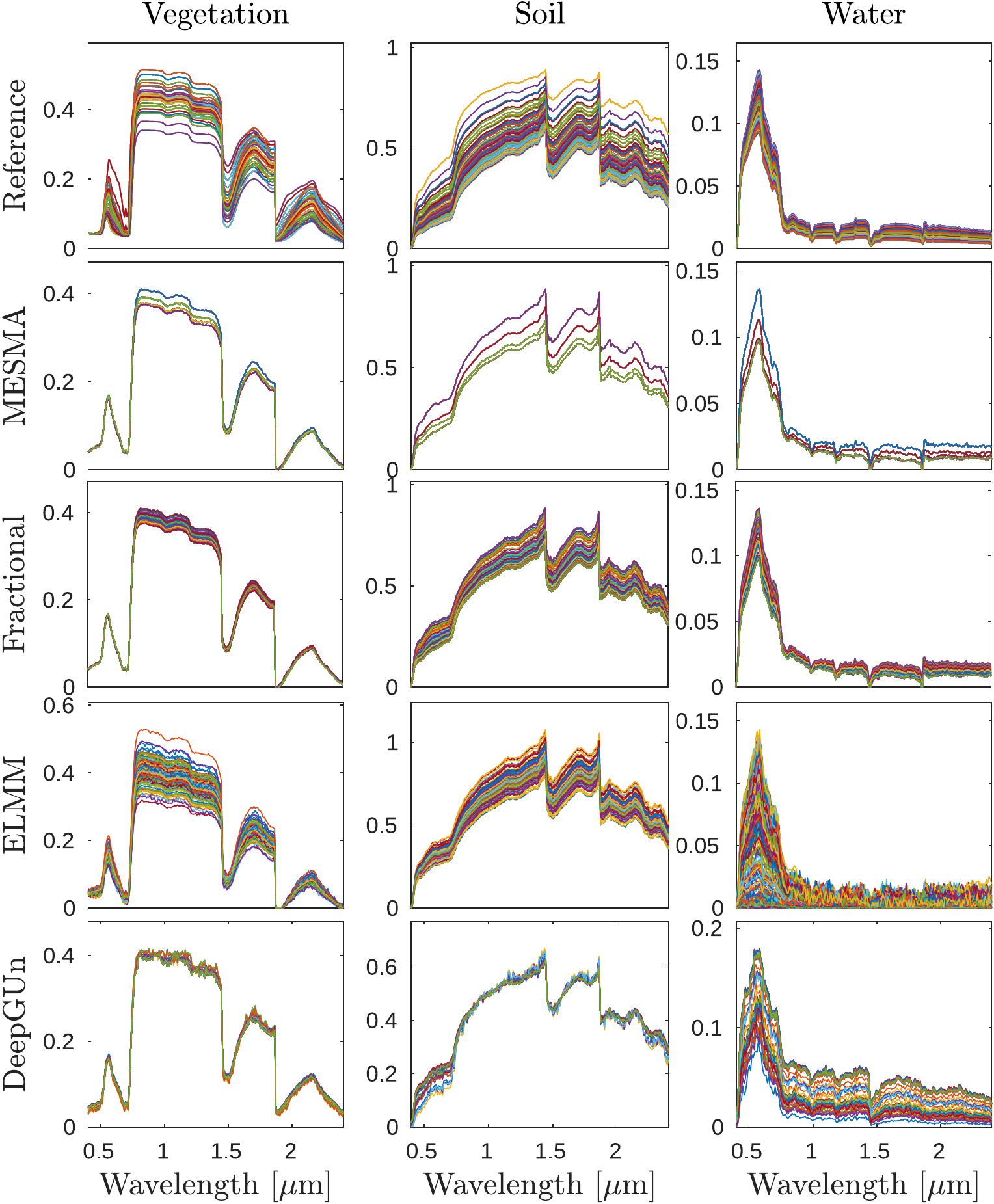}
    \caption{Spectral signatures returned by the algorithms that estimate the endmember spectra for each image pixel.}
    \label{fig:example_estim_endmembers}
\end{figure}

\subsubsection{\textbf{Discussion}}

The quantitative results are shown in Table~\ref{tab:results_synthData}, while the estimated abundance maps and endmembers are depicted in Figs.~\ref{fig:example_estim_abundances} and~\ref{fig:example_estim_endmembers}, respectively. Note that the $\operatorname{RMSE}_{\bM}$ and $\operatorname{SAM}_{\bM}$ are not available for the FCLS, RUSAL, NCM-E and for the BCM, since these algorithms do not estimate the spectral signatures of the endmembers present in each pixel of the image.
All methods that considered spectral variability led to better abundance reconstruction results than the FCLS baseline. In particular, the library-based methods (MESMA and fractional-based sparse SU) obtained a very good performance, which likely occurred due to the image-extracted spectral library accurately representing the typical EM variability contained in this scene. Moreover, sparse SU with fractional norms performed similarly and slightly better than MESMA.

The methods based on parametric EM models (ELMM and DeepGUn) also led to considerable improvements when compared to FCLS, especially considering that the EMs are estimated directly from the image. The EM-model-free method (RUSAL), which takes general variability and mismodelings into account, also provided an improvement over FCLS, albeit smaller when compared to ELMM and DeepGUn. However, the sensitivity of these techniques to the selection of the regularization parameters can negatively impact their performance when Monte Carlo simulations are considered.

Among the Bayesian methods (NCM-E and BCM), BCM provided a considerable performance improvement over FCLS, especially when taking into account the unsupervised nature of the method (i.e., no parameter has to be adjusted). The \mbox{NCM-E} results, on the other hand, were virtually identical to those of the FCLS, which indicates that the isotropic Gaussian EM hypothesis may not be appropriate for this dataset.

The performance of the different methods can be visually distinguished in Fig.~\ref{fig:example_estim_abundances}, especially from the soil endmember, in which the similarity between the reconstructions and the reference abundance maps reflects the general behavior of the quantitative results from Table~\ref{tab:results_synthData}.

The endmember reconstruction metrics in Table~\ref{tab:results_synthData} indicate that the EMs selected by library based approaches (MESMA and fractional sparse SU) are close to the reference ones, especially in terms of $\text{SAM}_{\bM}$, while the model-based approaches (ELMM and DeepGUn) provided slightly worse results in general, except for DeepGUn's $\text{RMSE}_{\bM}$. The visual assessment of the estimated signatures in Fig.~\ref{fig:example_estim_endmembers} shows an interesting pattern, since despite the quantitative metrics, the amount of variability (i.e., the variance) estimated by ELMM seems closer to the reference spectra. This shows that identifying the correct spectral signatures present in each pixel is very difficult.

We also note that smaller image reconstruction errors $\text{RMSE}_{\bY}$ did not correlate very well with better abundance estimation results. Since some SU methods that take spectral variability into account adopt flexible models, they can represent the hyperspectral image pixels in $\bY$ very closely without necessarily improving the abundance estimation.

The execution times show a considerable difference between the methods. Library-based approaches were able to run very fast (even for MESMA) since the spectral library contained few signatures. This shows that the construction of the library can significantly impact the run-time performance of these techniques. The methods based on parametric models (ELMM and DeepGUn) provided intermediate execution times, while RUSAL was very fast. Finally, Bayesian methods took the longest to run, with NCM-E taking significantly more than all the remaining techniques.

Finally, we note that this example is merely illustrative and not an in depth evaluation of these methods. Thus, their performance can be different for other datasets and scenarios.

\section{Discussion, Conclusions and Future Directions}
\label{sec:conclusions}

Significant advances have been made to mitigate spectral variability in SU during the last decade, encompassing contributions with both experimental and theoretical motivations. Recent work has, for instance, allowed spectral libraries to be directly extracted from observed hyperspectral images, provided more accurate or flexible models to represent the endmembers (e.g., in statistical or parametric methods), and included different kinds of \textit{a priori} external information in order to alleviate the ill-posedness of the problem, such as the locally correlated characteristics of the EMs and abundances. This was performed either explicitly, by means of regularization approaches or in the definition of statistical models, or implicitly in the design of the algorithms (e.g., in local SU). Other methods leveraged the spectral characteristics of EM variability to design improved algorithms (e.g., in spectral transformations or robust SU methods).

However, there is still a noticeable dependence between the the quality of the unmixing solutions and the necessary amount of user supervision in the algorithms. 
Many recent techniques need considerable or intricate tuning in order to reach their full potential, with a significant portion of algorithm design being left to the user.
The lack of more extensive data with reliable ground truths have also made the evaluation of the algorithms somewhat difficult.
In the following, we detail some aspects which we think deserve further consideration.

\begin{itemize}
    \item %
    As discussed above, one important research direction is to improve the robustness of the methods to the selection of their parameters, or to develop informed adjustment methodologies. This could be performed, for instance, by leveraging metadata (e.g., external classification maps) that is available in many applications. This point applies to the majority of SU algorithms reviewed in this paper, and would make those methods more readily employable as out-of-the-box solutions in practical scenarios.
    
    \item Most SU algorithms that address spectral variability depend strongly on spectral libraries or on reference endmember signatures known \textit{a priori} or extracted from the observed hyperspectral image. Improving the robustness of these methods to the selection of this data is important to guarantee a more reliable SU performance in practice.

    \item The vast majority of work reviewed in this paper \cmag{uses} the LMM to describe the interaction between incident light and the materials in the scene, even though nonlinear mixtures are common in many applications~\cite{Dobigeon-2014-ID322}. However, as shown in~\cite{drumetz2017relationshipsBilinearELMM}, a general nonlinear mixture model is closely related to a spatially varying version of the LMM, which indicates that linear unmixing with spectral variability is able to address the nonlinear unmixing problem to some extent. Nevertheless, the relationship between these two models deserves to be further investigated. Especially, deciding whether variations in the observed pixel spectra originate from spectral variability, from nonlinear interactions or from slight abundance variations can be very difficult.

    \item An aspect that induces difficulties to the evaluation of SU methods is the lack of more extensive data with reliable ground truth. However, there is no clear approach to reliably collect ground truth for abundance values. This problem is more pronounced when spectral variability is considered. Particularly, there is not a clearly agreed-upon protocol to generate realistic synthetic data. A larger, publicly available dataset would strengthen the validation of the methods.

    \item Although many ways have been proposed to model spectral variability, there is still a distinction between restrained models inspired by specific, concrete applications and mathematically flexible ones that aim for a more generic representation. Combining insight from the practical applications with a mathematically thorough treatment may lead to improved ways to represent spectral variability in a given scene.

    \item Many of the methods discussed in this paper rely, explicitly or implicitly, on the solution to complex, non-convex optimization problems which are often solved only approximately to achieve a computationally tractable algorithm. Investigating the use of more reliable approaches to solve those problems can help to evaluate the potential accuracy of the models by reducing the influence from the use of such approximations.

    \item Many algorithms (such as, e.g., MESMA and some statistical approaches) are computationally expensive and do not scale very well for large images. Considering the large amount of data currently in need of processing, it is important to have fast alternatives to solve this problem.

    \item %
    Traditional SU can be readily interpreted as a non-negative matrix factorization problem. This allows us to understand many of the limitations of the SU problem, as well as to identify conditions under which it can be solved exactly. However, such understanding is generally not available when endmember variability is considered, except for the particular case of illumination-based spectral variability~\cite{drumetz2019EMsDirectionalDataU_journal}. A deeper theoretical insight would be valuable to clearly define limiting conditions under which this problem can, or cannot be solved.

\end{itemize}

Initially motivated from Earth observation applications, spectral variability is now considered one of the main challenges of SU.
Although we have already seen a wealth of contributions from both application- and theoretically-oriented researchers, it is expected that the further exchange of ideas between these two areas will help to advance the field even further.

\bibliographystyle{IEEEtran}
\bibliography{references}

\end{document}